\DocumentMetadata{uncompress} 
\documentclass[
  10pt,
  twocolumn,
  twoside,
]{article}

\usepackage[numbers,round,comma,sort&compress]{natbib}

\usepackage[
  left=0.65in,
  right=0.65in,
  top=0.65in,
  bottom=0.65in,
]{geometry}

\usepackage{titlesec}
\titleformat{\section}{\raggedright\large\bfseries}{\thesection}{1em}{}
\titleformat{\subsection}{\raggedright\large}{\thesubsection}{1em}{}
\usepackage{ragged2e}

\usepackage[font=small,labelfont=bf]{caption}

\usepackage{tabularx}
\usepackage{longtable}
\usepackage{array}
\usepackage{array}
\newcolumntype{Y}{>{\raggedright\arraybackslash}X}

\usepackage{fancyhdr}
\usepackage[english]{babel}

\usepackage{amssymb}
\usepackage{amsmath}
\usepackage{newunicodechar}
\usepackage{amsfonts}  
\usepackage{array}
\usepackage{nameref}    
\usepackage{hyperref}   
\usepackage{placeins}

\usepackage{verbatim}

\usepackage{environ}

\usepackage{url}
\PassOptionsToPackage{hyphens}{url} 
\makeatletter
\g@addto@macro{\UrlBreaks}{\UrlOrds}
\makeatother

\usepackage[table]{xcolor}   
\usepackage{colortbl}        

\definecolor{goodblue}{RGB}{0, 91, 187}
\hypersetup{
  colorlinks=true,
  allcolors=goodblue,
  urlcolor=goodblue,
  citecolor=goodblue,
  pdfborder={0 0 0},
  breaklinks=true,
}

\usepackage[normalem]{ulem}

\usepackage{textcomp}

\usepackage[export]{adjustbox}

\makeatletter

\def\CT@@do@color{%
  \global\let\CT@do@color\relax
  \@tempdima\wd\z@
  \advance\@tempdima\@tempdimb
  \advance\@tempdima\@tempdimc
  \advance\@tempdimb\tabcolsep
  \advance\@tempdimc\tabcolsep
  \advance\@tempdima2\tabcolsep
  \kern-\@tempdimb
  \leaders\vrule
  \hskip\@tempdima\@plus  1fill
  \kern-\@tempdimc
  \hskip-\wd\z@ \@plus -1fill }
\makeatother

\newcommand{\done}[1]{}

\usepackage{titlesec}
\titleformat*{\paragraph}{\bfseries}

\usepackage{changepage}

\usepackage{graphicx}
\usepackage{epsfig}
\usepackage{verbatim}
\usepackage{enumerate}
\usepackage{enumitem}

\usepackage{ifthen}

\usepackage{longtable}

\usepackage{mathtools}

\usepackage{tabularx}

\newboolean{twocolswitch}

\newcommand{\PreserveBackslash}[1]{\let\temp=\\#1\let\\=\temp}

\newcommand{\sindex}[1]{}
\newcommand{\nindex}[1]{}

\newcommand{\www}[1]{\url{#1}}

\usepackage{lettrine}

\usepackage{changepage}

\usepackage{footmisc}

\usepackage{refcount}

\makeatletter
\newcommand{\footnotemarklabel}[1]{%
  \protect\footnotemark
  \begingroup
    \edef\@currentlabel{\thefootnote}
    \label{#1}%
  \endgroup
}
\makeatother

\renewcommand{\footnoterule}{%
  \kern 3pt
  \hrule width 0.4\columnwidth height 0.4pt
  \kern 6pt
}
\NewEnviron{excerpt}{
    \begin{quote}
      \medskip
      \BODY
      \medskip
    \end{quote}
}

\NewEnviron{editnote}{
    \begin{quote}
      \color{rose}
      $\blacksquare$
      \BODY
      \medskip
    \end{quote}
}

\newcommand{\command}[1]{
  \lstinline[language={[LaTeX]TeX},basicstyle=\ttfamily]{#1}
}

\newcommand{\editbox}[2]{
}

\newcommand{\editboxwithlatex}[2]{
}

\usepackage{fancyvrb}

\usepackage{tcolorbox}
\tcbuselibrary{skins,breakable,listings,breakable}

\usepackage{tikz}
\usetikzlibrary{shapes}

\tikzstyle{mybox} = [draw=lightblue!70, fill=lightblue!7, very thick,
    rectangle, rounded corners, inner sep=10pt, inner ysep=20pt]

\tikzstyle{editortitle} =[draw=archetyperowcoloralt, fill=archetyperowcoloralt, text=black]

\newcommand\Loadedframemethod{default}
\usepackage[framemethod=\Loadedframemethod]{mdframed}

\mdfsetup{skipabove=\topskip,skipbelow=\topskip}

\tikzstyle{loglinetitle} =[draw=icedark, fill=icemedium!50, text=black]

\newenvironment{loglinebox}[1][]{

  \ifstrempty{#1}%
  {\mdfsetup{%
    frametitle={%
       \tikz[baseline=(current bounding box.east),outer sep=0pt]
        \node[loglinetitle, anchor=east,rectangle]
        {\strut~~#1~~\strut};}}
  }%
  {\mdfsetup{%
     frametitle={%
       \tikz[baseline=(current bounding box.east),outer sep=0pt]
        \node[loglinetitle,anchor=east,rectangle]
        {\strut~~#1~~\strut};}}%
   }%
   \mdfsetup{innertopmargin=5pt,linecolor=icedark,%
             linewidth=0.5pt,topline=true,
             frametitleaboveskip=\dimexpr-\ht\strutbox\relax,}
   \begin{mdframed}[backgroundcolor=icelight,nobreak=true]\relax%
     \raggedright
}{\end{mdframed}}

\tikzstyle{abstracttitle} =[draw=magmadark!75, fill=magmamedium!75, text=black]

\newenvironment{abstractbox}[1][]{

  \ifstrempty{#1}%
  {\mdfsetup{%
    frametitle={%
       \tikz[baseline=(current bounding box.east),outer sep=0pt]
        \node[abstracttitle, anchor=east,rectangle]
        {\strut~~#1~~\strut};}}
  }%
  {\mdfsetup{%
     frametitle={%
       \tikz[baseline=(current bounding box.east),outer sep=0pt]
        \node[abstracttitle,anchor=east,rectangle]
        {\strut~~#1~~\strut};}}%
   }%
   \mdfsetup{innertopmargin=5pt,linecolor=magmadark,%
             linewidth=0.5pt,topline=true,
             frametitleaboveskip=\dimexpr-\ht\strutbox\relax,}
   \begin{mdframed}[backgroundcolor=magmalight,nobreak=true]\relax%
     \raggedright
}{\end{mdframed}}

\tikzstyle{infotitle} =[draw=darkgrey, fill=lightgrey!50, text=black]

\newenvironment{infobox}[1][]{

  \ifstrempty{#1}%
  {\mdfsetup{%
    frametitle={%
       \tikz[baseline=(current bounding box.east),outer sep=0pt]
        \node[infotitle, anchor=east,rectangle]
        {\strut~~#1~~\strut};}}
  }%
  {\mdfsetup{%
     frametitle={%
       \tikz[baseline=(current bounding box.east),outer sep=0pt]
        \node[infotitle,anchor=east,rectangle]
        {\strut~~#1~~\strut};}}%
   }%
   \mdfsetup{innertopmargin=5pt,linecolor=grey,%
             linewidth=0.5pt,topline=true,
             frametitleaboveskip=\dimexpr-\ht\strutbox\relax,}
   \begin{mdframed}[backgroundcolor=lightgrey!25,nobreak=true]\relax%
     \raggedright
}{\end{mdframed}}

\tikzstyle{essencetitle} = [draw=magmadark!75, fill=magmamedium!75, text=black]

\usepackage{enumitem}

\tikzstyle{changelogtitle} =[draw=darkgrey, fill=lightgrey!50, text=black]

\usepackage[table]{xcolor}

\definecolor{olivegreen}{rgb}{0.33333,.41961,0.18431}
\definecolor{forestgreen}{rgb}{0.13333,.5451,0.13333}

\definecolor{lightgrey}{rgb}{0.7,0.7,0.7}
\definecolor{verylightgrey}{rgb}{0.90,0.90,0.90}
\definecolor{veryverylightgrey}{rgb}{0.95,0.95,0.95}
\definecolor{grey}{rgb}{0.5,0.5,0.5}
\definecolor{darkgrey}{rgb}{0.3,0.3,0.3}
\definecolor{verydarkgrey}{rgb}{0.15,0.15,0.15}

\definecolor{headerblue}{HTML}{33367E}
\definecolor{unitednationsblue}{HTML}{4D88FF}

\definecolor{charcoal}{HTML}{36454F}
\definecolor{cinerous}{HTML}{98817B}
\definecolor{feldgrau}{HTML}{4D5D53}
\definecolor{glaucous}{HTML}{6082B6}
\definecolor{arsenic}{HTML}{3B444B}
\definecolor{xanadu}{HTML}{738678}

\definecolor{firebrick}{HTML}{B22222}
\definecolor{orangered}{HTML}{FF4500}
\definecolor{tomato}{HTML}{FF6347}

\definecolor{orange}{RGB}{255,116,0}

\definecolor{purpletaupe}{HTML}{3B444B}

\definecolor{rose}{HTML}{E3242B}

\colorlet{editnotecolor}{rose}

\definecolor{headerorange}{RGB}{255,116,0}
\definecolor{headergray}{RGB}{230,230,230}

\definecolor{headerpop}{RGB}{230,230,230}

\definecolor{magmalight}{RGB}{252,251,195}
\definecolor{magmalightalt}{RGB}{250,240,184}
\definecolor{magmamedium}{RGB}{245,200,146}
\definecolor{magmadark}{RGB}{224,106,98}

\definecolor{icelight}{RGB}{223,242,244}
\definecolor{icelightalt}{RGB}{189,222,226}
\definecolor{icemedium}{RGB}{132,184,204}
\definecolor{icedark}{RGB}{103,153,191}

\definecolor{traitrowcolor}{RGB}{223,242,244}
\definecolor{traitrowcoloralt}{RGB}{189,222,226}

\definecolor{characterrowcolor}{RGB}{252,251,195}
\definecolor{characterrowcoloralt}{RGB}{250,240,184}

\definecolor{archetyperowcolor}{RGB}{255,213,212} 
\definecolor{archetyperowcoloralt}{RGB}{255,182,179} 

\definecolor{datasetrowcolor}{RGB}{232,244,234}
\definecolor{datasetrowcoloralt}{RGB}{210,231,214}

\newcommand{\Ncharacters}{2000}
\newcommand{\Ntraits}{464}
\newcommand{\Nstories}{341}

\newcommand{\Ncharactersbase}{2000}

\newcommand{\Ncharactersmain}{2000}
\newcommand{\Ntraitsmain}{464}
\newcommand{\Nstoriesmain}{341}

\newcommand{\Ncharactersmainone}{0800}
\newcommand{\Ntraitsmainone}{235}
\newcommand{\Nstoriesmainone}{090}

\newcommand{\Ncharactersmaintwo}{1600}
\newcommand{\Ntraitsmaintwo}{364}
\newcommand{\Nstoriesmaintwo}{241}

\newcommand{\onlinesiteshort}{compstorylab.org/archetypometrics}
\newcommand{\onlinesite}{https://\onlinesiteshort}

\newcommand{\semdiffsign}{\Leftrightarrow}
\newcommand{\semdiffsignleft}{\Leftarrow}
\newcommand{\semdiffsignright}{\Rightarrow}

\newcommand{\semdiff}[2]{\{#1\,$\semdiffsign$\,#2\}}

\newcommand{\semdiffright}[2]{\{#1\,$\semdiffsignright$\,#2\}}

\newcommand{\semdiffbold}[2]{\{\textbf{#1}\,$\semdiffsign$\,\textbf{#2}\}}
\newcommand{\semdiffboldleft}[2]{\{\textbf{#1}\,$\semdiffsignleft$\,#2\}}
\newcommand{\semdiffboldright}[2]{\{#1\,$\semdiffsignright$\,\textbf{#2}\}}

\newcommand{\semdiffmath}[2]{\{\textnormal{\textbf{#1}}\!\semdiffsign\!{\textnormal{\textbf{#2}}\}}}
\newcommand{\semdiffmathleft}[2]{\{\textnormal{\textbf{#1}}\!\semdiffsignleft\!{\textnormal{#2}\}}}
\newcommand{\semdiffmathright}[2]{\{\textnormal{#1}\!\semdiffsignright\!{\textnormal{\textbf{#2}}\}}}

\newcommand{\datasetsymbol}{\colorbox{datasetrowcolor}{\textcolor{black}{\stackanchor{\scalebox{\babaisyouboxscale}{DA}}{\scalebox{\babaisyouboxscale}{TA}}}}}

\newcommand{\dataset}[1]{\mbox{\datasetsymbol}_{#1}}

\newcommand{\datasetbase}[1]{
    \IfEqCase{#1}{
        {0800}{\datasetsymbol{1}}
        {1600}{\datasetsymbol{2}}
        {2000}{\datasetsymbol{3}}
    }[\PackageError{datasetbase}{Undefined option to datasetbase: #1}{}]%
}%

\newcommand{\datasetNcharacters}[1]{
    \IfEqCase{#1}{
        {1}{800}
        {2}{1600}
        {3}{2000}
    }[\PackageError{datasetNcharacters}{Undefined option to datasetNcharacters: #1}{}]%
}%

\newcommand{\datasetNtraits}[1]{
    \IfEqCase{#1}{
        {1}{235}
        {2}{364}
        {3}{464}
    }[\PackageError{datasetNtraits}{Undefined option to datasetNtraits: #1}{}]%
}%

\newcommand{\datasetNstories}[1]{
    \IfEqCase{#1}{
        {1}{90}
        {2}{241}
        {3}{341}
    }[\PackageError{datasetNstories}{Undefined option to datasetNstories: #1}{}]%
}%

\newcommand{\padzero}[1]{\ifnum #1 < 10 0\fi #1}

\usepackage{siunitx}
\usepackage{multirow}
\usepackage{booktabs} 

\newcommand\zeropad[2]{%
  \ifnum#2<0\relax%
    {\ensuremath-}\zeropadA{#1}{\the\numexpr#2*-1\relax}%
  \else%
    \zeropadA{#1}{#2}%
  \fi%
}
\def\zeropadA#1#2{%
  \ifnum1#2<1#1
    \zeropadA{#1}{0#2}%
  \else%
    #2%
  \fi%
}

\usepackage{xstring}

\newcommand{\archetypesemdiff}[1]{
    \IfEqCase{#1}{
        {1}{\semdiffbold{\archetypelinkbase{Fool}}{\archetypelinkbase{Hero}}}                  
        {2}{\semdiffbold{\archetypelinkbase{Angel}}{\archetypelinkbase{Demon}}}                
        {3}{\semdiffbold{\archetypelinkbase{Traditionalist}}{\archetypelinkbase{Adventurer}}}  
        {4}{\semdiffbold{\archetypelinksimple{Lone-Wolf}{Lone~Wolf}}{\archetypelinkbase{Diva}}}
        {5}{\semdiffbold{\archetypelinkbase{Outcast}}{\archetypelinkbase{Sophisticate}}}       
        {6}{\semdiffbold{\archetypelinkbase{Brute}}{\archetypelinkbase{Geek}}}                 
    }[\PackageError{archetypesemdiff}{Undefined option to archetypesemdiff: #1}{}]%
}%

\newcommand{\archetypesemdiffleft}[1]{
    \IfEqCase{#1}{
        {1}{\semdiffboldleft{\archetypelinkbase{Fool}}{\archetypelinkbase{Hero}}}                  
        {2}{\semdiffboldleft{\archetypelinkbase{Angel}}{\archetypelinkbase{Demon}}}                
        {3}{\semdiffboldleft{\archetypelinkbase{Traditionalist}}{\archetypelinkbase{Adventurer}}}  
        {4}{\semdiffboldleft{\archetypelinksimple{Lone-Wolf}{Lone~Wolf}}{\archetypelinkbase{Diva}}}
        {5}{\semdiffboldleft{\archetypelinkbase{Outcast}}{\archetypelinkbase{Sophisticate}}}       
        {6}{\semdiffboldleft{\archetypelinkbase{Brute}}{\archetypelinkbase{Geek}}}                 
    }[\PackageError{archetypesemdiff}{Undefined option to archetypesemdiff: #1}{}]%
}%

\newcommand{\archetypesemdiffright}[1]{
    \IfEqCase{#1}{
        {1}{\semdiffboldright{\archetypelinkbase{Fool}}{\archetypelinkbase{Hero}}}                  
        {2}{\semdiffboldright{\archetypelinkbase{Angel}}{\archetypelinkbase{Demon}}}                
        {3}{\semdiffboldright{\archetypelinkbase{Traditionalist}}{\archetypelinkbase{Adventurer}}}  
        {4}{\semdiffboldright{\archetypelinksimple{Lone-Wolf}{Lone~Wolf}}{\archetypelinkbase{Diva}}}
        {5}{\semdiffboldright{\archetypelinkbase{Outcast}}{\archetypelinkbase{Sophisticate}}}       
        {6}{\semdiffboldright{\archetypelinkbase{Brute}}{\archetypelinkbase{Geek}}}                 
    }[\PackageError{archetypesemdiff}{Undefined option to archetypesemdiff: #1}{}]%
}%

\newcommand{\archetypesemdiffmath}[1]{
  \IfEqCase{#1}{
    {1}{\semdiffmath{\archetypelinkbase{Fool}}{\archetypelinkbase{Hero}}}
    {2}{\semdiffmath{\archetypelinkbase{Angel}}{\archetypelinkbase{Demon}}}
    {3}{\semdiffmath{\archetypelinkbase{Traditionalist}}{\archetypelinkbase{Adventurer}}}
    {4}{\semdiffmath{\archetypelinksimple{Lone-Wolf}{Lone~Wolf}}{\archetypelinkbase{Diva}}}
    {5}{\semdiffmath{\archetypelinkbase{Outcast}}{\archetypelinkbase{Sophisticate}}}
    {6}{\semdiffmath{\archetypelinkbase{Brute}}{\archetypelinkbase{Geek}}}                 
  }[\PackageError{archetypesemdiffmath}{Undefined option to archetypesemdiffmath: #1}{}]%
}%

\newcommand{\archetypesemdiffmathswap}[1]{
  \IfEqCase{#1}{
    {1}{\semdiffmath{\archetypelinkbase{Hero}}{\archetypelinkbase{Fool}}}
    {2}{\semdiffmath{\archetypelinkbase{Demon}}{\archetypelinkbase{Angel}}}
    {3}{\semdiffmath{\archetypelinkbase{Adventurer}}{\archetypelinkbase{Traditionalist}}}
    {4}{\semdiffmath{\archetypelinkbase{Diva}}{\archetypelinksimple{Lone-Wolf}{Lone~Wolf}}}
    {5}{\semdiffmath{\archetypelinkbase{Sophisticate}}{\archetypelinkbase{Outcast}}}
    {6}{\semdiffmath{{\archetypelinkbase{Geek}}\archetypelinkbase{Brute}}}
  }[\PackageError{archetypesemdiffmathswap}{Undefined option to archetypesemdiffmathswap: #1}{}]%
}%

\newcommand{\archetypesemdiffmathleft}[1]{
  \IfEqCase{#1}{
    {1}{\semdiffmathleft{\archetypelinkbase{Fool}}{\archetypelinkbase{Hero}}}                  
    {2}{\semdiffmathleft{\archetypelinkbase{Angel}}{\archetypelinkbase{Demon}}}                
    {3}{\semdiffmathleft{\archetypelinkbase{Traditionalist}}{\archetypelinkbase{Adventurer}}}  
    {4}{\semdiffmathleft{\archetypelinksimple{Lone-Wolf}{Lone~Wolf}}{\archetypelinkbase{Diva}}}
    {5}{\semdiffmathleft{\archetypelinkbase{Outcast}}{\archetypelinkbase{Sophisticate}}}       
    {6}{\semdiffmathleft{\archetypelinkbase{Brute}}{\archetypelinkbase{Geek}}}                 
  }[\PackageError{archetypesemdiffmathleft}{Undefined option to archetypesemdiffmathleft: #1}{}]%
}%

\newcommand{\archetypesemdiffmathright}[1]{
  \IfEqCase{#1}{
    {1}{\semdiffmathright{\archetypelinkbase{Fool}}{\archetypelinkbase{Hero}}}                  
    {2}{\semdiffmathright{\archetypelinkbase{Angel}}{\archetypelinkbase{Demon}}}                
    {3}{\semdiffmathright{\archetypelinkbase{Traditionalist}}{\archetypelinkbase{Adventurer}}}  
    {4}{\semdiffmathright{\archetypelinksimple{Lone-Wolf}{Lone~Wolf}}{\archetypelinkbase{Diva}}}
    {5}{\semdiffmathright{\archetypelinkbase{Outcast}}{\archetypelinkbase{Sophisticate}}}       
    {6}{\semdiffmathright{\archetypelinkbase{Brute}}{\archetypelinkbase{Geek}}}                 
  }[\PackageError{archetypesemdiffmathright}{Undefined option to archetypesemdiffmathright: #1}{}]%
}%

\newcommand{\essentialsemdiff}[1]{
  \IfEqCase{#1}{
    {1}{\semdiff{\essentialtraitlinknegative{1}{weak/incompetent/lazy/stupid}}{\essentialtraitlinkpositive{1}{powerful/capable/purposeful/intelligent}}}
    {2}{\semdiff{\essentialtraitlinknegative{2}{safe/pure/virtuous/humble}}{\essentialtraitlinkpositive{2}{dangerous/depraved/corrupt/arrogant}}}
    {3}{\semdiff{\essentialtraitlinknegative{3}{serious/predictable/humorless/uncreative}}{\essentialtraitlinkpositive{3}{playful/unpredictable/funny/creative}}}
    {4}{\semdiff{\essentialtraitlinknegative{4}{rugged/stoic/independent/blunt}}{\essentialtraitlinkpositive{4}{refined/dramatic/dependent/sensitive}}}
    {5}{\semdiff{\essentialtraitlinknegative{5}{unlucky/unsophisticated/traumatized}}{\essentialtraitlinkpositive{5}{fortunate/sophisticated/confident}}}
    {6}{\semdiff{\essentialtraitlinknegative{6}{physical/mainstream/simple-minded}}{\essentialtraitlinkpositive{6}{intellectual/weird/complex}}}
    {7}{\semdiff{\essentialtraitlinknegative{7}{dramatic/attractive/young}}{\essentialtraitlinkpositive{7}{comedic/ugly/old}}}
    {8}{\semdiff{\essentialtraitlinknegative{8}{spiritual/rural/historical}}{\essentialtraitlinkpositive{8}{skeptical/urban/modern}}}
    {9}{\semdiff{\essentialtraitlinknegative{9}{old/historical/low-tempo}}{\essentialtraitlinkpositive{9}{young/modern/high-tempo}}}
    {10}{\semdiff{\essentialtraitlinknegative{10}{feminine/luddite}}{\essentialtraitlinkpositive{10}{masculine/technophile}}}
    {11}{\semdiff{\essentialtraitlinknegative{11}{secondary/street-wise}}{\essentialtraitlinkpositive{11}{primary/sheltered}}}
  }[\PackageError{essentialsemdiff}{Undefined option to essentialsemdiff: #1}{}]%
}%

\newcommand{\essentialsemdiffloose}[1]{
  \IfEqCase{#1}{
    {1}{\semdiff{\essentialtraitlinknegative{1}{weak, incompetent, lazy, stupid}}{\essentialtraitlinkpositive{1}{powerful, capable, purposeful, intelligent}}}
    {2}{\semdiff{\essentialtraitlinknegative{2}{safe, pure, virtuous, humble}}{\essentialtraitlinkpositive{2}{dangerous, depraved, corrupt, arrogant}}}
    {3}{\semdiff{\essentialtraitlinknegative{3}{serious, predictable, humorless, uncreative}}{\essentialtraitlinkpositive{3}{playful, unpredictable, funny, creative}}}
    {4}{\semdiff{\essentialtraitlinknegative{4}{rugged, stoic, independent, blunt}}{\essentialtraitlinkpositive{4}{refined, dramatic, dependent, sensitive}}}
    {5}{\semdiff{\essentialtraitlinknegative{5}{unlucky, unsophisticated, traumatized}}{\essentialtraitlinkpositive{5}{fortunate, sophisticated, confident}}}
    {6}{\semdiff{\essentialtraitlinknegative{6}{physical, mainstream, simple-minded}}{\essentialtraitlinkpositive{6}{intellectual, weird, complex}}}
    {7}{\semdiff{\essentialtraitlinknegative{7}{dramatic, attractive, young}}{\essentialtraitlinkpositive{7}{comedic, ugly, old}}}
    {8}{\semdiff{\essentialtraitlinknegative{8}{spiritual, rural, historical}}{\essentialtraitlinkpositive{8}{skeptical, urban, modern}}}
    {9}{\semdiff{\essentialtraitlinknegative{9}{old, historical, low-tempo}}{\essentialtraitlinkpositive{9}{young, modern, high-tempo}}}
    {10}{\semdiff{\essentialtraitlinknegative{10}{feminine, luddite}}{\essentialtraitlinkpositive{10}{masculine, technophile}}}
    {11}{\semdiff{\essentialtraitlinknegative{11}{secondary, street-wise}}{\essentialtraitlinkpositive{11}{primary, sheltered}}}
  }[\PackageError{essentialsemdiffloose}{Undefined option to essentialsemdiffloose: #1}{}]%
}%

\newcommand{\essentialsemdifflooseleft}[1]{
    \IfEqCase{#1}{
        {1}{\semdiff{\textbf{\essentialtraitlinknegative{1}{weak, incompetent, lazy, stupid}}}{\essentialtraitlinkpositive{1}{powerful, capable, purposeful, intelligent}}}
        {2}{\semdiff{\textbf{\essentialtraitlinknegative{2}{safe, pure, virtuous, humble}}}{\essentialtraitlinkpositive{2}{dangerous, depraved, corrupt, arrogant}}}
        {3}{\semdiff{\textbf{\essentialtraitlinknegative{3}{serious, predictable, humorless, uncreative}}}{\essentialtraitlinkpositive{3}{playful, unpredictable, funny, creative}}}
        {4}{\semdiff{\textbf{\essentialtraitlinknegative{4}{rugged, stoic, independent, blunt}}}{\essentialtraitlinkpositive{4}{refined, dramatic, dependent, sensitive}}}
        {5}{\semdiff{\textbf{\essentialtraitlinknegative{5}{unlucky, unsophisticated, traumatized}}}{\essentialtraitlinkpositive{5}{fortunate, sophisticated, confident}}}
        {6}{\semdiff{\textbf{\essentialtraitlinknegative{6}{physical, mainstream, simple-minded}}}{\essentialtraitlinkpositive{6}{intellectual, weird, complex}}}
        {7}{\semdiff{\textbf{\essentialtraitlinknegative{7}{dramatic, attractive, young}}}{\essentialtraitlinkpositive{7}{comedic, ugly, old}}}
        {8}{\semdiff{\textbf{\essentialtraitlinknegative{8}{spiritual, rural, historical}}}{\essentialtraitlinkpositive{8}{skeptical, urban, modern}}}
        {9}{\semdiff{\textbf{\essentialtraitlinknegative{9}{old, historical, low-tempo}}}{\essentialtraitlinkpositive{9}{young, modern, high-tempo}}}
        {10}{\semdiff{\textbf{\essentialtraitlinknegative{10}{feminine, luddite}}}{\essentialtraitlinkpositive{10}{masculine, technophile}}}
        {11}{\semdiff{\textbf{\essentialtraitlinknegative{11}{secondary, street-wise}}}{\essentialtraitlinkpositive{11}{primary, sheltered}}}
    }[\PackageError{essentialsemdifflooseleft}{Undefined option to essentialsemdifflooseleft: #1}{}]%
 }%

\newcommand{\essentialsemdifflooseright}[1]{
  \IfEqCase{#1}{
    {1}{\semdiff{\essentialtraitlinknegative{1}{weak, incompetent, lazy, stupid}}{\textbf{\essentialtraitlinkpositive{1}{powerful, capable, purposeful, intelligent}}}}
    {2}{\semdiff{\essentialtraitlinknegative{2}{safe, pure, virtuous, humble}}{\textbf{\essentialtraitlinkpositive{2}{dangerous, depraved, corrupt, arrogant}}}}
    {3}{\semdiff{\essentialtraitlinknegative{3}{serious, predictable, humorless, uncreative}}{\textbf{\essentialtraitlinkpositive{3}{playful, unpredictable, funny, creative}}}}
    {4}{\semdiff{\essentialtraitlinknegative{4}{rugged, stoic, independent, blunt}}{\textbf{\essentialtraitlinkpositive{4}{refined, dramatic, dependent, sensitive}}}}
    {5}{\semdiff{\essentialtraitlinknegative{5}{unlucky, unsophisticated, traumatized}}{\textbf{\essentialtraitlinkpositive{5}{fortunate, sophisticated, confident}}}}
    {6}{\semdiff{\essentialtraitlinknegative{6}{physical, mainstream, simple-minded}}{\textbf{\essentialtraitlinkpositive{6}{intellectual, weird, complex}}}}
    {7}{\semdiff{\essentialtraitlinknegative{7}{dramatic, attractive, young}}{\textbf{\essentialtraitlinkpositive{7}{comedic, ugly, old}}}}
    {8}{\semdiff{\essentialtraitlinknegative{8}{spiritual, historical, rural}}{\textbf{\essentialtraitlinkpositive{8}{skeptical, urban, modern}}}}
    {9}{\semdiff{\essentialtraitlinknegative{9}{old, historical, low-tempo}}{\textbf{\essentialtraitlinkpositive{9}{young, modern, high-tempo}}}}
    {10}{\semdiff{\essentialtraitlinknegative{10}{feminine, luddite}}{\textbf{\essentialtraitlinkpositive{10}{masculine, technophile}}}}
    {11}{\semdiff{\essentialtraitlinknegative{11}{secondary, street-wise}}{\textbf{\essentialtraitlinkpositive{11}{primary, sheltered}}}}
  }[\PackageError{essentialsemdifflooseright}{Undefined option to essentialsemdifflooseright: #1}{}]%
}%

\newcommand{\essentialsemdiffmathleft}[1]{
    \IfEqCase{#1}{
        {1}{\semdiffmathleft{weak/incompetent/lazy/stupid}{powerful/capable/purposeful/intelligent}}
        {2}{\semdiffmathleft{safe/pure/virtuous/humble}{dangerous/depraved/corrupt/arrogant}}
        {3}{\semdiffmathleft{serious/predictable/humorless/uncreative}{playful/unpredictable/funny/creative}}
        {4}{\semdiffmathleft{rugged/stoic/independent/blunt}{refined/dramatic/dependent/sensitive}}
        {5}{\semdiffmathleft{unlucky/unsophisticated/traumatized}{fortunate/sophisticated/confident}}
        {6}{\semdiffmathleft{physical/mainstream/simple-minded}{intellectual/weird/complex}}
        {7}{\semdiffmathleft{dramatic/attractive/young}{comedic/ugly/old}}
        {8}{\semdiffmathleft{spiritual/rural/historical}{skeptical/urban/modern}}
        {9}{\semdiffmathleft{old/historical/low-tempo}{young/modern/high-tempo}}
        {10}{\semdiffmathleft{feminine/luddite}{masculine/technophile}}
        {11}{\semdiffmathleft{secondary/street-wise}{primary/sheltered}}
    }[\PackageError{essentialsemdiffmathleft}{Undefined option to essentialsemdiffmathleft: #1}{}]%
}%

\newcommand{\essentialsemdiffmathright}[1]{
    \IfEqCase{#1}{
        {1}{\semdiffmathright{weak/incompetent/lazy/stupid}{powerful/capable/purposeful/intelligent}}
        {2}{\semdiffmathright{safe/pure/virtuous/humble}{dangerous/depraved/corrupt/arrogant}}
        {3}{\semdiffmathright{serious/predictable/humorless/uncreative}{playful/unpredictable/funny/creative}}
        {4}{\semdiffmathright{rugged/stoic/independent/blunt}{refined/dramatic/dependent/sensitive}}
        {5}{\semdiffmathright{unlucky/unsophisticated/traumatized}{fortunate/sophisticated/confident}}
        {6}{\semdiffmathright{physical/mainstream/simple-minded}{intellectual/weird/complex}}
        {7}{\semdiffmathright{dramatic/attractive/young}{comedic/ugly/old}}
        {8}{\semdiffmathright{spiritual/rural/historical}{skeptical/urban/modern}}
        {9}{\semdiffmathright{old/historical/low-tempo}{young/modern/high-tempo}}
        {10}{\semdiffmathright{feminine/luddite}{masculine/technophile}}
        {11}{\semdiffmathright{secondary/street-wise}{primary/sheltered}}
    }[\PackageError{essentialsemdiffmathright}{Undefined option to essentialsemdiffmathright: #1}{}]%
}%

\newcommand{\essentialsemdiffmath}[1]{
    \IfEqCase{#1}{
        {1}{\semdiffmath{weak/incompetent/lazy/stupid}{powerful/capable/purposeful/intelligent}}
        {2}{\semdiffmath{safe/pure/virtuous/humble}{dangerous/depraved/corrupt/arrogant}}
        {3}{\semdiffmath{serious/predictable/humorless/uncreative}{playful/unpredictable/funny/creative}}
        {4}{\semdiffmath{rugged/stoic/independent/blunt}{refined/dramatic/dependent/sensitive}}
        {5}{\semdiffmath{unlucky/unsophisticated/traumatized}{fortunate/sophisticated/confident}}
        {6}{\semdiffmath{physical/mainstream/simple-minded}{intellectual/weird/complex}}
        {7}{\semdiffmath{dramatic/attractive/young}{comedic/ugly/old}}
        {8}{\semdiffmath{spiritual/rural/historical}{skeptical/urban/modern}}
        {9}{\semdiffmath{old/historical/low-tempo}{young/modern/high-tempo}}
        {10}{\semdiffmath{feminine/luddite}{masculine/technophile}}
        {11}{\semdiffmath{secondary/street-wise}{primary/sheltered}}
    }[\PackageError{essentialsemdiffmath}{Undefined option to essentialsemdiffmath: #1}{}]%
}%

\newcommand{\ousiometricsemdiff}[1]{
    \IfEqCase{#1}{
        {1}{\semdiffbold{weak}{powerful}}
        {2}{\semdiffbold{safe}{dangerous}}
        {3}{\semdiffbold{structured}{unstructured}}
    }[\PackageError{ousiometricsemdiff}{Undefined option to ousiometricsemdiff: #1}{}]%
}%

\newcommand{\ousiometricsemdiffmath}[1]{
    \IfEqCase{#1}{
        {1}{\semdiffmath{weak}{powerful}}
        {2}{\semdiffmath{safe}{dangerous}}
        {3}{\semdiffmath{structured}{unstructured}}
    }[\PackageError{ousiometricsemdiffmath}{Undefined option to ousiometricsemdiffmath: #1}{}]%
}%

\newcommand{\ousiometricsemdiffmathleft}[1]{
    \IfEqCase{#1}{
        {1}{\semdiffmathleft{weak}{powerful}}
        {2}{\semdiffmathleft{safe}{dangerous}}
        {3}{\semdiffmathleft{structured}{unstructured}}
    }[\PackageError{ousiometricsemdiffmathleft}{Undefined option to ousiometricsemdiffmathleft: #1}{}]%
}%

\newcommand{\ousiometricsemdiffmathright}[1]{
    \IfEqCase{#1}{
        {1}{\semdiffmathright{weak}{powerful}}
        {2}{\semdiffmathright{safe}{dangerous}}
        {3}{\semdiffmathright{structured}{unstructured}}
    }[\PackageError{ousiometricsemdiffmathright}{Undefined option to ousiometricsemdiffmathright: #1}{}]%
}%

\newcommand{\dimensiontype}[1]{
    \IfEqCase{#1}{
        {1}{Primary archetype}
        {2}{Primary archetype}
        {3}{Primary archetype}
        {4}{Secondary Archetype}
        {5}{Secondary Archetype}
        {6}{Secondary Archetype}
        {7}{Complex Trait}
        {8}{Complex Trait}
        {9}{Complex Trait}
        {10}{Complex Trait}
        {11}{Complex Trait}
    }[\PackageError{dimensiontype}{Undefined option to dimensiontype: #1}{}]%
}%

\usepackage{stackengine}
\setstackgap{S}{1pt}
\newcommand{\babaisyouboxscale}{0.48}

\newcommand{\archetyperatiosymbol}[2]{{\colorbox{#1}{\textcolor{#2}{\stackanchor{\scalebox{\babaisyouboxscale}{AR}}{\scalebox{\babaisyouboxscale}{CH}}}}}}

\newcommand{\appendixsymbol}[2]{{\colorbox{#1}{\textcolor{#2}{\stackanchor{\scalebox{\babaisyouboxscale}{AP}}{\scalebox{\babaisyouboxscale}{DX}}}}}}

\newcommand{\archetypometricshome}{\onlinesite}
\newcommand{\cardsdir}{\archetypometricshome/cards}

\newcommand{\onlinelinksymbol}{\nnearrow}
\newcommand{\externallinksymbol}{{\tiny$^{{}_{\onlinelinksymbol}}$\!\!}}
\newcommand{\internallinksymbol}{{$^{\Rsh}$\!\!}}
\newcommand{\paperlinksymbol}{\externallinksymbol}

\usepackage{xspace}

\usepackage{stmaryrd}

\newcommand{\characterlinksimple}[2]{\href{\cardsdir/#1-\Ncharacters-\Ntraits-\Nstories.pdf}{\textcolor{verydarkgrey}{#2\paperlinksymbol}}}

\newcommand{\characterlinksimpledataset}[3]{
  \IfEqCase{#3}{
    {1}{\href{\cardsdir/#1-\Ncharactersmainone-\Ntraitsmainone-\Nstoriesmainone.pdf}{\textcolor{verydarkgrey}{#2\colorbox{datasetrowcolor}{$\dataset{#3}$}\paperlinksymbol}}}
    {2}{\href{\cardsdir/#1-\Ncharactersmaintwo-\Ntraitsmaintwo-\Nstoriesmaintwo.pdf}{\textcolor{verydarkgrey}{#2\colorbox{datasetrowcolor}{$\dataset{#3}$}\paperlinksymbol}}}
    {3}{\href{\cardsdir/#1-\Ncharactersmain-\Ntraitsmain-\Nstoriesmain.pdf}{\textcolor{verydarkgrey}{#2\colorbox{datasetrowcolor}{$\dataset{#3}$}\paperlinksymbol}}}
  }[\PackageError{characterlinksimpledataset}{Undefined option to characterlinksimpledataset: #1}{}]%
}

\newcommand{\traitlinksimple}[2]{\textcolor{verydarkgrey}{\semdiff{\href{\cardsdir/#2--#1-\Ncharacters-\Ntraits-\Nstories.pdf}{#1\paperlinksymbol}}{\href{\cardsdir/#1--#2-\Ncharacters-\Ntraits-\Nstories.pdf}{#2\paperlinksymbol}}}}

\newcommand{\traitlinksimpledataset}[3]{
  \IfEqCase{#3}{
    {1}{\textcolor{verydarkgrey}{\semdiff{\href{\cardsdir/#2--#1-\Ncharactersmainone-\Ntraitsmainone-\Nstoriesmainone.pdf}{#1\colorbox{datasetrowcolor}{$\dataset{#3}$}\paperlinksymbol}}{\href{\cardsdir/#1--#2-\Ncharactersmainone-\Ntraitsmainone-\Nstoriesmainone.pdf}{#2\colorbox{datasetrowcolor}{$\dataset{#3}$}\paperlinksymbol}}}}
    {2}{\textcolor{verydarkgrey}{\semdiff{\href{\cardsdir/#2--#1-\Ncharactersmaintwo-\Ntraitsmaintwo-\Nstoriesmaintwo.pdf}{#1\colorbox{datasetrowcolor}{$\dataset{#3}$}\paperlinksymbol}}{\href{\cardsdir/#1--#2-\Ncharactersmaintwo-\Ntraitsmaintwo-\Nstoriesmaintwo.pdf}{#2\colorbox{datasetrowcolor}{$\dataset{#3}$}\paperlinksymbol}}}}
    {3}{\textcolor{verydarkgrey}{\semdiff{\href{\cardsdir/#2--#1-\Ncharactersmain-\Ntraitsmain-\Nstoriesmain.pdf}{#1\colorbox{datasetrowcolor}{$\dataset{#3}$}\paperlinksymbol}}{\href{\cardsdir/#1--#2-\Ncharactersmain-\Ntraitsmain-\Nstoriesmain.pdf}{#2\colorbox{datasetrowcolor}{$\dataset{#3}$}\paperlinksymbol}}}}
  }[\PackageError{traitlinksimpledataset}{Undefined option to traitlinksimpledataset: #1}{}]%
}

\newcommand{\traitlinksimpleright}[2]{\href{\cardsdir/#1--#2-\Ncharacters-\Ntraits-\Nstories.pdf}{\textcolor{verydarkgrey}{\semdiffright{#1}{\textbf{#2}}\paperlinksymbol}}}

\newcommand{\traitlinksimplerightalt}[4]{\href{\cardsdir/#1--#2-\Ncharacters-\Ntraits-\Nstories.pdf}{\textcolor{verydarkgrey}{\semdiffright{#3}{\textbf{#4}}\paperlinksymbol}}}

\newcommand{\traitlinksimpledatasetalt}[5]{
  \IfEqCase{#5}{
    {1}{\textcolor{verydarkgrey}{\semdiff{\href{\cardsdir/#2--#1-\Ncharactersmainone-\Ntraitsmainone-\Nstoriesmainone.pdf}{#3\colorbox{datasetrowcolor}{$\dataset{#5}$}\paperlinksymbol}}{\href{\cardsdir/#1--#2-\Ncharactersmainone-\Ntraitsmainone-\Nstoriesmainone.pdf}{#4\colorbox{datasetrowcolor}{$\dataset{#5}$}\paperlinksymbol}}}}
    {2}{\textcolor{verydarkgrey}{\semdiff{\href{\cardsdir/#2--#1-\Ncharactersmaintwo-\Ntraitsmaintwo-\Nstoriesmaintwo.pdf}{#3\colorbox{datasetrowcolor}{$\dataset{#5}$}\paperlinksymbol}}{\href{\cardsdir/#1--#2-\Ncharactersmaintwo-\Ntraitsmaintwo-\Nstoriesmaintwo.pdf}{#4\colorbox{datasetrowcolor}{$\dataset{#5}$}\paperlinksymbol}}}}
    {3}{\textcolor{verydarkgrey}{\semdiff{\href{\cardsdir/#2--#1-\Ncharactersmain-\Ntraitsmain-\Nstoriesmain.pdf}{#3\colorbox{datasetrowcolor}{$\dataset{#5}$}\paperlinksymbol}}{\href{\cardsdir/#1--#2-\Ncharactersmain-\Ntraitsmain-\Nstoriesmain.pdf}{#4\colorbox{datasetrowcolor}{$\dataset{#5}$}\paperlinksymbol}}}}
  }[\PackageError{traitlinksimpledatasetalt}{Undefined option to traitlinksimpledatasetalt: #1}{}]%
}

\newcommand{\storylinksimple}[2]{\href{\cardsdir/#1-\Ncharacters-\Ntraits-\Nstories.pdf}{\textcolor{darkgrey}{#2\paperlinksymbol}}}

\newcommand{\storylinksimpledataset}[3]{
  \IfEqCase{#3}{
    {1}{\href{\cardsdir/#1-\Ncharactersmainone-\Ntraitsmainone-\Nstoriesmainone.pdf}{\textcolor{verydarkgrey}{#2\colorbox{datasetrowcolor}{$\dataset{#3}$}\paperlinksymbol}}}
    {2}{\href{\cardsdir/#1-\Ncharactersmaintwo-\Ntraitsmaintwo-\Nstoriesmaintwo.pdf}{\textcolor{verydarkgrey}{#2\colorbox{datasetrowcolor}{$\dataset{#3}$}\paperlinksymbol}}}
    {3}{\href{\cardsdir/#1-\Ncharactersmain-\Ntraitsmain-\Nstoriesmain.pdf}{\textcolor{verydarkgrey}{#2\colorbox{datasetrowcolor}{$\dataset{#3}$}\paperlinksymbol}}}
  }[\PackageError{storylinksimpledataset}{Undefined option to storylinksimpledataset: #1}{}]%
}

\newcommand{\grouplinksimpledataset}[3]{
  \IfEqCase{#3}{
    {1}{\href{\cardsdir/#1-\Ncharactersmainone-\Ntraitsmainone-\Nstoriesmainone.pdf}{\textcolor{verydarkgrey}{#2\colorbox{datasetrowcolor}{$\dataset{#3}$}\paperlinksymbol}}}
    {2}{\href{\cardsdir/#1-\Ncharactersmaintwo-\Ntraitsmaintwo-\Nstoriesmaintwo.pdf}{\textcolor{verydarkgrey}{#2\colorbox{datasetrowcolor}{$\dataset{#3}$}\paperlinksymbol}}}
    {3}{\href{\cardsdir/#1-\Ncharactersmain-\Ntraitsmain-\Nstoriesmain.pdf}{\textcolor{verydarkgrey}{#2\colorbox{datasetrowcolor}{$\dataset{#3}$}\paperlinksymbol}}}
  }[\PackageError{grouplinksimpledataset}{Undefined option to grouplinksimpledataset: #1}{}]%
}

\newcommand{\archetypelinkbase}[1]{\href{\cardsdir/Archetype-#1-component-size-\Ncharacters-\Ntraits-\Nstories.pdf}{\textcolor{verydarkgrey}{#1\paperlinksymbol}}}

\newcommand{\archetypelinksimple}[2]{\href{\cardsdir/Archetype-#1-component-size-\Ncharacters-\Ntraits-\Nstories.pdf}{\textcolor{verydarkgrey}{#2\paperlinksymbol}}}

\newcommand{\archetypelinksimpledataset}[3]{
  \IfEqCase{#3}{
    {1}{\href{\cardsdir/Archetype-#1-component-size-\Ncharactersmainone-\Ntraitsmainone-\Nstoriesmainone.pdf}{\textcolor{verydarkgrey}{#2\colorbox{datasetrowcolor}{$\dataset{#3}$}\paperlinksymbol}}}
    {2}{\href{\cardsdir/Archetype-#1-component-size-\Ncharactersmaintwo-\Ntraitsmaintwo-\Nstoriesmaintwo.pdf}{\textcolor{verydarkgrey}{#2\colorbox{datasetrowcolor}{$\dataset{#3}$}\paperlinksymbol}}}
    {3}{\href{\cardsdir/Archetype-#1-component-size-\Ncharactersmain-\Ntraitsmain-\Nstoriesmain.pdf}{\textcolor{verydarkgrey}{#2\colorbox{datasetrowcolor}{$\dataset{#3}$}\paperlinksymbol}}}
  }[\PackageError{archetypelinksimpledataset}{Undefined option to archetypelinksimpledataset: #1}{}]%
}

\newcommand{\archetypelinkratiosimpleappendix}[2]{{\hypersetup{allcolors=.}\hyperref[page:N\Ncharactersbase_archetypometrics.archetypeclass-#1]{#2\archetyperatiosymbol{archetyperowcolor}{black}\,\appendixsymbol{verydarkgrey}{white}\internallinksymbol}}}

\newcommand{\essentialtraitlinknegative}[2]{\href{\cardsdir/Essential-Trait-\zeropad{000}{#1}-negative-component-size-\Ncharacters-\Ntraits-\Nstories.pdf}{\textcolor{verydarkgrey}{#2\paperlinksymbol}}}

\newcommand{\essentialtraitlinkpositive}[2]{\href{\cardsdir/Essential-Trait-\zeropad{000}{#1}-positive-component-size-\Ncharacters-\Ntraits-\Nstories.pdf}{\textcolor{verydarkgrey}{#2\paperlinksymbol}}}

\newcommand{\characterlink}[1]{
  \IfEqCase{#1}{
    }[\PackageError{characterlink}{Undefined option to characterlink: #1}{}]%
}%

\newcommand{\characterlinkinsert}[2]{
  \IfEqCase{#1}{
    }[\PackageError{characterlinkinsert}{Undefined option to characterlinkinsert: #1}{}]%
}%

\newcommand{\innerproduct}[2]{\bigl \langle \vec{#1}_{1}, \vec{#2}_{2} \bigr \rangle}

\newcommand{\rawtraitscore}{x}
\newcommand{\normtraitscore}{\chi} 

\newcommand{\norminnerproduct}{\innerproduct{\normtraitscore}{\normtraitscore}}

\newcommand{\rawtraitscoreessential}{y}
\newcommand{\normtraitscoreessential}{\psi} 

\newcommand{\norminnerproductessential}{\innerproduct{\normtraitscoreessential}{\normtraitscoreessential}}

\setboolean{twocolswitch}{true}

\usepackage{natbib}
\setcitestyle{square}
\raggedright

\usepackage{pdfpages}
\usepackage{newpax}
\newpaxsetup{usefileattributes=true}

\usepackage{authblk}

\begin{document}

\title{\protect
Archetypometrics of `Friends'

}
\onecolumn


\renewcommand*{\Authsep}{, }
\renewcommand*{\Authand}{, }
\renewcommand*{\Authands}{, }
\renewcommand*{\Affilfont}{\normalsize\normalfont}
\renewcommand*{\Authfont}{\bfseries}
\setlength{\affilsep}{2em}

\renewcommand*{\Authsep}{, }
\renewcommand*{\Authand}{, }
\renewcommand*{\Authands}{, }
\renewcommand*{\Affilfont}{\normalsize\normalfont}
\renewcommand*{\Authfont}{\bfseries}
\setlength{\affilsep}{2em}

\author[1\thanks{shun.zhang@uvm.edu}]{Shun~Zhang}
\author[1,2]{Tabia~Tanzin~Prama}
\author[1,3]{Christopher~M.~Danforth}
\author[1,2,4,5\thanks{peter.dodds@uvm.edu}]{Peter~Sheridan~Dodds}

\affil[1]{
  Computational Story Lab,
  Vermont Complex Systems Institute,
  MassMutual Center of Excellence for Complex Systems and Data Science,
  Vermont Advanced Computing Center,
  University of Vermont,
  Burlington,
  VT 05405,
  US
}
\affil[2]{
  Department of Computer Science,
  University of Vermont,
  Burlington,
  VT 05405,
  US
}
\affil[3]{
  Department of Mathematics \& Statistics,
  University of Vermont,
  Burlington,
  VT 05405,
  US
}
\affil[4]{
  Santa Fe Institute,
  1399 Hyde Park Rd,
  Santa Fe,
  NM 87501,
  US
}
\affil[5]{
  Complexity Science Hub,
  Metternichgasse 8,
  1030 Vienna,
  Austria
}

\date{\today}

\maketitle


\mbox{}

\bigskip
\bigskip
\bigskip

\hspace*{50pt}
\begin{minipage}{300pt}

  \begin{tabular}{>{\raggedright\arraybackslash}p{180pt}p{30pt}p{180pt}}

 \parbox{180pt}{    
   \begin{loglinebox}[Logline]
     \raggedright
     We perform an extensive archetypometric analysis 
of the ensemble sitcom `Friends',
both for individual characters and 
for all pairwise relationships.
Our computational approach positions 
the show's six main characters 
within a larger context of 2,000 characters
from 341 stories, 
all evaluated on 464 semantic differential traits.
Our treatment is repeatable 
for the quantification of 
any story-ensemble geometry.

     \smallskip
   \end{loglinebox}
 }

 &
    
    &
    
     \parbox{180pt}{    
    \includegraphics[width=180pt,valign=m,frame]{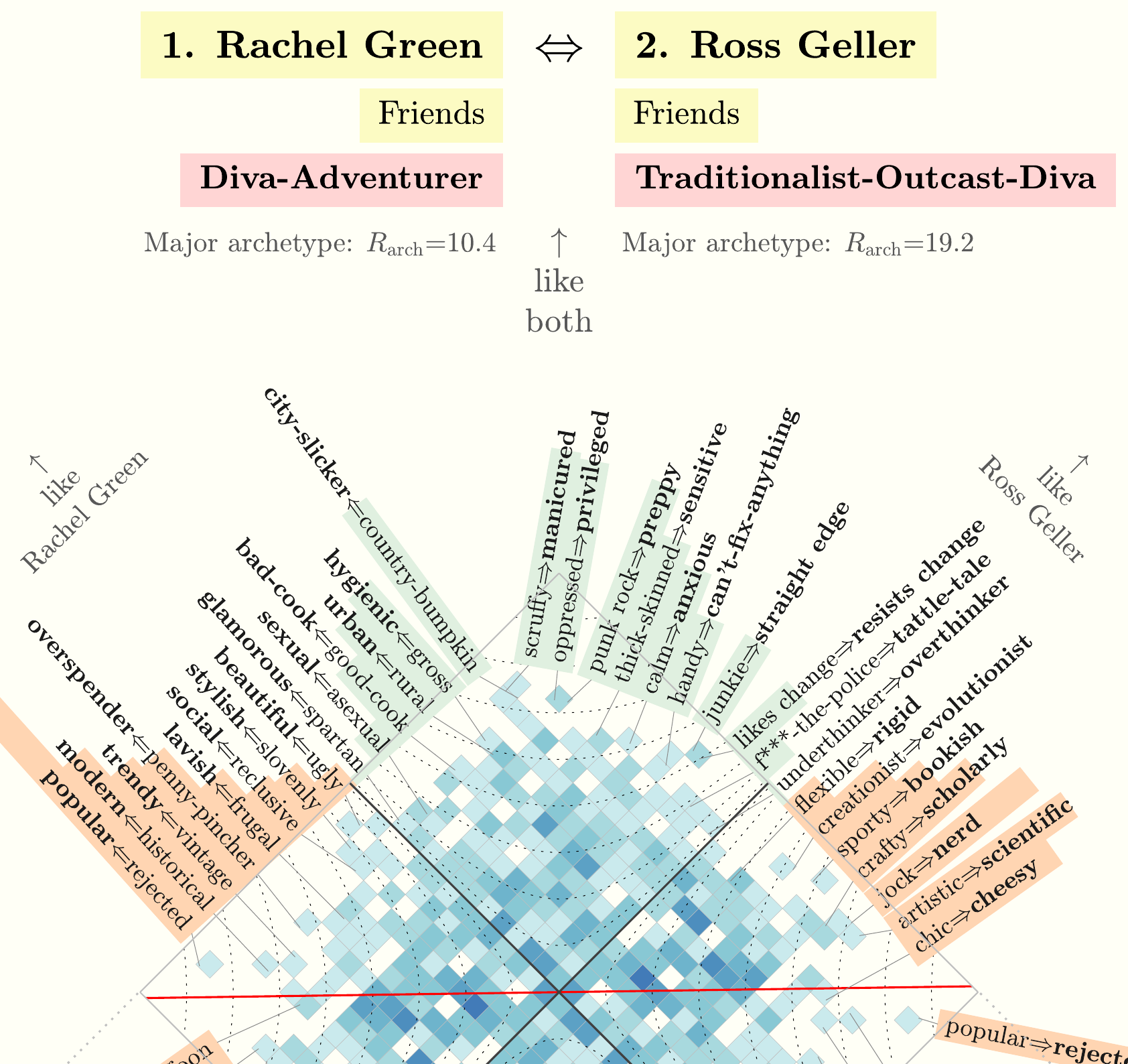}
 }
    
  \end{tabular}
  
\end{minipage}

\clearpage

\newgeometry{
  left=2in,
  right=2in,
  top=0.65in,
  bottom=0.65in,
  }

\onecolumn

\renewcommand{\baselinestretch}{1.2}
\selectfont


\begin{abstractbox}[Abstract]
  \raggedright
  Storytelling inherently revolves around characters.
Using the television sitcom `Friends' as a case study, we investigate how well archetype vectors capture both individual characterization and the relational structure of a specific ensemble.
Our work is based on the archetypometrics framework, which locates 2,000 fictional characters from 341 stories in a continuous space derived from 464 bipolar traits.

We proceed in three stages:
interpreting each character's archetypal profile against narrative evidence,
projecting the ensemble onto ousiograms of the six essential dimensions,
and
measuring pairwise similarity with vector inner products.

We show that the six characters of `Friends' occupy distinct archetypal positions that accord with their established identities, while the projections expose ensemble structure invisible in individual profiles,
including the collapse of the Angel--Demon dimension, a signature of the sitcom's uniformly sympathetic cast.

Based on inner products, we construct a similarity matrix
that resolves three main kinds of relational structure:
alignment (e.g., Phoebe--Joey),
contrast (e.g., Phoebe--Ross),
and
orthogonality (e.g., Rachel--Ross and Monica--Chandler).
The orthogonality of the romantic pairings
affords a detailed view of relationships
built on complementary rather than overlapping
character traits.

Overall, our case study suggests that for ensemble-based stories
the archetypometric geometry is fully interpretable in narrative terms, from individual identities to the structure of the group's relationships.
  \smallskip
\end{abstractbox}


\begin{infobox}[Keywords]
  \centering
  stories,
characters,
television,
sitcoms,
ensemble sitcoms, 
comedy,
archetypometrics, 
character archetypes, 
fictional characters, 
friendship,
romance,
friends

 \smallskip
\end{infobox}

\renewcommand{\baselinestretch}{1}
\selectfont

\twocolumn

\restoregeometry

\clearpage

\tableofcontents

\clearpage

\section{Introduction}

Fictional characters are central to narrative comprehension because they drive plot development and provide coherence across events~\cite{Eder2010CharactersIF, Margolin1986THEDA}. Beyond describing individual personalities, narratives are fundamentally relational. Characters acquire meaning through their interactions within a shared narrative structure~\cite{Bamman2013LearningLP}. Consequently, understanding relationships among characters has become a central problem in computational narrative analysis~\cite{Elson2010ExtractingSN, Brahman2021LetYC}, with recent work increasingly focusing on interaction patterns and relational structures~\cite{Jahan2021InducingSC, Christou2025RelationalAA}, particularly for ensemble television series where a stable cast repeatedly interacts across many episodes~\cite{Mills2009TheS}. Addressing this problem requires computational character representations that place multiple characters within a common analytical space, enabling systematic comparison of their narrative roles and relationships~\cite{Gurung2024CHIRONRC, Piper2021NarrativeTF}. Character archetypes provide a natural representation for this problem because they describe recurring narrative functions that can be compared across stories and within character ensembles~\cite{Jung1968TheAA, Sutton1958AnatomyOC}. 

However, most practical applications of archetypes still represent characters as discrete categories~\cite{Pearson2012AwakeningTH, Vogler2007TheWJ}. While useful for describing recurring character types, such discrete classifications provide limited ability to quantify relational configurations among characters within the same narrative~\cite{Bowman2010TheFO}. In contrast, continuous representations can both enable quantitative comparison of individual characters and explain relationships among multiple characters through their geometric organization~\cite{Inoue2022LearningAE}. Recent work has addressed  this limitation by representing characters as continuous vectors. Dodds et al.~\cite{dodds2021archetypometrics} developed a large-scale archetypometric framework that derives essential archetypal dimensions from empirical survey data~\cite{openpsychometrics, dodds2025archetypometrics_dataset}, enabling characters to be represented in an archetype space. This representation makes it possible to quantify geometric relationships among characters, while the narrative significance of these relationships has remained largely unexplored.

Ensemble narratives provide an ideal setting for investigating this question because recurring interactions among a stable cast make relational character structures both observable and comparable across episodes~\cite{OMeara2015ChangingTW, Mills2009TheS}. Among such narratives, \storylinksimple{Friends}{Friends}~\cite{friends_wikipedia} is particularly suitable because its six main characters maintain stable identities while exhibiting diverse and recurring interaction patterns across the show, providing a controlled setting for evaluating computational archetype representations.

This study thus addresses the following research questions through a case study of the television series \storylinksimple{Friends}{Friends}, building on the six essential archetypal dimensions derived by Dodds et al.~\cite{dodds2021archetypometrics}. 

\textbf{RQ1: Do computationally derived archetypes provide an interpretable representation of the six main characters in \storylinksimple{Friends}{Friends}?}
We examine whether the computationally derived archetypes are consistent with the established narrative characterization of the main characters.

\textbf{RQ2: Do the geometric positions of characters along these orthogonal axes align with their narrative roles as understood from the series?} To answer this question, we project characters from the six-dimensional archetypal space into a series of two-dimensional subspaces defined by different pairs of archetypal dimensions, enabling visual comparison of their relative narrative positions.

\textbf{RQ3: Can relationships between characters be characterized through the geometric relationships among their archetype representations?} We extend the archetypometric framework from individual classification to relational analysis by using pairwise inner products to assess similarity among characters and to determine whether these structural patterns correspond to narrative features such as romantic pairings.

\section{Literature Review}

\subsection{What Is an Archetype?}
The concept of archetype has been studied across several disciplines, including analytical psychology and myth theory~\cite{Jung1968TheAA, Frye2020AnatomyOC, Luomala1949TheHW}. In analytical psychology, Jung~\cite{Jung1968TheAA} defined archetypes as recurrent symbolic motifs that structure human imagination. Extending this idea to literary theory, Frye~\cite{Frye2020AnatomyOC} argued that archetypes organize recurring narrative structures across literary genres, while Sutton~\cite{Sutton1958AnatomyOC} emphasized their role in shaping recognizable patterns of characterization. These definitions converge on a common view: archetypes describe recurring narrative functions embodied by characters. Unlike personality traits, which describe relatively stable psychological dispositions of individuals~\cite{McCrae1997PersonalityTS, Herman2009BasicEO}, archetypes describe the narrative functions that characters repeatedly fulfill within stories~\cite{Frye2020AnatomyOC}. Contemporary narrative theory further argues that archetypal characters recur because they perform recognizable narrative functions that audiences repeatedly encounter across stories~\cite{Booker2024TheSB}.
As archetypes recur across narratives, they provide a common framework for comparing characters that originate from different stories and genres~\cite{Frye2020AnatomyOC}. This property makes archetypes particularly suitable for computational character representation, where characters from different stories must be represented within a common analytical framework.

\subsection{Computational Character Representation}

A fundamental problem in computational narrative analysis is how to represent fictional characters in a form that enables systematic comparison and analysis~\cite{Gurung2024CHIRONRC, Piper2021NarrativeTF}. Existing approaches have represented fictional characters through explicit semantic attributes and continuous latent personality traits~\cite{Bamman2013LearningLP, Yang2024EvaluatingCR, Bourgois2026TowardAO}.

One line of work represented characters through explicit semantic attributes extracted from narrative context. For instance, Brahman et al.~\cite{Brahman2021LetYC} developed a dataset and a modeling framework for character-centric narrative understanding, encoding characters using discrete feature sets extracted from stories. While effective for describing individual properties, such representations are difficult to compare across characters in a continuous and relational manner. Characters assigned the same attribute labels may differ substantially in how those attributes are expressed within the narrative. An alternative approach represents characters using variables learned directly from narrative data. Bamman et al.~\cite{Bamman2013LearningLP} learned latent personas of film characters from screenplays using probabilistic models, showing that recurring behavioral patterns can be represented as shared latent character types. Continuous representations enable quantitative comparisons among characters and the positioning of multiple characters within a shared space. 

Another line of research represents fictional characters through psychological characteristics inferred from narrative text~\cite{Rashkin2018ModelingNP, Flekova2015PersonalityPO}. Flekova et al.~\cite{Flekova2015PersonalityPO} introduced one of the first datasets for automatic personality profiling of fictional characters, demonstrating that semantic information extracted from narrative text supports personality prediction. Building on this idea, Pizzolli et al.~\cite{Pizzolli2019PersonalityTR} showed that dialogue alone provides sufficient linguistic evidence for predicting personality. More recently, Sang et al.~\cite{Sang2022MBTIPP} further incorporated narrative descriptions alongside dialogue, improving personality modeling through the Story2Personality benchmark. These studies show that personality traits are useful for computationally modeling fictional characters~\cite{Flekova2015PersonalityPO, Pizzolli2019PersonalityTR, Sang2022MBTIPP}. However, personality profiles mainly capture stable psychological dispositions rather than a character's narrative role or relationships within the story. As a result, personality similarity does not necessarily imply similarity in narrative function. This motivates archetype-based computational representations, which encode recurring narrative roles in a continuous and comparable form.

\subsection{Computational Archetype Modeling}

Building on psychological and narrative representations, archetypes offer a complementary framework for capturing recurring narrative functions rather than story-specific identities~\cite{Green2019ArchetypesAN}. Recent computational work has begun to operationalize archetypal structure through data-driven methods. For example, Lima et al.~\cite{Lima2026RevisitingNF} used large language models to infer archetypal roles grounded in Jungian and Fryean theory, demonstrating that such recurring functions can be computationally identified across diverse literary genres.

Dodds et al.~\cite{dodds2021archetypometrics} developed a large-scale computational framework for character archetypes. Instead of treating archetypes as categorical labels, their model represents each character as a weighted combination of archetypal dimensions. The resulting representation embeds thousands of fictional characters into a continuous space, enabling quantitative comparison between characters. Building on these embeddings, the framework introduces ousiograms, low-dimensional projections that organize archetypal space into interpretable dimension pairs such as \archetypesemdiff{1} and \archetypesemdiff{2}. These projections provide an interpretable means of visualizing similarities among characters while preserving the underlying continuous representation.

The archetypometric framework has also supported downstream analyses of fictional characters~\cite{Beauregard2026BuffyVB}, including the study of gendered patterns in archetypal expression~\cite{Beauregard2024MisrepresentationOO}. However, little work has examined whether these dimensions align with characters' narrative roles or explain the relational structure of character ensembles. This study therefore evaluates both the narrative interpretability and relational utility of computational archetype representations.

\subsection{Ensemble Narratives}

Ensemble narratives, especially long-running sitcoms, provide a suitable setting for evaluating computational character representations~\cite{Nan2015SocialNA}, because they contain a stable cast of recurring characters whose characterizations remain consistent across repeated appearances~\cite{OMeara2015ChangingTW}. Unlike single-character narratives where character development focuses on one individual, ensemble narratives derive much of their meaning from the interactions among multiple characters~\cite{Mills2009TheS}. This relational structure has motivated computational studies that analyze television narratives through social networks~\cite{Nan2015SocialNA}. The episodic structure of sitcoms repeatedly places the same characters in comparable situations, allowing recurring narrative functions to be expressed consistently across episodes~\cite{OMeara2015ChangingTW, Bednarek2011Chapter1T}. 

Among ensemble sitcoms, \storylinksimple{Friends}{Friends} provides a suitable case study for evaluating computational archetype representations. The series spans ten seasons and follows a fixed ensemble of six principal characters whose identities remain relatively stable while interacting across a wide range of recurring everyday situations~\cite{imdb_friends, klika2021ensemble}. It creates a relatively controlled narrative environment in which differences among characters are more likely to reflect recurring narrative functions than changes in genre~\cite{Bazzan2018IWB}. This stability makes \storylinksimple{Friends}{Friends} an appropriate benchmark for evaluating computational archetype representations.

\section{Data and Methods}

\subsection{Data}

Our study builds on the archetypometric dataset~\cite{dodds2021archetypometrics}. The dataset consists of personality traits for 2,000 fictional characters drawn from 341 stories from films and television~\cite{dodds2021archetypometrics}. Each character is represented by evaluations on 464 bipolar personality traits collected through the Open Source Psychometrics Project~\cite{openpsychometrics}. Character evaluations are collected through semantic differential scales. Each item pairs a fictional character with a bipolar trait (e.g., \traitlinksimple{uninspiring}{charismatic}). Traits are presented as opposing descriptors along a 101-point integer scale, anchored at two extremes (100\% to 0\%). 
We normalize these scores to be between $-$1/2 and $+$1/2,
and denote
the raw trait vector
for character by
$
\vec{\rawtraitscore}_{i}
\in 
\mathbb{R}^{464}.
$
Using singular value decomposition (SVD) and extensive 
story and concept analysis, 
Dodds et~al. obtained an orthogonal basis for
traits and characters of which the first six dimensions
and their various combinations
could be viewed as being of archetype level (i.e., 
representing whole characters).
Specifically, the first six essential dimensions 
account for 74.6\% of the total variance, and 99.2\% of characters have one of these as their leading direction~\cite{dodds2021archetypometrics}. 
This compact basis describes the dominant archetypal structure observed across the dataset, which are labeled as \archetypesemdiff{1}, \archetypesemdiff{2}, \archetypesemdiff{3}, \archetypesemdiff{4}, \archetypesemdiff{5}, and \archetypesemdiff{6}~\cite{dodds2025archetypes_lists} (only the first six components are interpreted and named). 
We write character $i$'s essential trait vector as
$
\vec{\rawtraitscoreessential}_{i}
\in 
\mathbb{R}^{464}.
$

All reported quantities are normalized relative to the full reference dataset, with characters and traits normalized separately (the two normalizations cannot be achieved simultaneously): Character vectors are scaled so that the largest magnitude across the 2,000 characters is 100, and trait scores are scaled so that the largest-magnitude trait score is 100. Our focus here is on character vectors.

\subsection{Methods}

\subsubsection{Individual Character Validation}
We begin by examining the interpretability of the archetypal representation of each main character using character cards, which summarize the contribution of the six essential character archetype dimensions for a given character, together with descriptive statistics including dominant traits, nearest neighbors, and the proportion of variance explained by the leading dimensions. This representation provides an interpretable overview of each character's archetypal composition before examining relationships among characters. The character cards of all six main characters of \storylinksimple{Friends}{Friends} are listed in Appendix~\ref{Appendix:Character_Cards} (Figs.~\ref{fig:Rachel}--\ref{fig:Ross}). 

\begin{figure*}[p]
      \centering
      \includegraphics[width=1.0\linewidth]{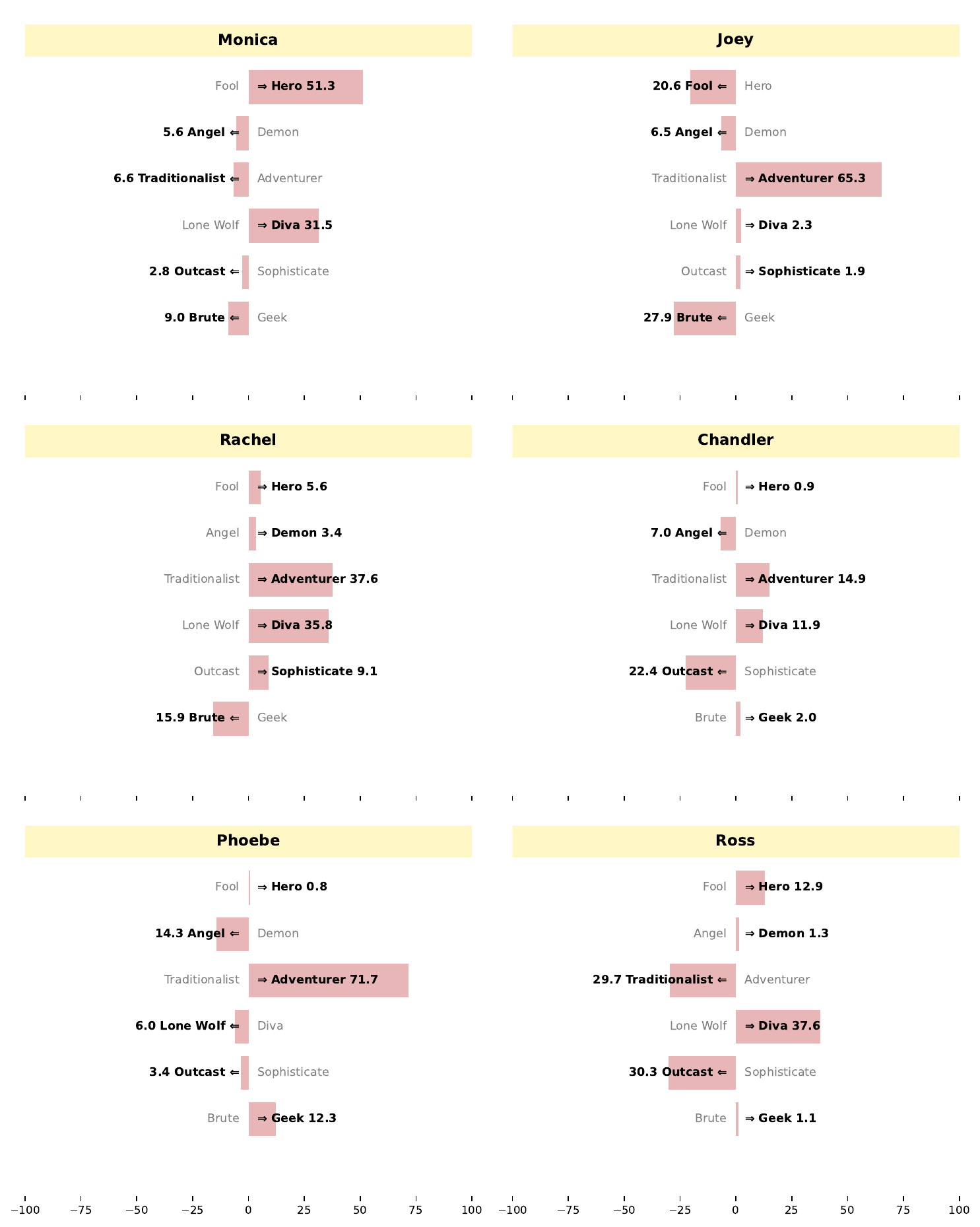}
      \caption{\textbf{Archetype dimension profiles of the six main characters in \storylinksimple{Friends}{Friends}.} Each subplot represents one character across six archetypal essential dimensions. Bars extending to the right indicate stronger alignment with the right-hand archetype (e.g., \archetypelinksimple{Hero}{Hero}, \archetypelinksimple{Demon}{Demon}, \archetypelinksimple{Adventurer}{Adventurer}), whereas bars extending to the left indicate stronger alignment with the opposing archetype (e.g., \archetypelinksimple{Fool}{Fool}, \archetypelinksimple{Angel}{Angel}, \archetypelinksimple{Traditionalist}{Traditionalist}). Bar lengths correspond to archetype scores normalized so that the largest-magnitude score across the full dataset is 100.}
      \label{fig:Friends_Main_Characters_6Dimensions}
    \end{figure*}

Fig.~\ref{fig:Friends_Main_Characters_6Dimensions} shows the six-dimensional archetype profiles of \characterlinksimple{Friends-Rachel-Green}{Rachel}, \characterlinksimple{Friends-Monica-Geller}{Monica}, \characterlinksimple{Friends-Phoebe-Buffay}{Phoebe}, \characterlinksimple{Friends-Joey-Tribbiani}{Joey}, \characterlinksimple{Friends-Chandler-Bing}{Chandler}, and \characterlinksimple{Friends-Ross-Geller}{Ross}. Each profile shows how a character is positioned across the six main archetypal dimensions. 

We compare these quantitative archetype profiles with the evidence from the show to assess whether the archetypal representation derived from the archetypometric framework remains consistent with detailed characterization in an ensemble sitcom. Specifically, we examine representative narrative evidence associated with each character to evaluate whether the archetype scores correspond to their established on-screen identities.

\subsubsection{Archetypal Dimension Projection}

The individual character analysis shows how the six characters are positioned relative to one another within the shared archetypal space. We therefore examine the ensemble structure of \storylinksimple{Friends}{Friends} through archetypal projection using an ousiogram. An ousiogram is a geometric visualization of the archetypal space~\cite{dodds2021archetypometrics}. Each plot represents one pair of archetype dimensions, allowing characters to be positioned according to their coordinates along the two corresponding axes. Characters are displayed as labeled points, with their locations determined by their archetype coordinates. The surrounding labels are the eight possible combinations of the two archetype dimensions, while the background density map summarizes the distribution of all 2,000 fictional characters in the archetypometric dataset. These figures are displayed as Figs.~\ref{fig:1-2}--\ref{fig:5-6}.

Each character is represented by six archetype coordinates and projected onto three pairwise two-dimensional planes corresponding to the first six archetype dimensions 
(\archetypesemdiff{1}
vs.\ 
\archetypesemdiff{2},
\archetypesemdiff{3}
vs.\ 
\archetypesemdiff{4},
and
\archetypesemdiff{5}
vs.\ 
\archetypesemdiff{6}).
We choose pairwise projections because these three projections constitute the most compact two-dimensional representation that covers all dimensions while preserving the original archetype coordinates. This allows us to examine the relative spatial organization of characters without introducing an additional dimensionality reduction step.

\subsubsection{Pairwise Relationship Analysis}

Ensemble narratives derive much of their structure from recurring interactions among a stable cast, so the analysis cannot stop at individual profiles. To quantify the relationships between pairs of characters, we examine pairwise archetypal relationships.

Unlike the archetypal projections in the previous section, which use the first six dominant archetype dimensions for interpretability, pairwise relationship analysis is performed on the complete 464-dimensional character representations provided by the archetypometric framework. This allows all available trait information to contribute to the similarity measure, whereas the six-dimensional projections introduced in the previous section are used only for visualization and qualitative interpretation. Several geometric measures can be used to compare character vectors, including Euclidean distance, cosine similarity, and the inner product. Here we use the inner product to capture both magnitude and directional similarities.

To quantify the relationship between two characters, we compute the inner product of their raw trait and essential trait vectors.
For inner products only, we include one more normalization step so that across
all pairs of characters the maximum absolute inner product is 100.
Per above, each character $i$ is represented by
$\vec{\rawtraitscore}_{i}$
and an essential trait vector
$\vec{\rawtraitscoreessential}_{i}$.
Introducing a dual notation for the appropriately normalized vectors,
we write the inner products as
\begin{equation}
R(i,j)
=
\norminnerproduct
=
\norminnerproductessential,
\end{equation}
which is true because inner products are independent of basis.
The normalization helps with presentation and comparison---on character cards and throughout our figures.
They make all reported values directly interpretable: An inner product of $+53.80$, for example, indicates a pair at 53.8\% of the strongest alignment observed between any two characters in the dataset.

Based on the sign and magnitude of the inner product, we distinguish three qualitatively different types of archetypal relationships. Positive values represent aligned archetypal profiles, suggesting that two characters share similar archetypal orientations. Negative values represent opposing archetypal profiles, which can indicate contrasting narrative tendencies. Values close to zero represent approximately orthogonal archetype vectors. We then compute the inner products for all character pairs in \storylinksimple{Friends}{Friends} and interpret the resulting geometric relationships together with narrative evidence from the series. 

The archetypometric framework provides two complementary representations of fictional characters. The six-dimensional archetype representation is used to interpret character identities, whereas the complete 464-dimensional representation is used to quantify relationships between characters. For each pairwise inner product computed in the full 464-dimensional trait space, we decompose the total value into dimension-wise contributions. Because the archetypal dimensions are ordered by their global variance explained, dimensions beyond the first six typically contribute little to the pairwise inner product. To maintain both interpretability and transparency, we adopt the following reporting rule for the character pairs:
\begin{itemize}
    \item 
    We first compute the full 464-dimensional inner product for every pair.
    \item 
    We then compute the proportion of the total absolute inner product contributed by each dimension (similarity contributions).
    \item 
    For dimensions beyond the sixth, we verify that their cumulative absolute contribution does not exceed 10\% of the total. This confirms that the exclusion of higher-order dimensions does not alter the interpretation. 
\end{itemize}

\section{Results and Discussion}
\subsection*{RQ1: Individual Character Validation}
We first evaluate whether the archetypal coordinates provide meaningful descriptions of the six main characters in \storylinksimple{Friends}{Friends}. 

\subsubsection{\characterlinksimple{Friends-Rachel-Green}{Rachel}}

\characterlinksimple{Friends-Rachel-Green}{Rachel Green}~\cite{wikipedia_Rachel_Green} is primarily associated with the \archetypelinksimple{Diva}{Diva}--\archetypelinksimple{Adventurer}{Adventurer} archetype. As shown in Fig.~\ref{fig:Friends_Main_Characters_6Dimensions}, her major archetype is \archetypelinksimple{Diva}{Diva} (35.8)--\archetypelinksimple{Adventurer}{Adventurer} (37.6), indicating a character who is both socially expressive and strongly oriented toward self-presentation and personal experience. Dominant traits such as \traitlinksimpleright{rejected}{popular}, \traitlinksimpleright{ugly}{beautiful}, \traitlinksimpleright{slovenly}{stylish}, \traitlinksimpleright{repulsive}{attractive}, and \traitlinksimpleright{spartan}{glamorous} characterize the \archetypelinksimple{Diva}{Diva} dimension, while the \archetypelinksimple{Adventurer}{Adventurer} dimension emphasizes openness to change and exploration (see Character Card in Fig.~\ref{fig:Rachel}). This archetypal configuration finds direct support in the narrative. In the pilot episode (S01E01), \characterlinksimple{Friends-Rachel-Green}{Rachel}'s decision to run out on her wedding to Barry---a dentist with a secure, conventional future---directly embodies the \archetypelinksimple{Adventurer}{Adventurer}'s rejection of stability in favor of uncertain self-discovery. Her decision to cut up her credit cards and take a waitressing job at Central Perk, despite having no prior work experience, further illustrates the \archetypelinksimple{Adventurer}{Adventurer}'s willingness to embrace risk for personal growth. At the same time, her \archetypelinksimple{Diva}{Diva} traits are consistently foregrounded: she is dramatic in her emotional outbursts, as in S04E01:
\begin{quote}
\textbf{Rachel:} And hey! Just so you know, it's not that common! It doesn't happen to every guy! And it is a big deal!!\\
\end{quote}
And she is also sensitive to perceived slights or failures, as seen in her devastated reaction to losing a promised promotion at Bloomingdale's in S04E09, when her boss dies before filing the paperwork.
Among characters in the broader archetypal corpus, \characterlinksimple{Friends-Rachel-Green}{Rachel} shows notable similarity to \characterlinksimple{Sex-and-the-City-Carrie-Bradshaw}{Carrie Bradshaw} from \storylinksimple{Sex-and-the-City}{Sex and the City}~\cite{imdb_sexandthecity}. They are both fashion-conscious, socially ambitious women whose \archetypelinksimple{Diva}{Diva}--\archetypelinksimple{Adventurer}{Adventurer} orientation drives their respective narratives of romantic life and careers in a metropolitan setting.

\subsubsection{\characterlinksimple{Friends-Monica-Geller}{Monica}}

\characterlinksimple{Friends-Monica-Geller}{Monica Geller}'s~\cite{wikipedia_Monica_Geller} profile is dominated by the \archetypelinksimple{Hero}{Hero} dimension, with a secondary alignment to the \archetypelinksimple{Diva}{Diva} archetype (Fig.~\ref{fig:Monica}). The strongest loadings in her profile are concentrated on traits related to order and control, including \traitlinksimpleright{sloppy}{fussy}, \traitlinksimpleright{gross}{hygienic}, \traitlinksimpleright{absentminded}{focused}, \traitlinksimpleright{disorganized}{self-disciplined}, \traitlinksimpleright{messy}{neat}, and \traitlinksimpleright{lazy}{diligent}. Several additional traits, such as \traitlinksimpleright{spontaneous}{scheduled}, \traitlinksimpleright{tardy}{on-time}, and \traitlinksimpleright{intuitive}{analytical}, reinforce the same pattern (see Character Card in Fig.~\ref{fig:Monica}).  \characterlinksimple{Friends-Monica-Geller}{Monica}'s \archetypelinksimple{Hero}{Hero} orientation can be observed in S03E09, where her obsessive competitiveness with \characterlinksimple{Friends-Ross-Geller}{Ross} over a childhood ``Geller Cup'' trophy transforms a casual Thanksgiving game into an all-out war. She refuses to concede even after injuring her knee. This will to win borders on the pathological. These behaviors align with her disciplined and diligent nature. She rises from a struggling cook to head chef at a Manhattan restaurant, which is a career progression that requires sustained effort and self-mastery~\cite{wikipedia_Monica_Geller}. The \archetypelinksimple{Diva}{Diva} archetype (31.5) manifests in her as an obsession with order and perfection. Neatness and hygiene are its defining traits, illustrated by the fact that \characterlinksimple{Friends-Monica-Geller}{Monica} categorizes her towels into eleven separate categories (mentioned in the lightning round of S04E12):
\begin{quote}
\textbf{Ross:} Monica categorizes her towels. How many categories are there?\\
\textbf{Joey:} Everyday use.\\
\textbf{Chandler:} Fancy.\\
\textbf{Joey:} Guest.\\
\textbf{Chandler:} Fancy guest.\\
\textbf{Ross:} Two seconds.\\
\textbf{Joey:} Uh, Eleven!\\
\textbf{Ross:} Eleven. Unbelievable. Eleven is correct.\\
\end{quote}
\characterlinksimple{Friends-Monica-Geller}{Monica} shows notable similarity to \characterlinksimple{The-Big-Bang-Theory-Bernadette-Rostenkowski-Wolowitz}{Bernadette Rostenkowski-Wolowitz} from \storylinksimple{The-Big-Bang-Theory}{The Big Bang Theory}~\cite{fandom_Bernadette}. Both serve the disciplinary and competitive role in their respective friend groups.

\subsubsection{\characterlinksimple{Friends-Phoebe-Buffay}{Phoebe}}

While \characterlinksimple{Friends-Monica-Geller}{Monica}'s profile emphasizes structure and control, \characterlinksimple{Friends-Phoebe-Buffay}{Phoebe Buffay}~\cite{wikipedia_Phoebe_Buffay} represents a markedly different orientation. She is characterized by a dominant \archetypelinksimple{Adventurer}{Adventurer} archetype, with a very strong alignment on the \archetypesemdiff{3} dimension (71.7) (see Character Card in Fig.~\ref{fig:Phoebe}), indicating a clear preference for unpredictability and creativity. For instance, \characterlinksimple{Friends-Phoebe-Buffay}{Phoebe} repeatedly affects ordinary events through unconventional beliefs, including insisting that a stray cat embodies the spirit of her deceased mother in S04E02. 
\begin{quote}
\textbf{Phoebe:} I just have this really strong feeling that this cat is my Mother.\\
\textbf{Rachel:} You mean the mom you met in Montauk. She was a cat?!\\
\textbf{Phoebe:} No, no-no, she was a human lady. This is the spirit of my Mom Lily, the one who killed herself.\\
\textbf{Ross:} Are you sure she's in the cat...?\\
\end{quote}
This example shows her imaginative and non-conforming approach to reality, consistent with her established characterization~\cite{wikipedia_Phoebe_Buffay}. At the trait level, this orientation is reflected in high scores on attributes such as \traitlinksimplerightalt{uncreative}{open-to-new-experiences}{uncreative}{open to new experiences}, \traitlinksimpleright{deliberate}{spontaneous}, \traitlinksimpleright{routine}{innovative}, \traitlinksimpleright{practical}{imaginative}, and \traitlinksimpleright{monotone}{expressive} (see Character Card in Fig.~\ref{fig:Phoebe}). Her unconventional worldview frequently places her in direct opposition to \characterlinksimple{Friends-Ross-Geller}{Ross}'s scientific rationalism. Most notably, when she dismisses Darwinian evolution as ``too easy'' in S02E03, she believes in a more whimsical version of the world~\cite{wikipedia_Phoebe_Buffay}. A similar character to \characterlinksimple{Friends-Phoebe-Buffay}{Phoebe} would be \characterlinksimple{Elizabethtown-Claire-Colburn}{Claire Colburn} from \storylinksimple{Elizabethtown}{Elizabethtown}~\cite{imdb_Elizabethtown}. Both are free-spirited, perceptive women who take a spontaneous, unconventional approach to life.

\subsubsection{\characterlinksimple{Friends-Joey-Tribbiani}{Joey}}
\characterlinksimple{Friends-Joey-Tribbiani}{Joey Tribbiani}'s~\cite{wikipedia_Joey_Tribbiani} archetypal profile is dominated by the \archetypelinksimple{Adventurer}{Adventurer} dimension (65.3), with a secondary orientation toward the \archetypelinksimple{Brute}{Brute} archetype (27.9), as shown in Fig.~\ref{fig:Friends_Main_Characters_6Dimensions}. His \archetypelinksimple{Adventurer}{Adventurer} score indicates a character driven by immediate gratification and sensory pleasure. These tendencies can be seen in traits such as \traitlinksimplerightalt{unenthusiastic-about-food}{foodie}{unenthusiastic about food}{foodie}, \traitlinksimpleright{negative}{positive}, and \traitlinksimpleright{asexual}{sexual}, alongside \traitlinksimpleright{serious}{playful} and \traitlinksimpleright{prudish}{flirtatious} (see Character Card in Fig.~\ref{fig:Joey}). \characterlinksimple{Friends-Joey-Tribbiani}{Joey} demonstrates the \archetypelinksimple{Adventurer}{Adventurer} dimension through impulsive decisions that prioritize experience over practicality. In S06E24, for example, he purchases a boat despite having neither a clear need for it nor a place to keep it. The decision is driven primarily by the appeal of a new experience instead of careful planning.  
\begin{quote}
\textbf{Joey:} No way! It's mine!!\\
\textbf{Rachel:} But Joey, you don't have \$20,000!\\
\textbf{Joey:} Who cares?! I'll make payments, whatever it takes, I want the Mr. Bowmont!!\\
\end{quote}
These traits align with his on-screen characterization~\cite{wikipedia_Joey_Tribbiani}. His playful and flirtatious nature is encapsulated in his signature pickup line, ``How you doin'?''. He is also loyal to his friends, once giving up meat entirely during \characterlinksimple{Friends-Phoebe-Buffay}{Phoebe}'s pregnancy so that ``no extra animals die'' on her behalf. A similar character to \characterlinksimple{Friends-Joey-Tribbiani}{Joey} is \characterlinksimple{The-Good-Place-Jason-Mendoza}{Jason Mendoza} from \storylinksimple{The-Good-Place}{The Good Place}~\cite{fandom_Jason_Mendoza}. Both are lovable, oblivious, optimistic people whose intellectual limitations are offset by unwavering loyalty and childlike joy.

\subsubsection{\characterlinksimple{Friends-Chandler-Bing}{Chandler}}
\characterlinksimple{Friends-Chandler-Bing}{Chandler Bing}~\cite{wikipedia_Chandler_Bing} is characterized by a more distributed archetypal configuration. As shown in Fig.~\ref{fig:Friends_Main_Characters_6Dimensions}, his profile combines elements of the \archetypelinksimple{Diva}{Diva} (11.9), \archetypelinksimple{Adventurer}{Adventurer} (14.9), and \archetypelinksimple{Outcast}{Outcast} (22.4) dimensions. At the trait level, attributes such as \traitlinksimpleright{anti-prank}{prankster}, \traitlinksimpleright{unfrivolous}{goofy}, \traitlinksimpleright{humorless}{funny}, \traitlinksimpleright{genuine}{sarcastic}, and \traitlinksimpleright{charming}{awkward} all connect to this configuration. Additional features, including resistance to change and a tendency toward self-consciousness, further contribute to a profile that blends social expression with underlying uncertainty (see Character Card in Fig.~\ref{fig:Chandler}). \characterlinksimple{Friends-Chandler-Bing}{Chandler}'s archetypal structure is more evenly distributed across dimensions, resulting in a less concentrated but more composite profile. His use of sarcasm and self-deprecating humor as a psychological defense mechanism is rooted in his childhood trauma---his parents' divorce~\cite{wikipedia_Chandler_Bing}. This origin story helps explain the \archetypelinksimple{Outcast}{Outcast} dimension: he consistently positions himself as an observer in social situations and thus uses jokes to maintain emotional distance while still seeking connection. His awkward and self-conscious traits are repeatedly foregrounded, most famously in his inability to break up with Janice effectively. 
\begin{quote}
\textbf{Chandler:} I hear ya. (Pause) But! Unfortunately, my company is transferring me overseas!\\
\textbf{Janice:} Oh no! Where to? (Gasps) To Paris?\\
\textbf{Chandler:} No! No! Not, Paris.\\
\textbf{Janice:} To London? No-no, Rome? Vienna? Ooh-ooh, Barcelona?\\
\textbf{Chandler:} Okay, could you just stop talking for a second? (Thinks) Yemen. That's right, yes, I'm being transferred to Yemen!\\
\end{quote}
His desperate lie of moving to Yemen shows both his social clumsiness and his resistance to direct emotional confrontation. \characterlinksimple{Friends-Chandler-Bing}{Chandler} shows notable similarity to \characterlinksimple{The-Big-Sick-Kumail}{Kumail} in \storylinksimple{The-Big-Sick}{The Big Sick}~\cite{imdb_The_Big_Sick}. Both deploy humor as a coping mechanism to navigate anxieties. Both characters' comedic personas function as a shield that gradually gives way to genuine emotional commitment as their respective narratives progress.

\subsubsection{\characterlinksimple{Friends-Ross-Geller}{Ross}}
\characterlinksimple{Friends-Ross-Geller}{Ross Geller} is characterized by a \archetypelinksimple{Traditionalist}{Traditionalist}--\archetypelinksimple{Outcast}{Outcast}--\archetypelinksimple{Diva}{Diva} configuration, with strong loadings on the \archetypelinksimple{Traditionalist}{Traditionalist} (29.7), \archetypelinksimple{Diva}{Diva} (37.6), and \archetypelinksimple{Outcast}{Outcast} (30.3) dimensions as shown in Fig.~\ref{fig:Friends_Main_Characters_6Dimensions}. At the trait level, \traitlinksimplerightalt{junkie}{straight-edge}{junkie}{straight edge}, \traitlinksimpleright{sporty}{bookish}, \traitlinksimpleright{jock}{nerd}, \traitlinksimpleright{artistic}{scientific}, \traitlinksimpleright{crafty}{scholarly}, \traitlinksimpleright{physical}{intellectual}, and \traitlinksimplerightalt{likes-change}{resists-change}{likes change}{resists change} align closely with the \archetypelinksimple{Traditionalist}{Traditionalist} dimension, emphasizing rule-following and predictability (see Character Card in Fig.~\ref{fig:Ross}). \characterlinksimple{Friends-Ross-Geller}{Ross}'s \archetypelinksimple{Traditionalist}{Traditionalist} orientation is most evident in his deep investment in intellectual authority and his compulsion to be right. For example, in S02E03, he tries to convince \characterlinksimple{Friends-Phoebe-Buffay}{Phoebe} to accept evolution:
\begin{quote}
\textbf{Ross:} You don't believe in evolution?\\
\textbf{Phoebe:} ...I just think it's a little too easy.\\
\textbf{Ross:} Too easy? The process of every living thing on this planet evolving over millions of years from single-celled organisms---too easy?\\
\textbf{Phoebe:} Yeah, I just don't buy it.\\
\textbf{Ross:} Evolution is not for you to buy, Phoebe. Evolution is scientific fact, like the air we breathe, like gravity.\\
\end{quote}
His \archetypelinksimple{Diva}{Diva} dimension manifests as emotional intensity and dramatic self-presentation. In S06E10, his cringe-worthy performance of a high-school dance routine on Dick Clark's New Year's Eve special suggests a need for recognition and attention that borders on the theatrical. Also, his \archetypelinksimple{Outcast}{Outcast} dimension is probably rooted in the experience of rejection. For example, in S02E14, the prom video episode, he had been secretly preparing to take \characterlinksimple{Friends-Rachel-Green}{Rachel} to the prom after her date stood her up. But he eventually was left standing alone and humiliated when her date finally appeared, which could explain some of his insecurity. Among characters in the broader archetypal corpus, \characterlinksimple{Friends-Ross-Geller}{Ross} shows notable similarity to \characterlinksimple{The-Hangover-Stu}{Stu} from \storylinksimple{The-Hangover}{The Hangover}~\cite{imdb_The_Hangover} (a near-100\% match). They are both educated professionals who consistently ``do everything by the book.'' They represent moral authority and social norms within their respective groups, yet repeatedly find themselves in situations beyond their control.

Across the six main characters, the archetypal profiles show a high degree of agreement with their well-established narrative identities. The dominant archetypes consistently correspond to recognizable social roles presented throughout the show. This consistency substantiates the claim that the archetypometric framework, while derived from a large corpus, retains its descriptive power when applied to specific textually grounded cases.

\subsection*{RQ2: Character Positioning in the Archetype Space}

\begin{figure*}[t!]
      \centering
      \includegraphics[width=0.7\linewidth]{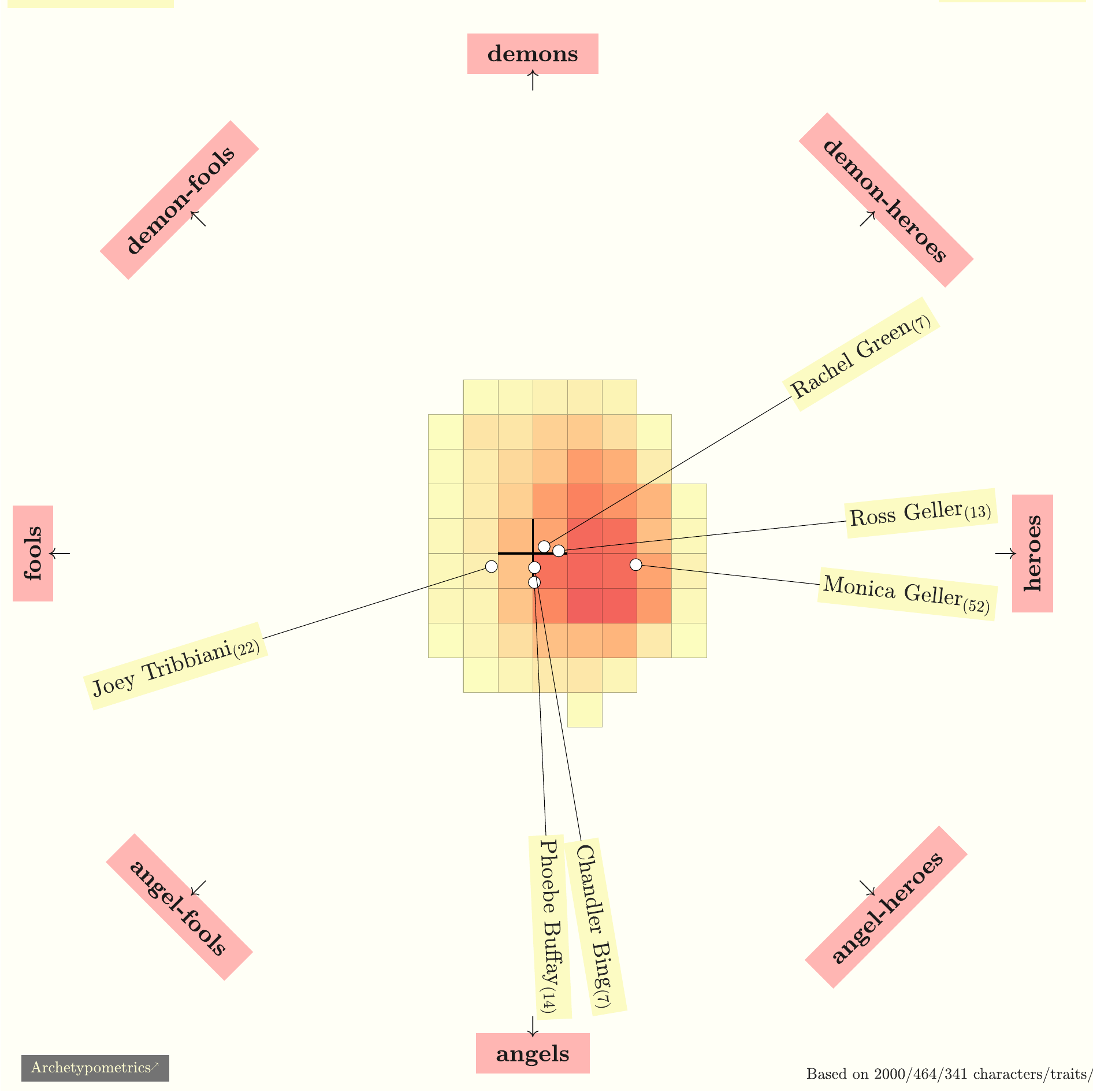}
      \caption{Two-dimensional projection of the six main characters onto the first and second essential dimensions, illustrating their directional alignment within archetypal space.}
      \label{fig:1-2}
    \end{figure*}

\begin{figure*}[t!]
      \centering
      \includegraphics[width=0.7\linewidth]{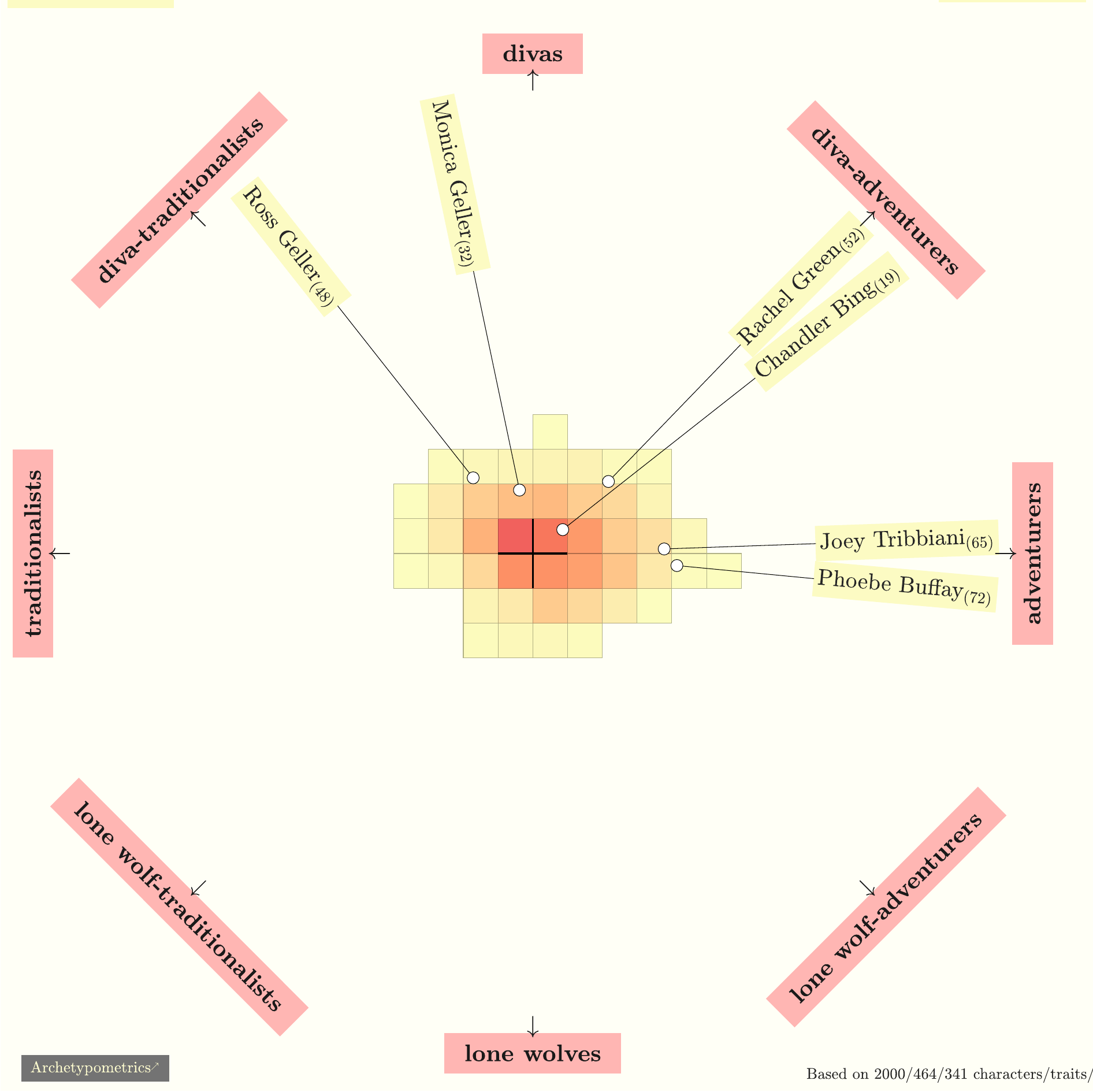}
      \caption{Two-dimensional projection of the six main characters onto the third and fourth essential dimensions, illustrating their directional alignment within archetypal space.}
      \label{fig:3-4}
    \end{figure*}

\begin{figure*}[t!]
      \centering
      \includegraphics[width=0.7\linewidth]{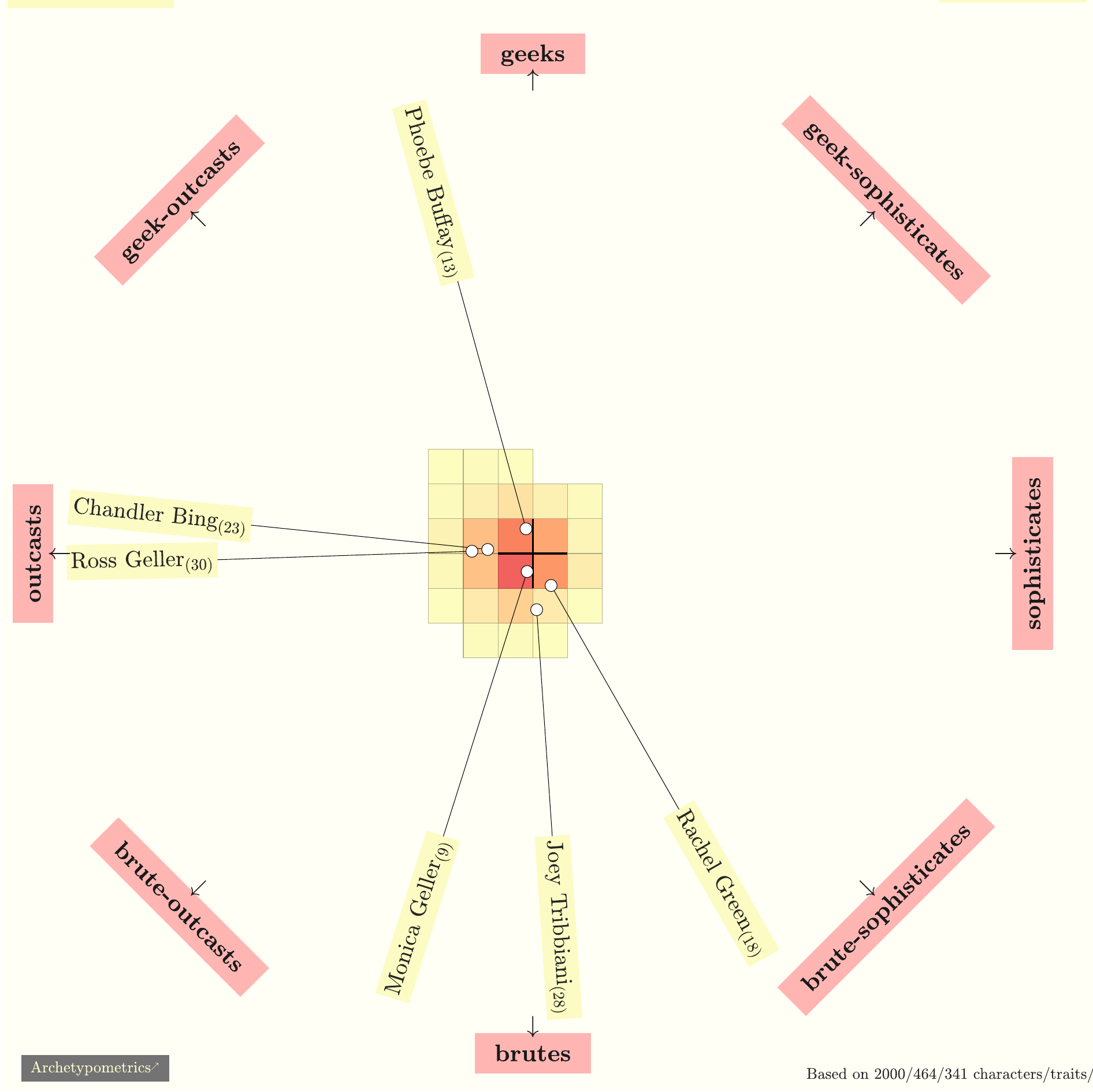}
      \caption{Two-dimensional projection of the six main characters onto the fifth and sixth essential dimensions, illustrating their directional alignment within archetypal space.}
      \label{fig:5-6}
    \end{figure*}

A single dimension shows how closely a character matches one archetype, but on its own, it cannot capture the characters' relative narrative positions. To examine this structure, we project the archetype vectors of the six characters onto three selected pairs of dimensions (Figs.~\ref{fig:1-2}, \ref{fig:3-4}, \ref{fig:5-6}). 

The first projection (Fig.~\ref{fig:1-2}) visualizes the characters along the \archetypesemdiff{1} and \archetypesemdiff{2} dimensions. The six characters occupy a compressed central band along the \archetypesemdiff{2} axis, exhibiting little vertical separation from one another. In sitcoms like \storylinksimple{Friends}{Friends}, all the core characters are essentially ``good people''. They may have flaws or moments of selfishness, but overall, they lack any truly ``demonic'' quality. Consequently, this \archetypesemdiff{2} dimension lacks sufficient variation to distinguish the characters from one another in sitcoms. Along the \archetypesemdiff{1} dimension, by contrast, \characterlinksimple{Friends-Monica-Geller}{Monica} occupies the most extreme position at the \archetypelinksimple{Hero}{Hero} end, \characterlinksimple{Friends-Joey-Tribbiani}{Joey} occupies the opposite \archetypelinksimple{Fool}{Fool} end, and the remaining four characters cluster in the intermediate region between them. \characterlinksimple{Friends-Monica-Geller}{Monica}'s extreme \archetypelinksimple{Hero}{Hero} positioning can be understood through her distinctive mode of goal-directed behavior, as illustrated before. The \archetypelinksimple{Hero}{Hero} here includes order and control, which represent her core identity~\cite{wikipedia_Monica_Geller}. \characterlinksimple{Friends-Joey-Tribbiani}{Joey}'s \archetypelinksimple{Fool}{Fool} positioning, by contrast, represents the inverse of this structure. His character operates on impulse; his goals are short-term and quickly satisfied (like food, social approval, or casual romance)~\cite{wikipedia_Joey_Tribbiani}. 

The second projection (Fig.~\ref{fig:3-4}) displays the characters on the \archetypesemdiff{3} and \archetypesemdiff{4} dimensions. As shown in the figure, the distribution along the vertical \archetypesemdiff{4} axis is asymmetrical: all six characters are positioned in the \archetypelinksimple{Diva}{Diva} hemisphere, with \characterlinksimple{Friends-Ross-Geller}{Ross}, \characterlinksimple{Friends-Monica-Geller}{Monica}, and \characterlinksimple{Friends-Chandler-Bing}{Chandler} extending to the most extreme \archetypelinksimple{Diva}{Diva} end, while no character occupies the \archetypelinksimple{Lone-Wolf}{Lone Wolf} region. This means that the social world of \storylinksimple{Friends}{Friends} is characterized by a high degree of interaction among its members, where the traits of all the main characters are defined by their roles within the group. This aligns with the essence of the sitcom form. Along the horizontal \archetypesemdiff{3} axis, by contrast, the characters are widely distributed. \characterlinksimple{Friends-Joey-Tribbiani}{Joey} and \characterlinksimple{Friends-Phoebe-Buffay}{Phoebe} occupy the \archetypelinksimple{Adventurer}{Adventurer} end, consistent with their shared narrative orientation toward spontaneity and unconventionality. As illustrated in the previous section, evidence includes \characterlinksimple{Friends-Joey-Tribbiani}{Joey}'s impulsive purchase of a boat and \characterlinksimple{Friends-Phoebe-Buffay}{Phoebe}'s conviction that a stray cat embodies her mother's spirit. \characterlinksimple{Friends-Ross-Geller}{Ross}, in sharp contrast, is positioned toward the \archetypelinksimple{Traditionalist}{Traditionalist} end, aligning with his characterization as a convention-driven academic who values structure and social recognition. Indeed, much of the comedic tension and conflict in sitcoms arises from the clash between \archetypelinksimple{Traditionalist}{Traditionalists} and \archetypelinksimple{Adventurer}{Adventurers}. As Bergson argues~\cite{BergsonLaughterAE}, comedy derives from the friction between rigid rule-following and flexible spontaneity---a dynamic that finds clear expression in the \archetypesemdiff{3} axis of our archetype space. 

The third projection (Fig.~\ref{fig:5-6}) displays the characters on the \archetypesemdiff{6} and \archetypesemdiff{5} dimensions. Along the vertical \archetypesemdiff{5} axis, \characterlinksimple{Friends-Chandler-Bing}{Chandler} and \characterlinksimple{Friends-Ross-Geller}{Ross} are positioned distinctly toward the \archetypelinksimple{Outcast}{Outcast} end, while the remaining four characters occupy a more central band. This separation reflects a shared narrative function: both \characterlinksimple{Friends-Chandler-Bing}{Chandler} and \characterlinksimple{Friends-Ross-Geller}{Ross} are characterized by forms of social awkwardness that set them apart from the group's social mainstream. \characterlinksimple{Friends-Ross-Geller}{Ross}'s \archetypelinksimple{Outcast}{Outcast} positioning derives from his pedantic academic inflexibility and obsessive attachment to intellectual correctness. \characterlinksimple{Friends-Chandler-Bing}{Chandler}'s \archetypelinksimple{Outcast}{Outcast} positioning, by contrast, comes from his defensive sarcasm and emotional guardedness. This distances him from genuine social connection until his later character development with \characterlinksimple{Friends-Monica-Geller}{Monica}. Along the horizontal \archetypesemdiff{6} axis, \characterlinksimple{Friends-Phoebe-Buffay}{Phoebe} and \characterlinksimple{Friends-Joey-Tribbiani}{Joey} occupy the opposing poles. \characterlinksimple{Friends-Phoebe-Buffay}{Phoebe}'s \archetypelinksimple{Geek}{Geek} alignment is closely linked to her eccentric intellectualism relative to the rest of the group. \characterlinksimple{Friends-Joey-Tribbiani}{Joey}'s \archetypelinksimple{Brute}{Brute} positioning reflects his direct, non-intellectual approach to problem-solving. This separation between \characterlinksimple{Friends-Phoebe-Buffay}{Phoebe} and \characterlinksimple{Friends-Joey-Tribbiani}{Joey} along the intellectual axis is not visible in the first two projections. The third projection thus reveals that their shared spontaneity and unconventionality mask a deep divergence: \characterlinksimple{Friends-Phoebe-Buffay}{Phoebe}'s eccentric intellectualism and \characterlinksimple{Friends-Joey-Tribbiani}{Joey}'s rejection of intellectual effort.

The three projections collectively demonstrate that the geometric configuration produced by the archetype framework is interpretable in narrative terms. This interpretability validates the use of archetypal projection as a tool for visualizing and exploring ensemble narrative structure, providing a spatial complement to the individual character profiles examined in RQ1. 

There is another implication for the applicability of archetype representations to narrative analysis. Not all archetypal dimensions contribute equally to character differentiation within a given narrative context. For instance, the \archetypesemdiff{2} axis proved uninformative in the \storylinksimple{Friends}{Friends} ensemble, indicating a generic constraint of the sitcom form, where characters are uniformly positioned as morally sympathetic people regardless of their individual flaws~\cite{rosser2022lawfriends}. In dramatic or villain-centered narratives, this dimension might become the primary axis of differentiation. Researchers applying archetype frameworks should therefore be mindful of the genre-specific activation of different dimensions.

\subsection*{RQ3: Archetypal Pairwise Comparison}
\begin{figure}[ht!]
      \centering
      \includegraphics[width=1.0\linewidth]{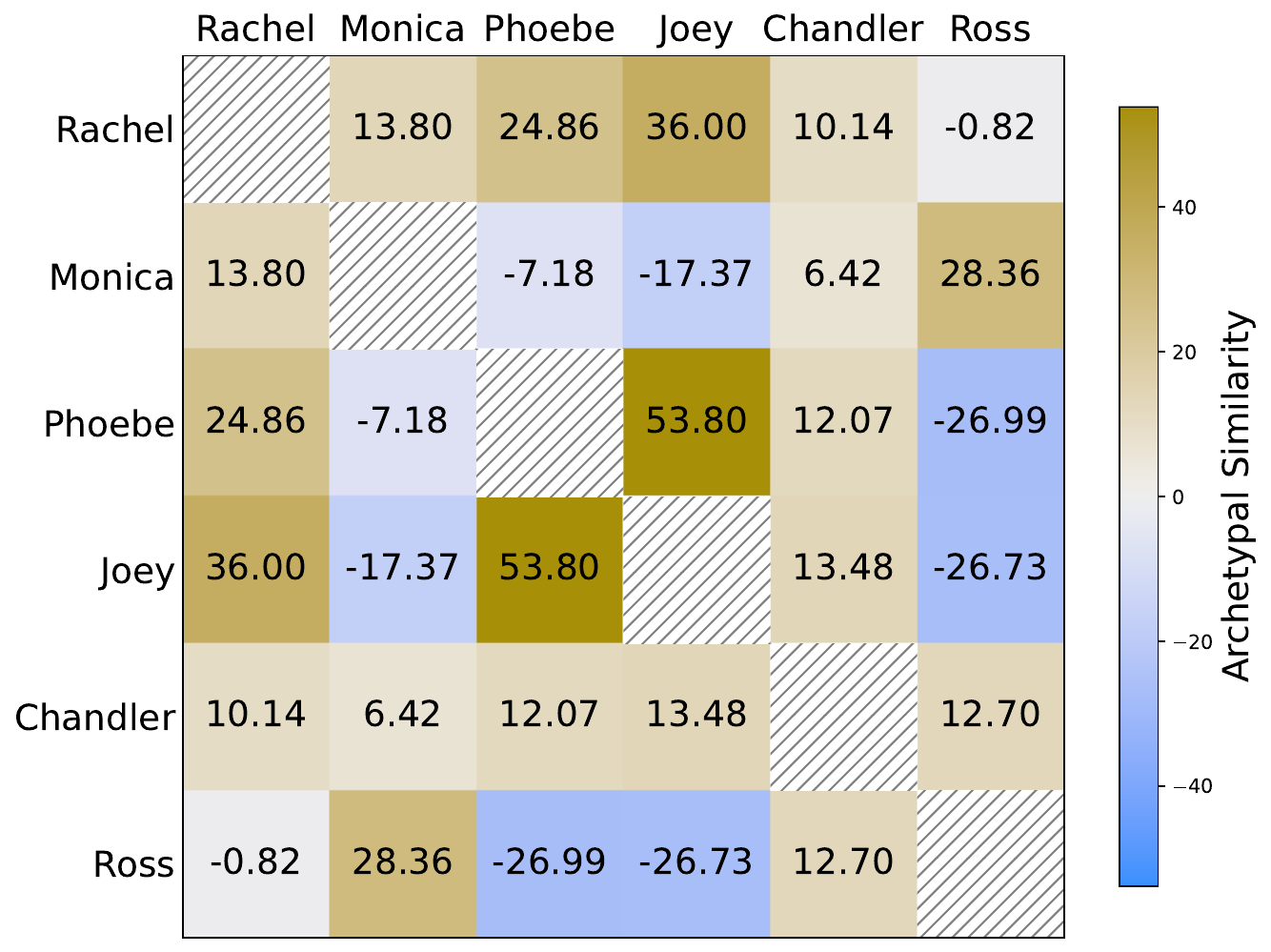}
      \caption{Pairwise archetypal similarity between the six main characters, computed as the inner product of their archetypal vectors.}
      \label{fig:6x6}
    \end{figure}

To visualize the pairwise relationships, we construct a $6 \times 6$ similarity matrix for the six main characters, where each entry represents the inner product between a pair of archetypal vectors. The resulting matrix is shown in Fig.~\ref{fig:6x6}, which provides a complete view of pairwise structural similarity within the ensemble. Positive values indicate alignment, negative values indicate contrast, and values close to zero indicate orthogonality. Trait- and archetype-based comparison figures for all 15 character pairs are provided in Appendix~\ref{Appendix:Pairwise_Comparison_Trait}.
As illustrated in Fig.~\ref{fig:6x6}, the similarity values span a wide range, from strongly positive (e.g., $53.80$) to strongly negative (e.g., $-26.99$), with several pairs clustering near zero (e.g., $-0.82$). To operationalize these similarity measures for narrative analysis, we interpret the sign and magnitude of each inner product as indicating distinct modes of character relationship. Positive similarity signals alignment. Characters with positive similarity tend to occupy similar narrative roles and respond to situations in comparable ways. Negative values signal contrast and structural tension. Such characters carry an inherent potential for dramatic conflict. Such tension is not necessarily the same as hostility. Instead, their archetypal differences create instability and negotiation in the narrative field. Between these two poles, values near zero suggest orthogonality, where the two characters occupy distinct regions of the archetypal space, sharing neither overlapping functions nor directly opposing orientations. This condition opens the possibility of complementarity, in which the characters' contributions to the narrative are distinct and non-redundant. This can support relationships grounded in differentiation rather than similarity or opposition. 

We now examine three representative pairs from the similarity matrix: one exhibiting strong positive correlation, one exhibiting strong negative correlation, and one from the orthogonal range. Together, they demonstrate how archetypal vectors capture distinct forms of narrative structuring across the six main characters.

\paragraph{Alignment.}
    \begin{figure*}[ht!]
      \centering
      \includegraphics[width=1.0\linewidth]{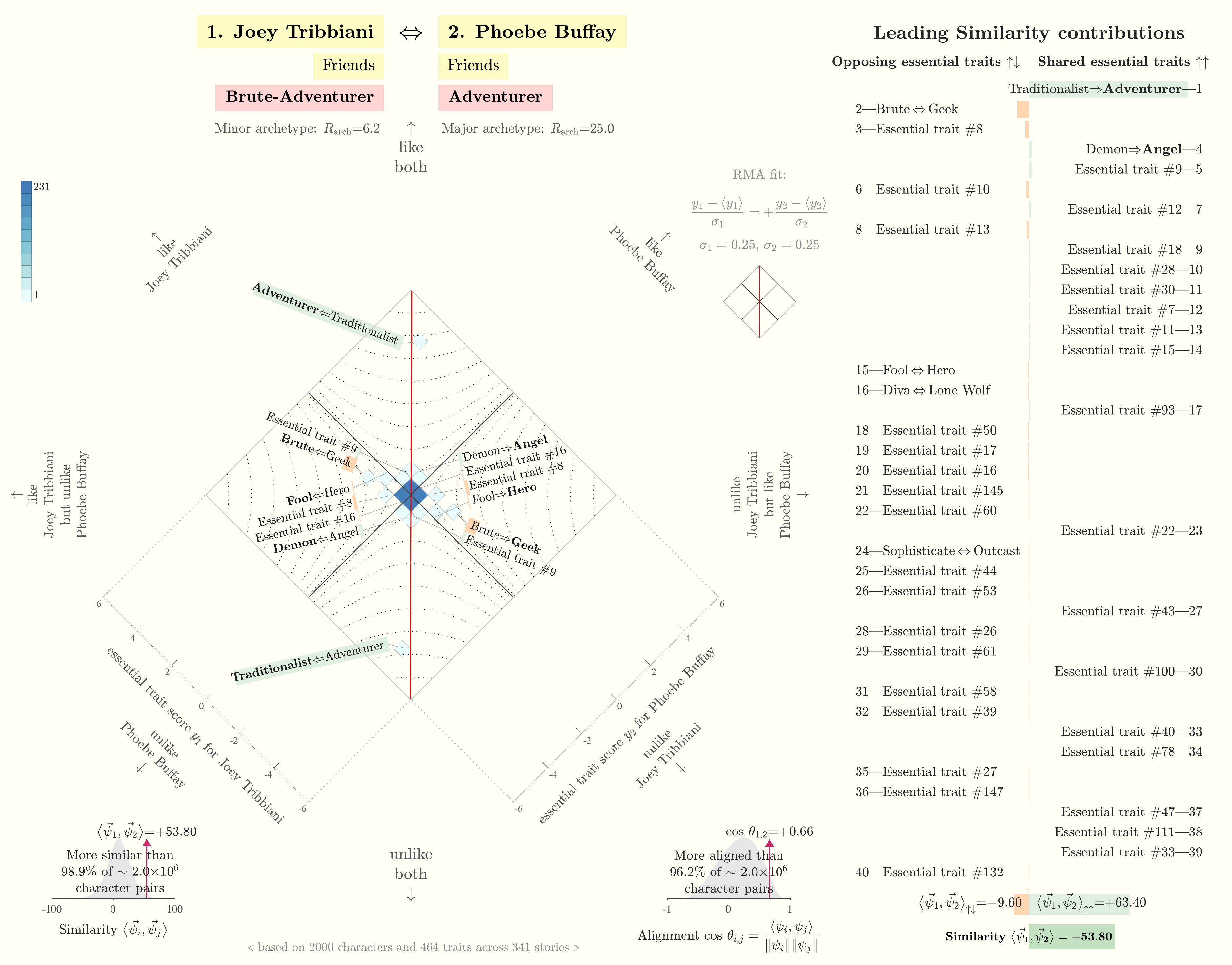}
      \caption{Archetypal comparison between \characterlinksimple{Friends-Joey-Tribbiani}{Joey Tribbiani} and \characterlinksimple{Friends-Phoebe-Buffay}{Phoebe Buffay} in the dominant essential subspace. The radial visualization shows their loadings across the leading essential dimensions. The right-hand panel lists the principal trait-level contributions to similarity. With an inner product of $+53.80$ and cosine alignment of $0.66$, the pair exhibits strong structural alignment within the ensemble.}
      \label{fig:joey_phoebe_archetypes}
    \end{figure*}

Among all character pairs in \storylinksimple{Friends}{Friends}, \characterlinksimple{Friends-Joey-Tribbiani}{Joey Tribbiani} and \characterlinksimple{Friends-Phoebe-Buffay}{Phoebe Buffay} show the highest similarity in the dominant archetype space (Fig.~\ref{fig:joey_phoebe_archetypes}), with an inner product of $+53.80$, exceeding 96.2\% of pairs in the reference dataset. This alignment is driven primarily by their shared \archetypelinksimple{Adventurer}{Adventurer} orientation: \characterlinksimple{Friends-Phoebe-Buffay}{Phoebe} strongly expresses this dimension, while \characterlinksimple{Friends-Joey-Tribbiani}{Joey} exhibits a \archetypelinksimple{Brute}{Brute}--\archetypelinksimple{Adventurer}{Adventurer} profile. Their overlap is reflected in traits such as \traitlinksimpleright{deliberate}{spontaneous}, \traitlinksimpleright{serious}{playful}, \traitlinksimpleright{monotone}{expressive}, \traitlinksimpleright{negative}{positive}, and \traitlinksimplerightalt{uncreative}{open-to-new-experiences}{uncreative}{open to new experiences} (see the trait-based comparison in Fig.~\ref{fig:phoebe_joey_traits}). Together, these traits indicate low structural constraint and a strong experiential orientation, placing them in a local cluster within the archetypal space.


This spatial proximity indicates structural similarity within a narrative system: these two characters reinforce a shared mode of variation within the ensemble. To understand what these functions are, we turn to a basic principle of narrative construction. Most narrative systems establish themselves through the contrast between `what is normal' and `what is abnormal'~\cite{GarlandThomson2006NarrativePD}. \characterlinksimple{Friends-Joey-Tribbiani}{Joey} and \characterlinksimple{Friends-Phoebe-Buffay}{Phoebe} jointly serve as a frame of reference for the `abnormal' within the \storylinksimple{Friends}{Friends} ensemble. They realize this function differently, however. In a comedy set in urban middle-class life, \characterlinksimple{Friends-Joey-Tribbiani}{Joey}'s understanding of the world remains at the level of the senses---immediate and concrete. \characterlinksimple{Friends-Phoebe-Buffay}{Phoebe}'s abnormality is of a different kind: her logic forms a self-contained system that operates according to its own internal rules, which is completely different from mainstream social norms. Both characters function as agents of unpredictability: they often interrupt the narrative precisely at moments when the plot is about to slide into predictability.

    \begin{figure*}[ht!]
      \centering
      \includegraphics[width=1.0\linewidth]{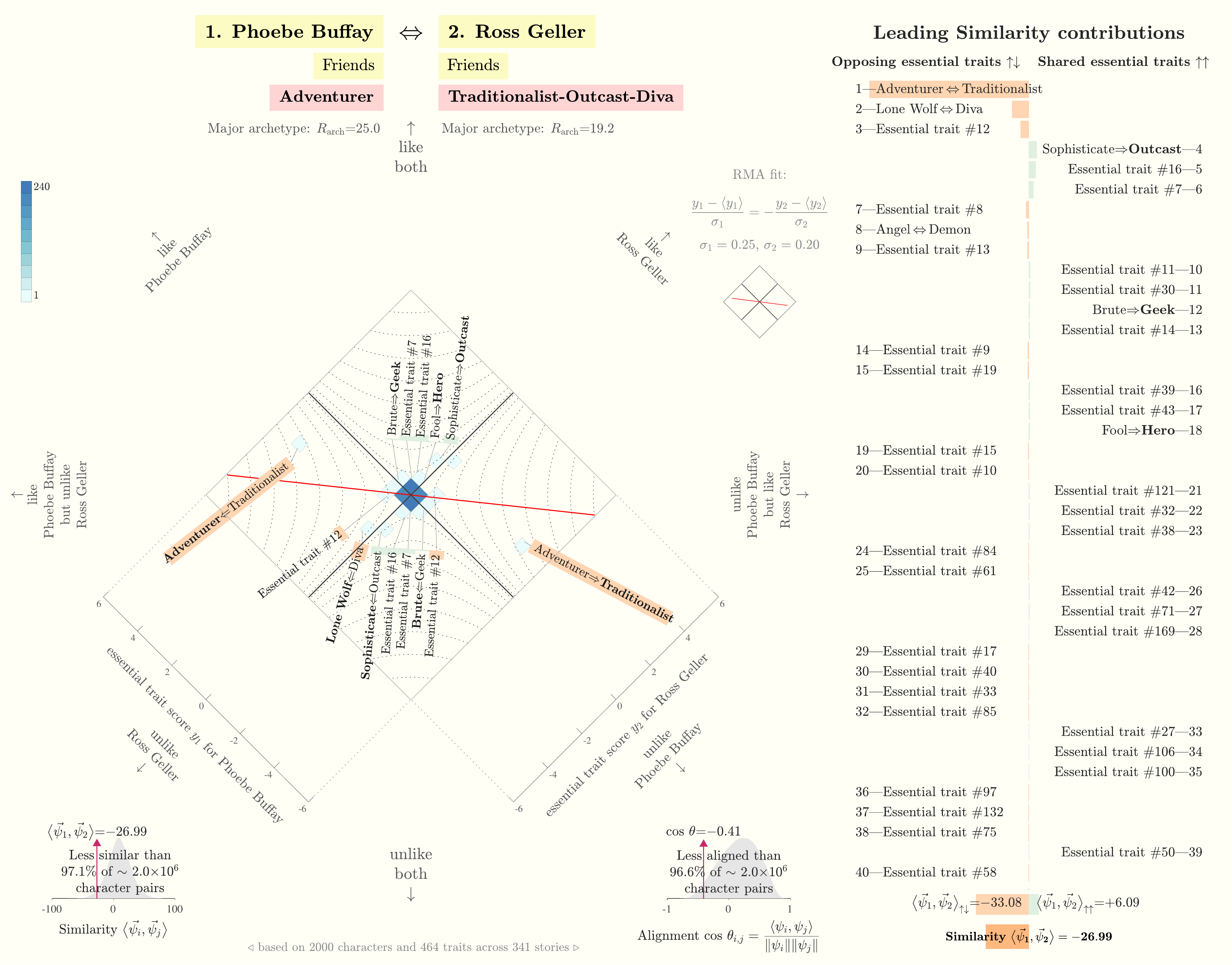}
      \caption{Archetypal comparison between \characterlinksimple{Friends-Phoebe-Buffay}{Phoebe Buffay} and \characterlinksimple{Friends-Ross-Geller}{Ross Geller}. The radial visualization displays their loadings across the dominant essential dimensions, while the right panel lists the leading trait-level contributions to similarity. The pair exhibits a negative inner product of $-26.99$ and cosine alignment of $-0.41$ (which is the cosine of the angle between these two character vectors).}
      \label{fig:phoebe_ross_archetype}
    \end{figure*}

\paragraph{Contrast.} The least similar pair in the ensemble is \characterlinksimple{Friends-Phoebe-Buffay}{Phoebe Buffay} and \characterlinksimple{Friends-Ross-Geller}{Ross Geller} (Fig.~\ref{fig:phoebe_ross_archetype}). Their inner product is $-26.99$, placing them below 96.6\% of the approximately two million character pairs in the reference distribution~\cite{dodds2021archetypometrics}. This divergence is driven primarily by the \archetypesemdiff{3} dimension. \characterlinksimple{Friends-Phoebe-Buffay}{Phoebe}'s profile reflects experiential openness and resistance to constraint, whereas \characterlinksimple{Friends-Ross-Geller}{Ross}'s is organized around order and intellectual authority. Together, they define opposing poles of the ensemble's narrative space, ranging from anti-structure to hyper-structure. The remaining characters occupy intermediate positions along this axis: \characterlinksimple{Friends-Joey-Tribbiani}{Joey} lies closer to \characterlinksimple{Friends-Phoebe-Buffay}{Phoebe}, \characterlinksimple{Friends-Monica-Geller}{Monica} closer to \characterlinksimple{Friends-Ross-Geller}{Ross}, while \characterlinksimple{Friends-Chandler-Bing}{Chandler} and \characterlinksimple{Friends-Rachel-Green}{Rachel} fall nearer the middle.


The narrative significance of this polarity is especially visible in ``The One Where Heckles Dies'' (S02E03), where \characterlinksimple{Friends-Ross-Geller}{Ross} and \characterlinksimple{Friends-Phoebe-Buffay}{Phoebe} debate evolutionary theory. \characterlinksimple{Friends-Ross-Geller}{Ross} treats evolution as settled scientific fact, while \characterlinksimple{Friends-Phoebe-Buffay}{Phoebe} resists the closure of alternative possibilities:

\begin{quote}
\textbf{Ross:} Evolution is scientific fact, like the air we breathe, like gravity.\\
\textbf{Phoebe:} Yeah, I just don't buy it.\\
\end{quote}

\characterlinksimple{Friends-Ross-Geller}{Ross} later presents fossils and insists that evolution is ``the only possibility,'' whereas \characterlinksimple{Friends-Phoebe-Buffay}{Phoebe} challenges his certainty:

\begin{quote}
\textbf{Phoebe:} Could you just open your mind like this much? [...] Are you telling me that you are so unbelievably arrogant that you can't admit that there's a teeny tiny possibility that you could be wrong about this?\\
\textbf{Ross:} There might be a teeny, tiny possibility.\\
\textbf{Phoebe:} I can't believe you caved.\\
\end{quote}

The scene dramatizes \characterlinksimple{Friends-Ross-Geller}{Ross}'s commitment to order, evidence, and epistemic certainty against \characterlinksimple{Friends-Phoebe-Buffay}{Phoebe}'s openness to ambiguity and resistance to authoritative closure. The other four characters observe the exchange from the margins, positioning themselves between these opposing poles. The audience is similarly invited to view \characterlinksimple{Friends-Ross-Geller}{Ross} and \characterlinksimple{Friends-Phoebe-Buffay}{Phoebe} as complementary extremes whose tension helps define the ensemble's narrative space.

    \begin{figure*}[ht!]
      \centering
      \includegraphics[width=1.0\linewidth]{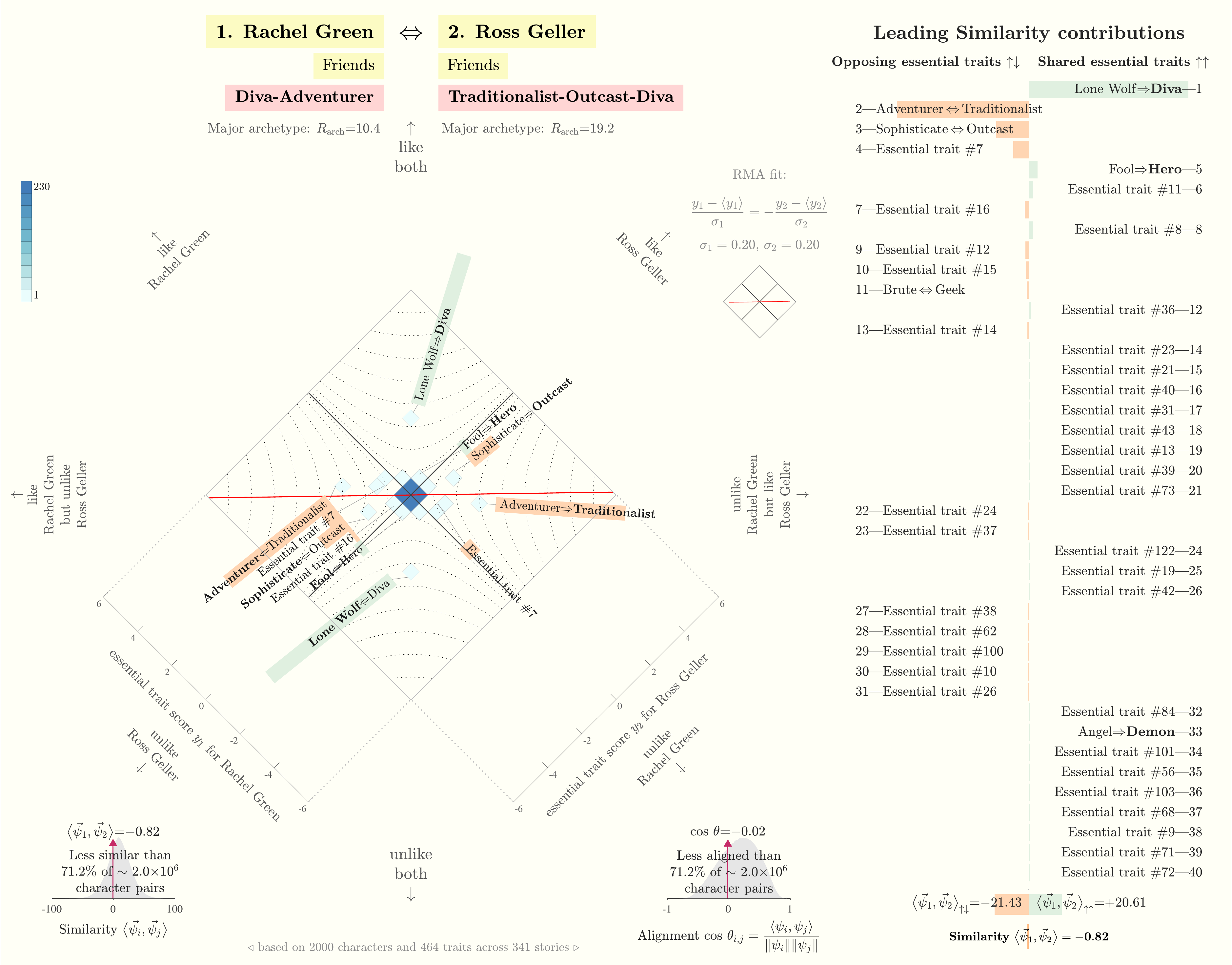}
      \caption{Archetypal comparison between \characterlinksimple{Friends-Rachel-Green}{Rachel Green} and \characterlinksimple{Friends-Ross-Geller}{Ross Geller}. The radial visualization displays their loadings across the dominant essential dimensions, while the right panel lists the leading trait-level contributions to similarity. The pair exhibits an inner product of $-0.82$ and cosine alignment of $-0.02$ (which is the cosine of the angle between these two character vectors).}
      \label{fig:rachel_ross_archetype}
    \end{figure*}

\paragraph{Orthogonality.}
Pairs with similarity values near zero reveal a distinct relational structure. \characterlinksimple{Friends-Rachel-Green}{Rachel Green} and \characterlinksimple{Friends-Ross-Geller}{Ross Geller} have an inner product of $-0.82$, placing them near statistical orthogonality within the dominant essential subspace~\cite{dodds2021archetypometrics}. As shown in Fig.~\ref{fig:rachel_ross_archetype}, this result reflects a positive shared contribution along the \archetypelinksimple{Diva}{Diva} dimension, associated with social awareness and sensitivity, offset by opposing positions on the \archetypesemdiff{3} axis: \characterlinksimple{Friends-Rachel-Green}{Rachel} aligns more strongly with the \archetypelinksimple{Adventurer}{Adventurer} pole, whereas \characterlinksimple{Friends-Ross-Geller}{Ross} aligns with the \archetypelinksimple{Traditionalist}{Traditionalist} pole.

This structure is reflected in their narrative trajectories. \characterlinksimple{Friends-Ross-Geller}{Ross} is oriented toward a stable life script centered on marriage, children, and long-term certainty, while \characterlinksimple{Friends-Rachel-Green}{Rachel} begins the series by abandoning her wedding and gradually constructs an independent identity, progressing from waitress to fashion executive. Their opposing orientations repeatedly prevent full synchronization of their life plans, yet their shared \archetypelinksimple{Diva}{Diva} traits sustain emotional legibility and attachment across repeated separations. Their relationship is therefore better understood not as alignment or direct contrast, but as the recurring intersection of distinct narrative dimensions: two trajectories that repeatedly cross without fully converging.

\begin{figure*}[ht!]
      \centering
      \includegraphics[width=1.0\linewidth]{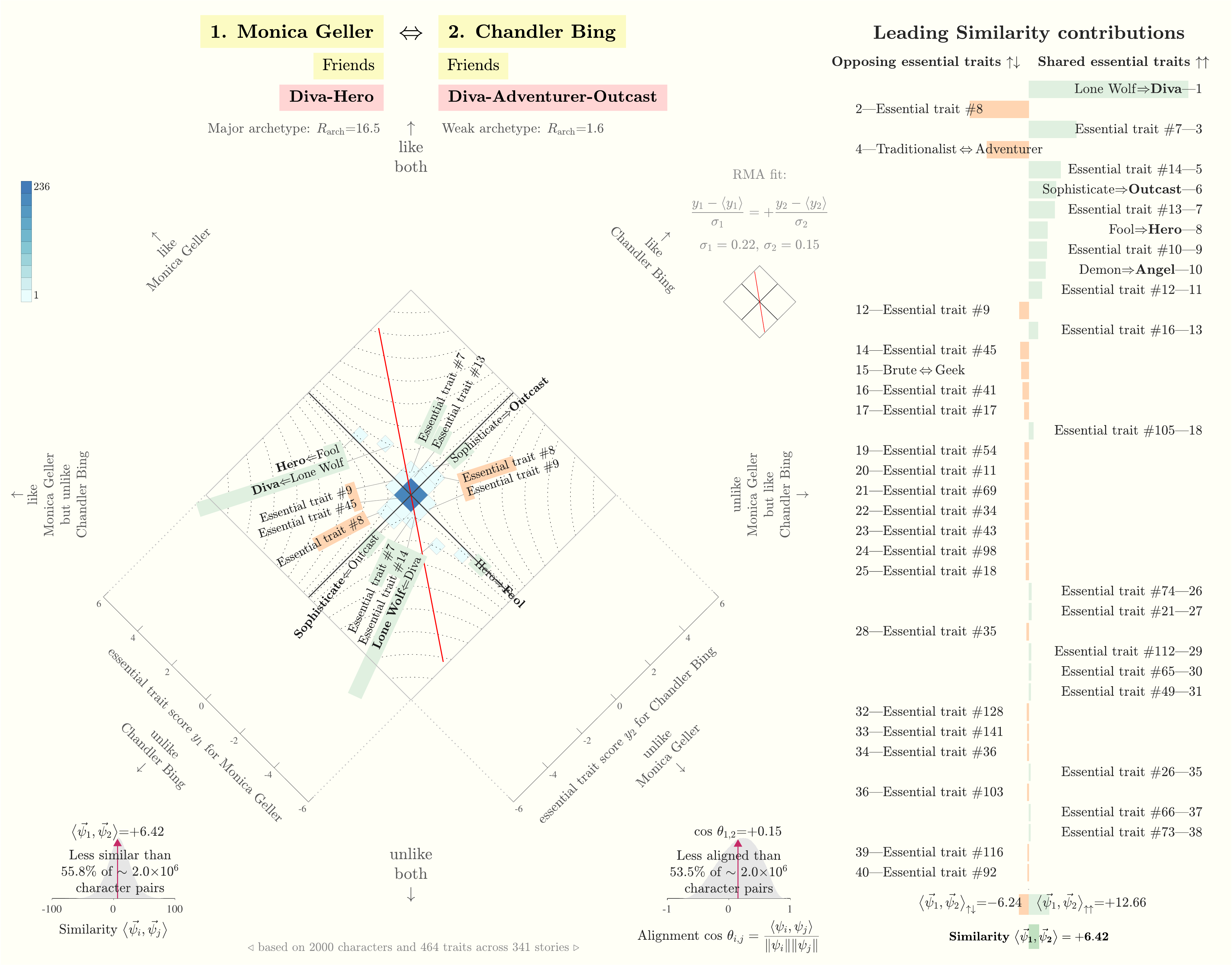}
      \caption{Archetypal comparison between \characterlinksimple{Friends-Monica-Geller}{Monica Geller} and \characterlinksimple{Friends-Chandler-Bing}{Chandler Bing}. The radial visualization displays their loadings across the dominant essential dimensions, while the right panel lists the leading trait-level contributions to similarity. The pair exhibits an inner product of $+6.42$.}
      \label{fig:monica_chandler_archetype}
    \end{figure*}

A similar pattern appears in the series' other central romantic pair. \characterlinksimple{Friends-Monica-Geller}{Monica Geller} and \characterlinksimple{Friends-Chandler-Bing}{Chandler Bing} have an inner product of $+6.42$, likewise placing them near orthogonality (Fig.~\ref{fig:monica_chandler_archetype}). They share positive contributions along the \archetypelinksimple{Diva}{Diva} dimension while differing on the \archetypesemdiff{3} axis, though less sharply than \characterlinksimple{Friends-Ross-Geller}{Ross} and \characterlinksimple{Friends-Rachel-Green}{Rachel}. \characterlinksimple{Friends-Monica-Geller}{Monica}'s \archetypelinksimple{Traditionalist}{Traditionalist} orientation emphasizes order, control, and stability, whereas \characterlinksimple{Friends-Chandler-Bing}{Chandler}'s \archetypelinksimple{Adventurer}{Adventurer} orientation is expressed through spontaneity, humor, and emotional flexibility. Their complementarity is articulated clearly in \characterlinksimple{Friends-Chandler-Bing}{Chandler}'s appeal to their prospective surrogate mother:

\begin{quote}
\textbf{Chandler:} My wife's an incredible woman. She's loving and devoted and caring. [...] I love my wife more than anything in this world. [...] She's a mother... without a baby. Please?\\
\end{quote}

Here, \characterlinksimple{Friends-Monica-Geller}{Monica} provides structure and devotion, while \characterlinksimple{Friends-Chandler-Bing}{Chandler} contributes flexibility and emotional openness. Unlike \characterlinksimple{Friends-Ross-Geller}{Ross} and \characterlinksimple{Friends-Rachel-Green}{Rachel}, whose orthogonality produces a prolonged asynchronous attachment, \characterlinksimple{Friends-Monica-Geller}{Monica} and \characterlinksimple{Friends-Chandler-Bing}{Chandler} gradually integrate their differences into a stable partnership.

Taken together, these two relationships suggest that romantic pairings in \storylinksimple{Friends}{Friends} are structured through complementarity rather than archetypal duplication. Each partner contributes distinct narrative qualities, generating both tension and compatibility. Such orthogonality may represent a broader sitcom logic: romantic partners must differ enough to sustain conflict and comedy, yet remain sufficiently compatible to support long-term emotional investment.

Overall, the three structural conditions identified in this analysis (alignment, contrast, and orthogonality) outline the relational geometry of the ensemble. Alignment, as illustrated by \characterlinksimple{Friends-Joey-Tribbiani}{Joey} and \characterlinksimple{Friends-Phoebe-Buffay}{Phoebe}, shows that two characters aligning along a shared dominant essential direction occupy similar positions within the archetypal space and reinforce a common narrative function. Contrast, shown by \characterlinksimple{Friends-Phoebe-Buffay}{Phoebe} and \characterlinksimple{Friends-Ross-Geller}{Ross}, represents structural divergence, emphasizing tension across essential dimensions and defining the polar boundaries of the ensemble's narrative. Orthogonality, observed in pairs such as \characterlinksimple{Friends-Ross-Geller}{Ross}--\characterlinksimple{Friends-Rachel-Green}{Rachel}, refers to structural independence, in which the characters occupy orthogonal positions and contribute distinct, non-overlapping components. 

This distributed architecture resonates with Propp's foundational observation~\cite{Propp1959MorphologyOT} that narrative structure emerges from the distribution of functions across character roles. While Propp focused on examining narrative functions through the successive actions of characters within a linear plot, our analysis reveals a distribution of archetypal functions within an ensemble, particularly in complementary pairs like \characterlinksimple{Friends-Ross-Geller}{Ross} and \characterlinksimple{Friends-Rachel-Green}{Rachel}. In addition, our analysis provides a quantitative operationalization of Mittell's observation~\cite{mittell2015complex} that complex ensemble television distributes the narrative-driving function across multiple characters. Narrative weight circulates among the six characters within the ensemble, with different pairs activating different relational modes depending on the demands of the scene or storyline.

\section{Conclusion}
This study aims to investigate whether computationally derived archetypes provide an interpretable representation of the six main characters in \storylinksimple{Friends}{Friends}, and more importantly, whether such representations can extend beyond individual character classification to capture relational dynamics within an ensemble cast. 

First, we find that the archetypal profiles generated by the archetypometric framework align consistently with narrative evidence. These correspondences demonstrate that the computational archetypometric framework retains descriptive traction when applied to specific textually grounded cases. 

Second, we project the six characters onto two-dimensional planes defined by pairs of archetypal dimensions. The resulting configurations consistently align with narrative evidence, confirming that archetypal projection captures interpretable aspects of ensemble structure. The projections further revealed that not all dimensions contribute equally to character differentiation within a given context: the \archetypesemdiff{2} axis proved uninformative in \storylinksimple{Friends}{Friends}. This implies that an archetypal dimension's significance is not intrinsic but genre-dependent---different narrative forms selectively activate different dimensions to organize their character systems.

Third, we demonstrate that archetypes can function as relational representations. Pairwise inner products of the six character vectors reveal three distinct structural configurations within the ensemble. \characterlinksimple{Friends-Joey-Tribbiani}{Joey} and \characterlinksimple{Friends-Phoebe-Buffay}{Phoebe} exhibit strong alignment. Their shared \archetypelinksimple{Adventurer}{Adventurer} orientation corresponds to a common narrative function that both introduce unpredictability to the plot when it tends to become routine. At the opposite extreme, \characterlinksimple{Friends-Phoebe-Buffay}{Phoebe} and \characterlinksimple{Friends-Ross-Geller}{Ross} display strong contrast. Their positions at opposing ends of the \archetypesemdiff{3} axis define the polar boundaries of the ensemble's narrative space, with the remaining characters distributed between them. Between these two poles, the two central romantic pairs (\characterlinksimple{Friends-Ross-Geller}{Ross} and \characterlinksimple{Friends-Rachel-Green}{Rachel}, and \characterlinksimple{Friends-Monica-Geller}{Monica} and \characterlinksimple{Friends-Chandler-Bing}{Chandler}) occupy a third configuration: near-orthogonality. In both cases, a shared \archetypelinksimple{Diva}{Diva} component provides a basis of emotional resonance, while opposing orientations along the \archetypesemdiff{3} axis introduce productive difference. This orthogonal structure indicates that romantic partnerships in \storylinksimple{Friends}{Friends} are grounded in cross-dimensional complementarity, where each partner contributes distinct narrative dimensions. Here, relationships balance stability with dramatic flexibility.

Overall, this study offers a methodological contribution to the computational analysis of narrative characters. By demonstrating that computationally derived archetypes can describe both individual characters and the relational organization of an ensemble, we suggest that archetype analysis is a lens through which the narrative role becomes quantifiable.

\section{Limitations and Future Work}

Several limitations should be acknowledged when interpreting the results of this study. First, the study is limited to a single ensemble series within a specific genre context. Sitcoms, and \storylinksimple{Friends}{Friends} in particular, exhibit strong structural conventions that may influence the observed patterns of archetypal alignment. Moreover, \storylinksimple{Friends}{Friends} represents a particular historical and cultural moment in American television---the ensemble comedy of the 1990s---whose narrative logic may not generalize to other genres or periods. Serialized genres, such as long-form fantasy, might display more diverse archetypes and relations. Future work could extend this framework to a broader range of narratives to evaluate the generalizability of these findings. 

Second, the analysis focuses on static character representations aggregated across the entire series, without explicitly modeling temporal dynamics. This static representation can capture the tendency of each character's archetypal orientation, but it cannot account for the possibility that a character's relational mode with another character might shift across seasons or episodes. The current approach does not capture such temporal variation.  

A third limitation concerns the methodological choices embedded in the archetypal framework itself. The essential dimensions we rely on are derived from a pre-existing computational model built from a particular corpus of character descriptions. While the \storylinksimple{Friends}{Friends} characters load meaningfully onto these dimensions, the framework's categories may not cover the narrative functions that matter most within the sitcom genre. Traits such as comedic timing, which are central to \storylinksimple{Friends}{Friends}, may not be fully captured by the archetypal dimensions available in the current model. Future research could use LLMs to generate archetypes from real texts to identify gaps in character archetypes that the computational model might have missed. 

Fourth, although we observe that romantic pairs in \storylinksimple{Friends}{Friends} are near-orthogonal, we do not claim that orthogonality causes romantic pairing. One possible explanation is that the observed orthogonality is an incidental byproduct of ensemble design, where creators simply seek to minimize character redundancy across all dimensions. Alternatively, this may be a characteristic of the sitcom genre: romantic partners must be sufficiently different to generate comedy, yet compatible enough to sustain long-term emotional investment.



\section*{Acknowledgements}

The authors are grateful for
National Science Foundation Award \#2242829
(Science of Online Corpora, Knowledge, and Stories),
foundational support from MassMutual,
and
an anonymous philanthropic gift.



\clearpage

\addcontentsline{toc}{section}{References}

\bibliography{archetypes-of-LLMs}


\onecolumn

\appendix
\section{Appendices}

\setcounter{page}{1}
\renewcommand{\thepage}{A\arabic{page}}
\renewcommand{\thefigure}{A\arabic{figure}}
\renewcommand{\thetable}{A\arabic{table}}
\setcounter{figure}{0}
\setcounter{table}{0}

\renewcommand{\thesection}{A\arabic{section}}
\setcounter{section}{0}

\section{Character Cards}
\label{Appendix:Character_Cards}

This appendix section includes character cards of the six main characters.

    \begin{figure*}[p]
      \centering
      \includegraphics[width=1.0\linewidth]{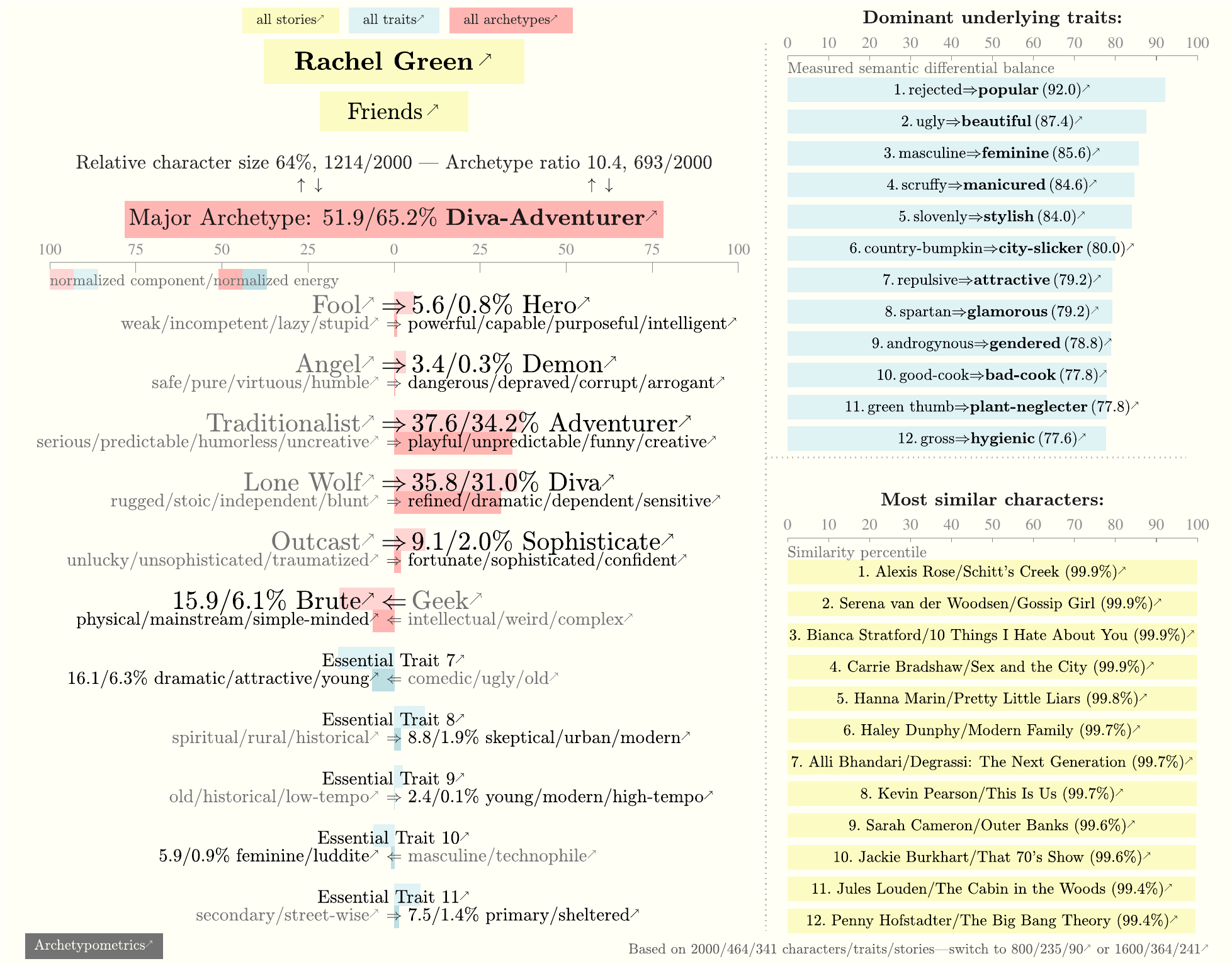}
      \caption{\characterlinksimple{Friends-Rachel-Green}{Rachel Green}'s position in archetypal space, highlighting both her dominant traits and her closest counterparts. On the left, \characterlinksimple{Friends-Rachel-Green}{Rachel}'s primary archetypal configuration is characterized as a combination of \archetypelinksimple{Diva}{Diva} (normalized energy of 31.0\%) and \archetypelinksimple{Adventurer}{Adventurer} (34.2\%). archetypes.
      }
      \label{fig:Rachel}
    \end{figure*}

    \begin{figure*}[p]
      \centering
      \includegraphics[width=1.0\linewidth]{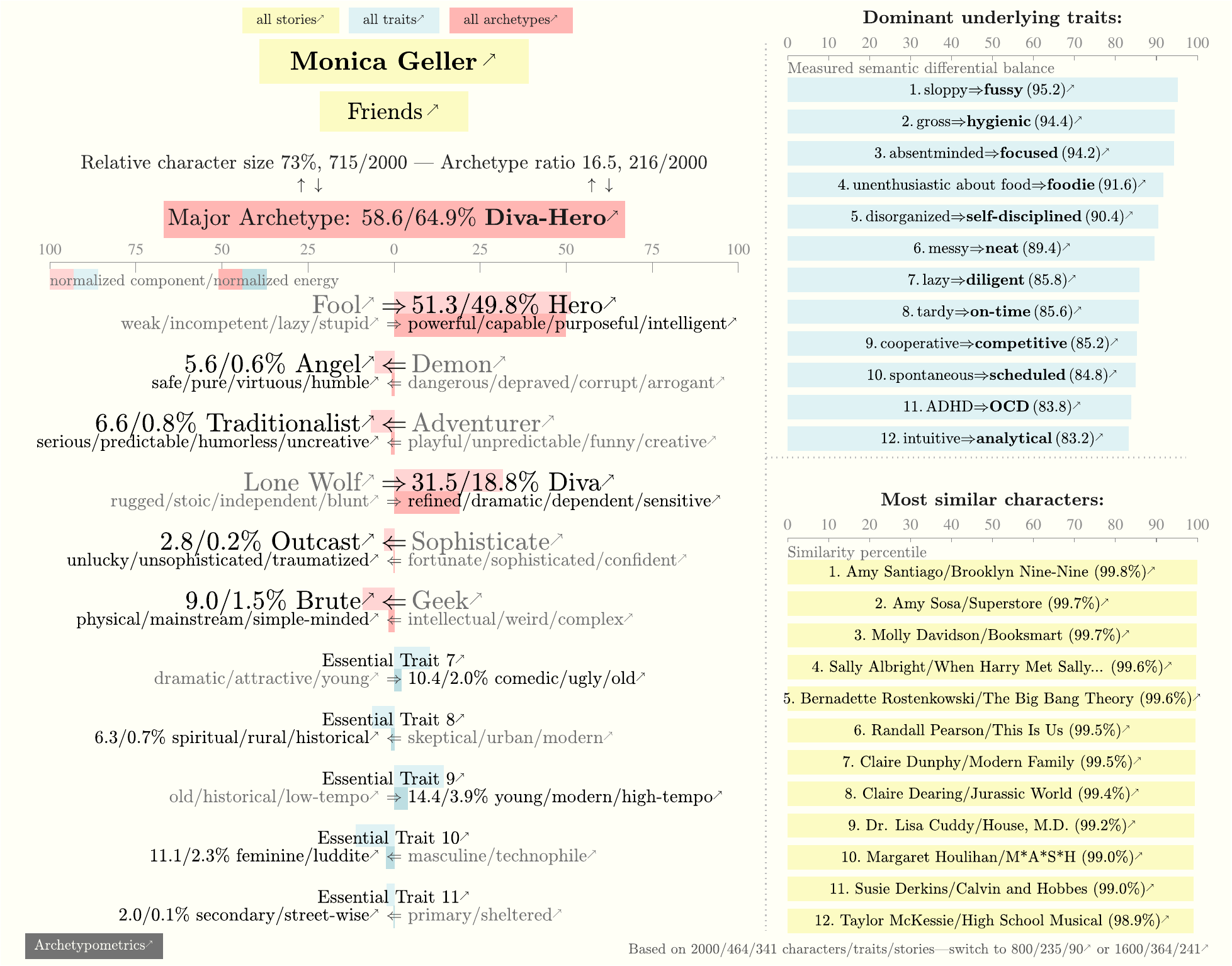}
      \caption{Archetypal profile of \characterlinksimple{Friends-Monica-Geller}{Monica Geller} in the essential dimension space, showing her directional alignment across archetypal axes along with dominant underlying traits.}
      \label{fig:Monica}
    \end{figure*}

    \begin{figure*}[p]
      \centering
      \includegraphics[width=1.0\linewidth]{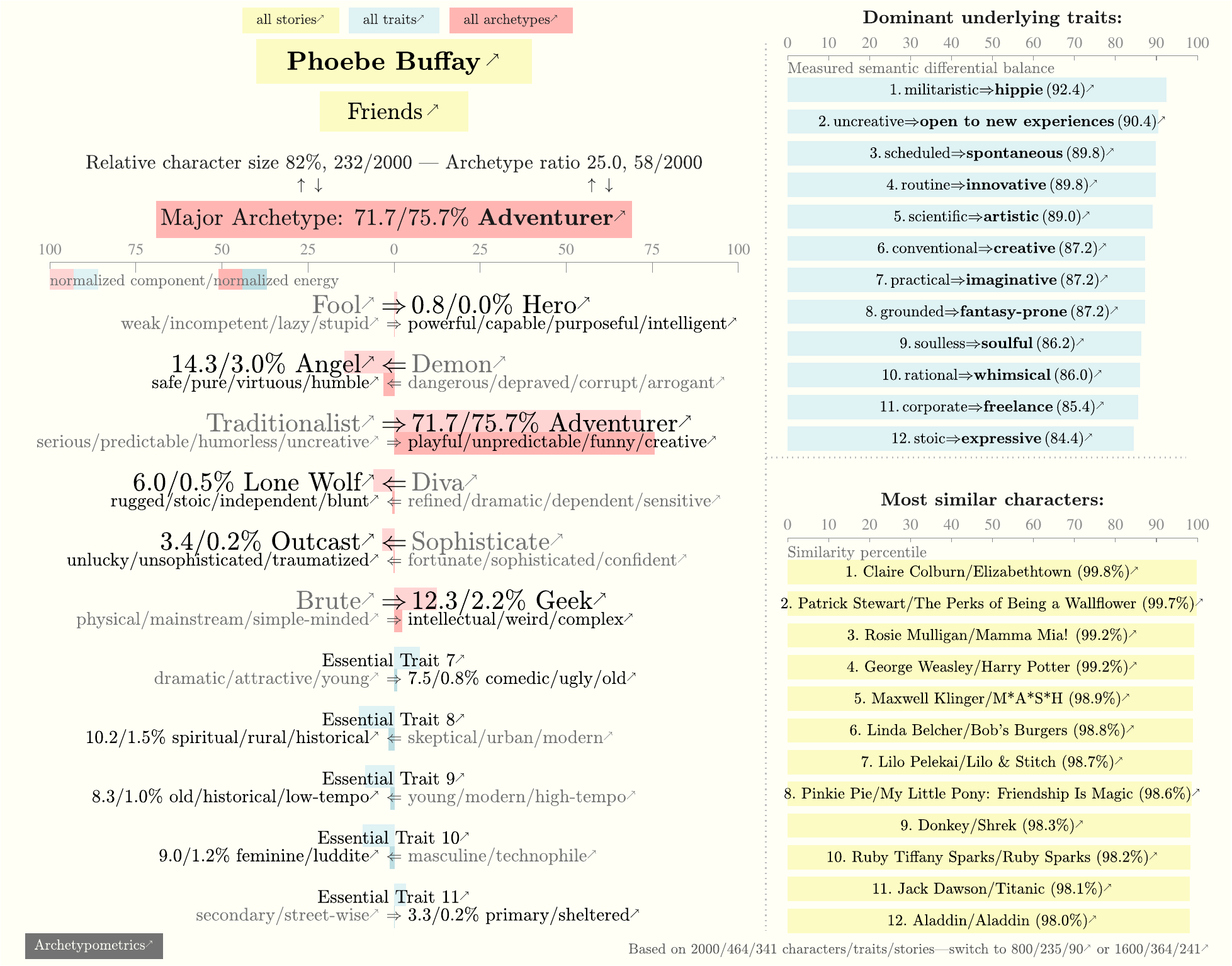}
      \caption{Archetypal profile of \characterlinksimple{Friends-Phoebe-Buffay}{Phoebe Buffay} in the essential dimension space, showing her directional alignment across archetypal axes along with dominant underlying traits.}
      \label{fig:Phoebe}
    \end{figure*}

    \begin{figure*}[p]
      \centering
      \includegraphics[width=1.0\linewidth]{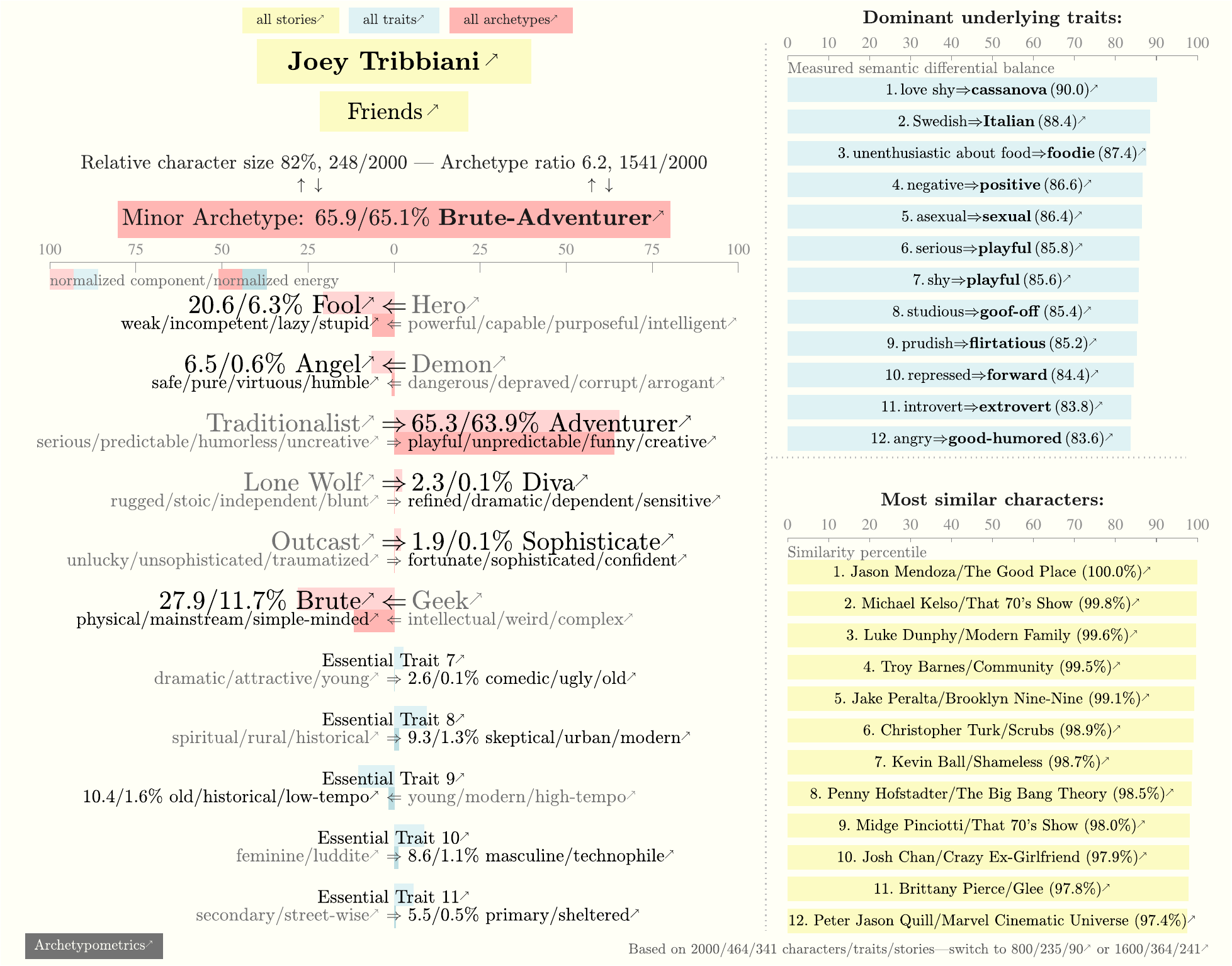}
      \caption{Archetypal profile of \characterlinksimple{Friends-Joey-Tribbiani}{Joey Tribbiani} in the essential dimension space, showing his directional alignment across archetypal axes along with dominant underlying traits.}
      \label{fig:Joey}
    \end{figure*} 

    \begin{figure*}[p]
      \centering
      \includegraphics[width=1.0\linewidth]{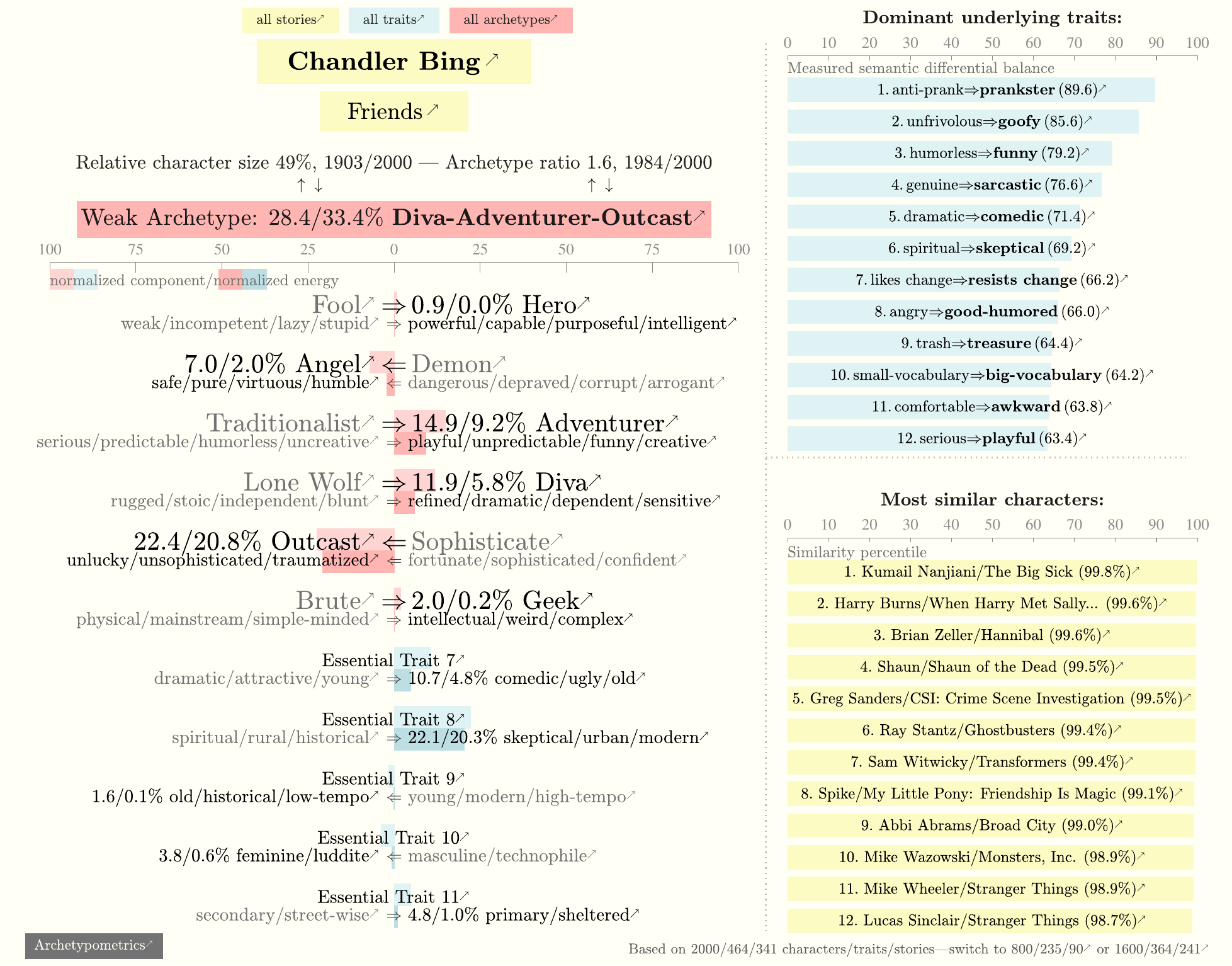}
      \caption{Archetypal profile of \characterlinksimple{Friends-Chandler-Bing}{Chandler Bing} in the essential dimension space, showing his directional alignment across archetypal axes along with dominant underlying traits.}
      \label{fig:Chandler}
    \end{figure*}

    \begin{figure*}[p]
      \centering
      \includegraphics[width=1.0\linewidth]{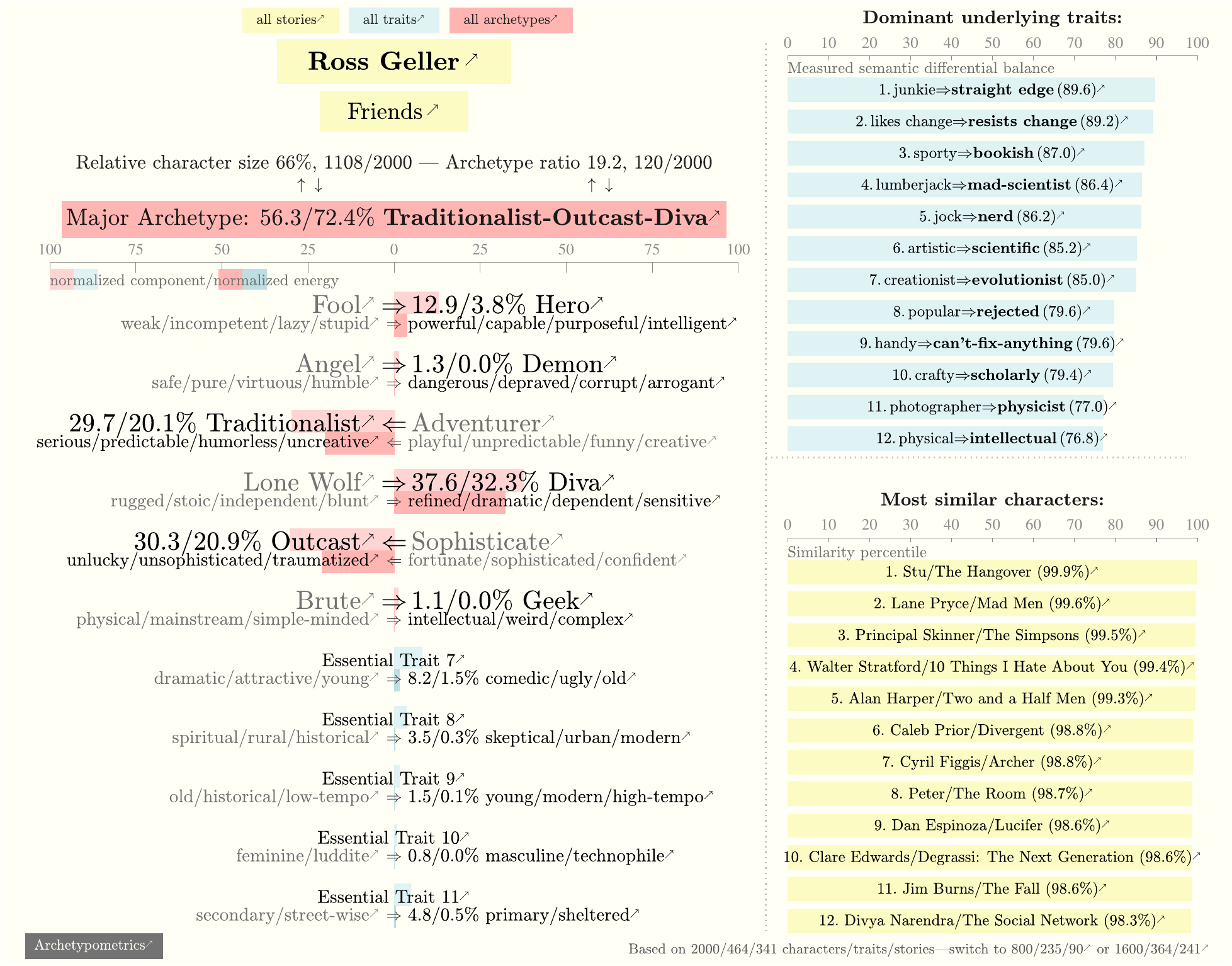}
      \caption{Archetypal profile of \characterlinksimple{Friends-Ross-Geller}{Ross Geller} in the essential dimension space, showing his directional alignment across archetypal axes along with dominant underlying traits.}
      \label{fig:Ross}
    \end{figure*}

\clearpage

\section{Pairwise Comparison}

This appendix section includes 30 pairwise comparison figures. There are 15 pairs of characters, and each pair has a trait-based figure and an archetype-based figure.

\label{Appendix:Pairwise_Comparison_Trait}
\label{Appendix:Pairwise_Comparison_Archetype}

\begin{figure*}[p]
      \centering
      \includegraphics[width=1.0\linewidth]{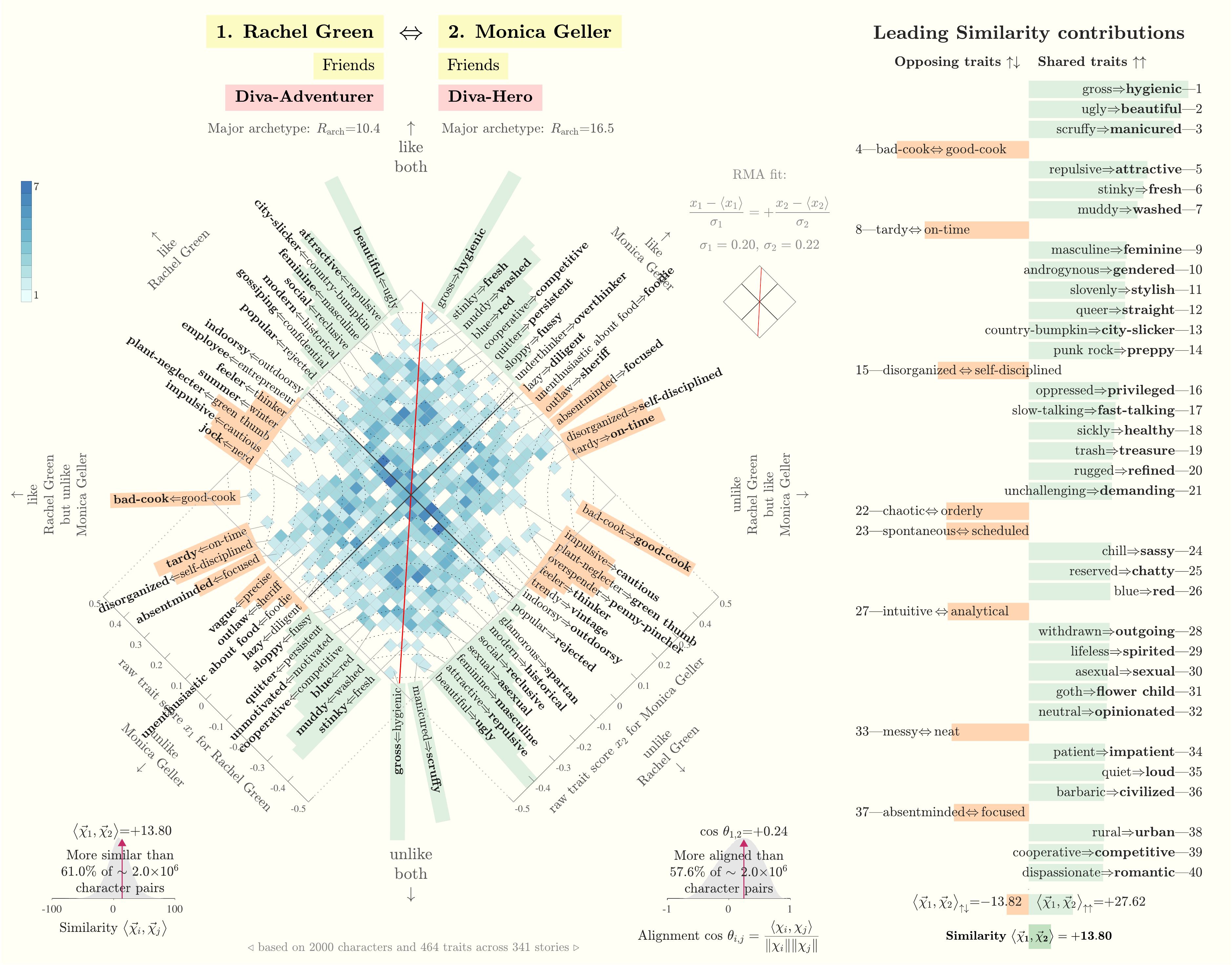}
      \caption{Trait-based comparison between \characterlinksimple{Friends-Rachel-Green}{Rachel Green} and \characterlinksimple{Friends-Monica-Geller}{Monica Geller}.}
      \label{fig:rachel_monica_traits}
    \end{figure*}
\begin{figure*}[p]
  \centering
  \includegraphics[width=1.0\linewidth]{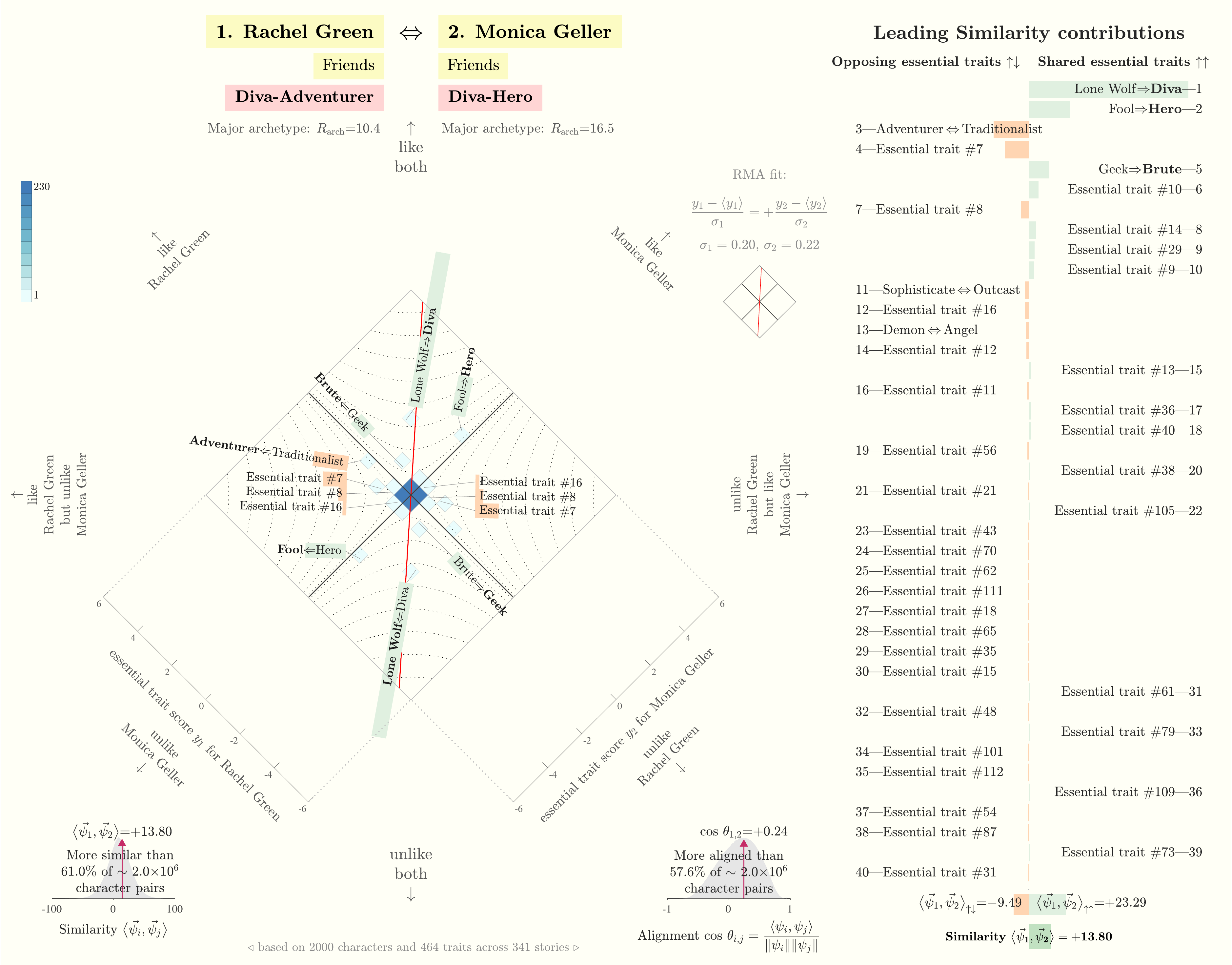}
  \caption{Archetype-based comparison between \characterlinksimple{Friends-Rachel-Green}{Rachel Green} and \characterlinksimple{Friends-Monica-Geller}{Monica Geller}.}
  \label{fig:rachel_monica_archetype}
\end{figure*}
\clearpage

\begin{figure*}[p]
      \centering
      \includegraphics[width=1.0\linewidth]{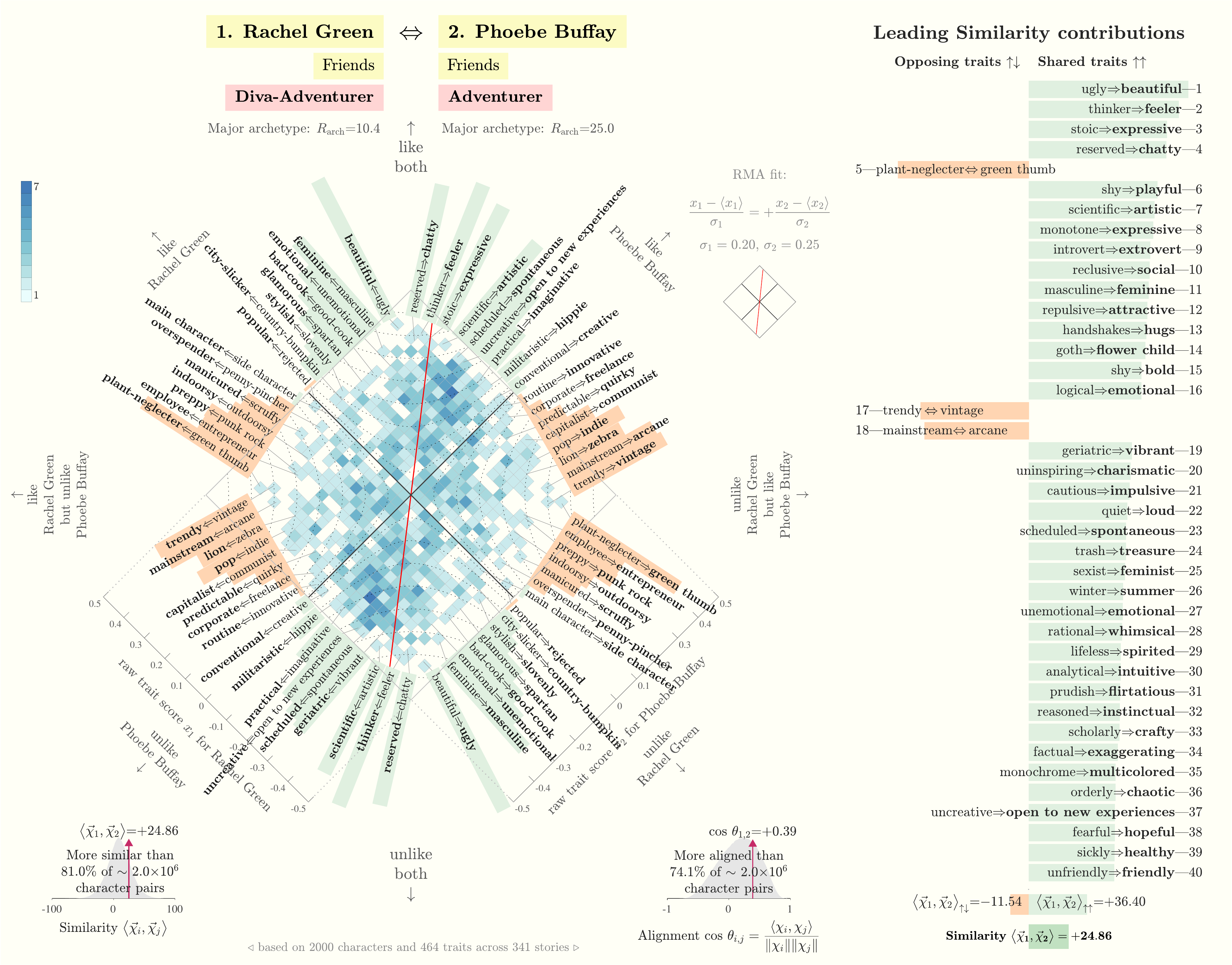}
      \caption{Trait-based comparison between \characterlinksimple{Friends-Rachel-Green}{Rachel Green} and \characterlinksimple{Friends-Phoebe-Buffay}{Phoebe Buffay}.}
      \label{fig:rachel_phoebe_traits}
    \end{figure*}
\begin{figure*}[p]
  \centering
  \includegraphics[width=1.0\linewidth]{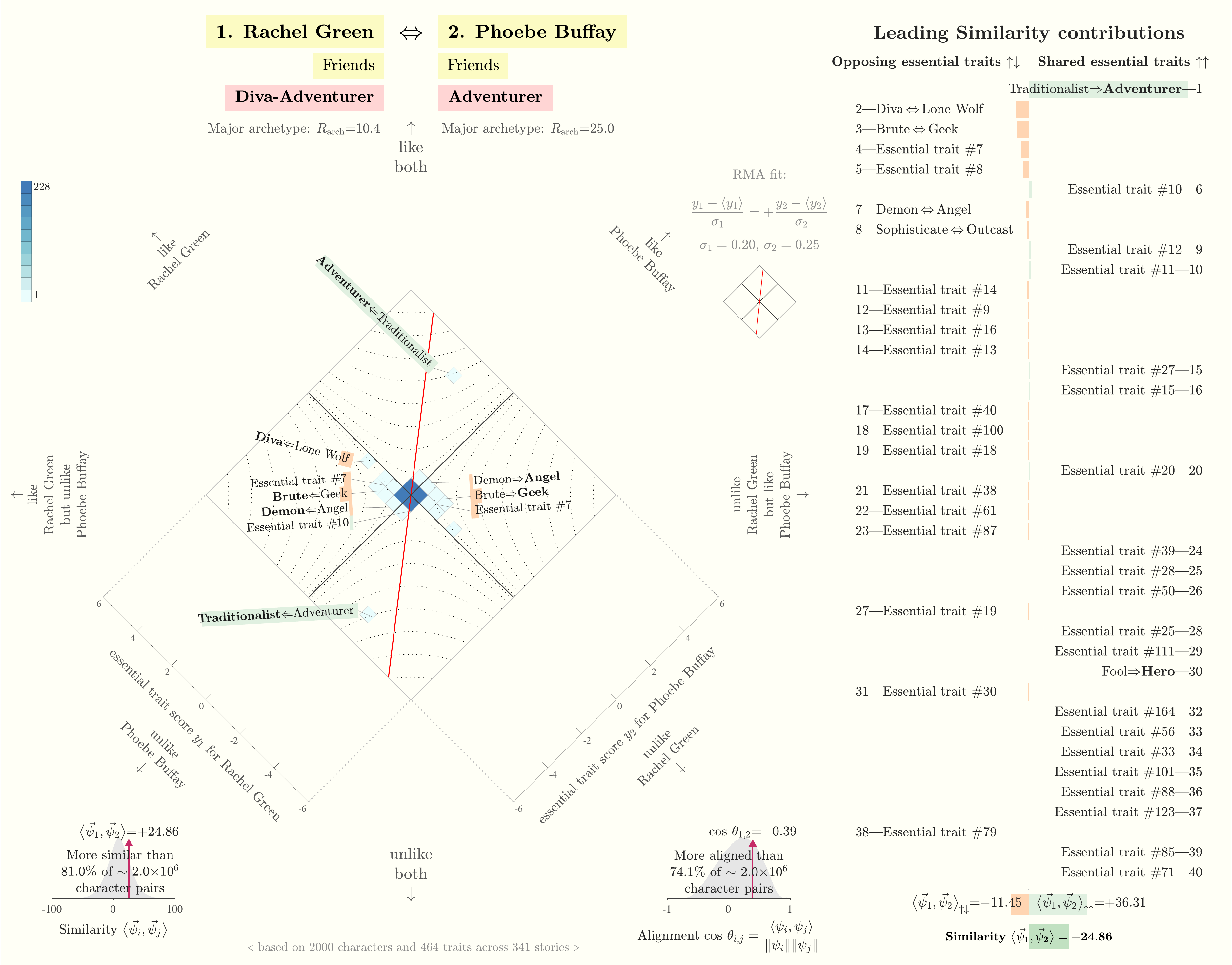}
  \caption{Archetype-based comparison between \characterlinksimple{Friends-Rachel-Green}{Rachel Green} and \characterlinksimple{Friends-Phoebe-Buffay}{Phoebe Buffay}.}
  \label{fig:rachel_phoebe_archetype}
\end{figure*}
\clearpage

\begin{figure*}[p]
      \centering
      \includegraphics[width=1.0\linewidth]{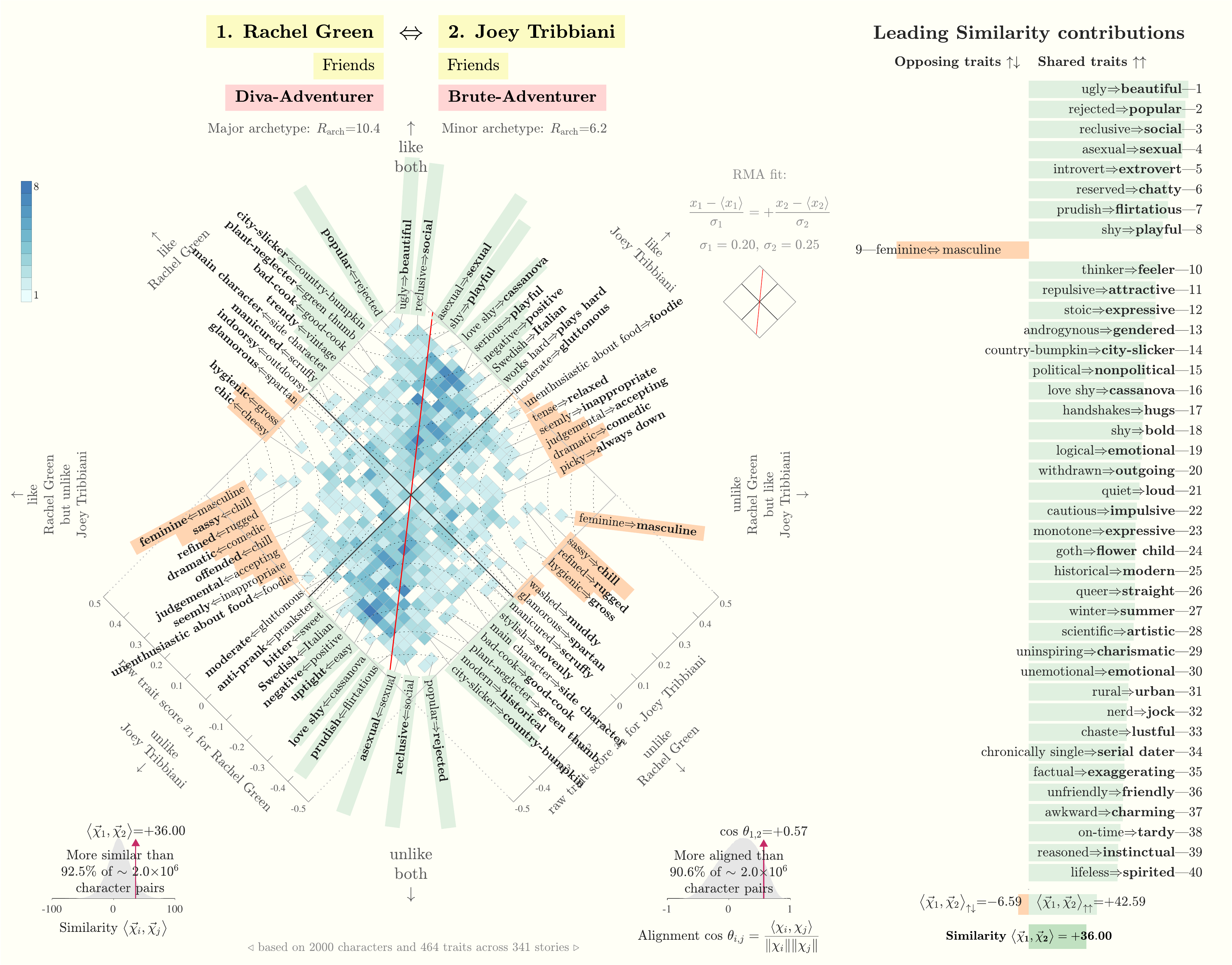}
      \caption{Trait-based comparison between \characterlinksimple{Friends-Rachel-Green}{Rachel Green} and \characterlinksimple{Friends-Joey-Tribbiani}{Joey Tribbiani}.}
      \label{fig:rachel_joey_traits}
    \end{figure*}
\begin{figure*}[p]
  \centering
  \includegraphics[width=1.0\linewidth]{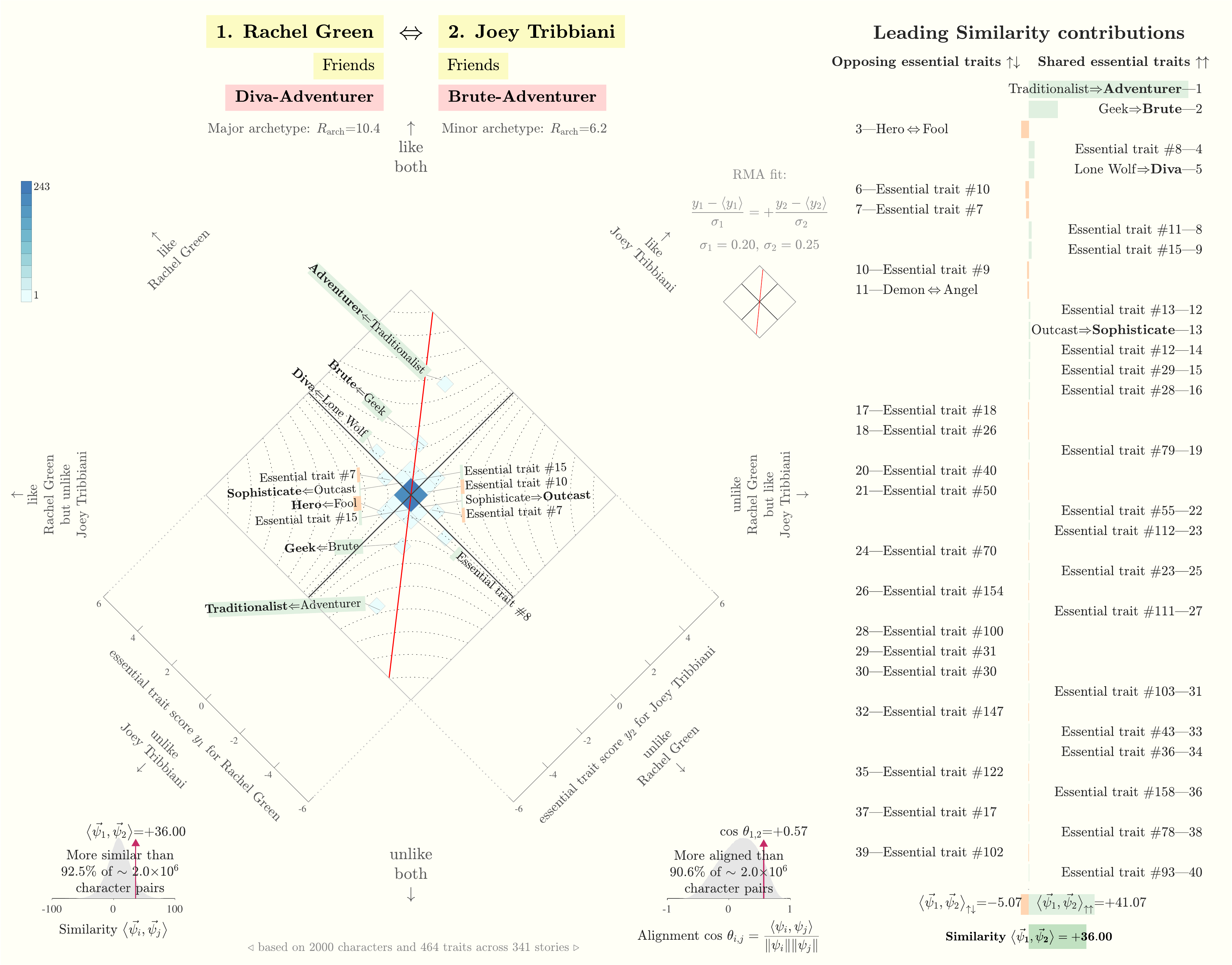}
  \caption{Archetype-based comparison between \characterlinksimple{Friends-Rachel-Green}{Rachel Green} and \characterlinksimple{Friends-Joey-Tribbiani}{Joey Tribbiani}.}
  \label{fig:rachel_joey_archetype}
\end{figure*}
\clearpage

\begin{figure*}[p]
      \centering
      \includegraphics[width=1.0\linewidth]{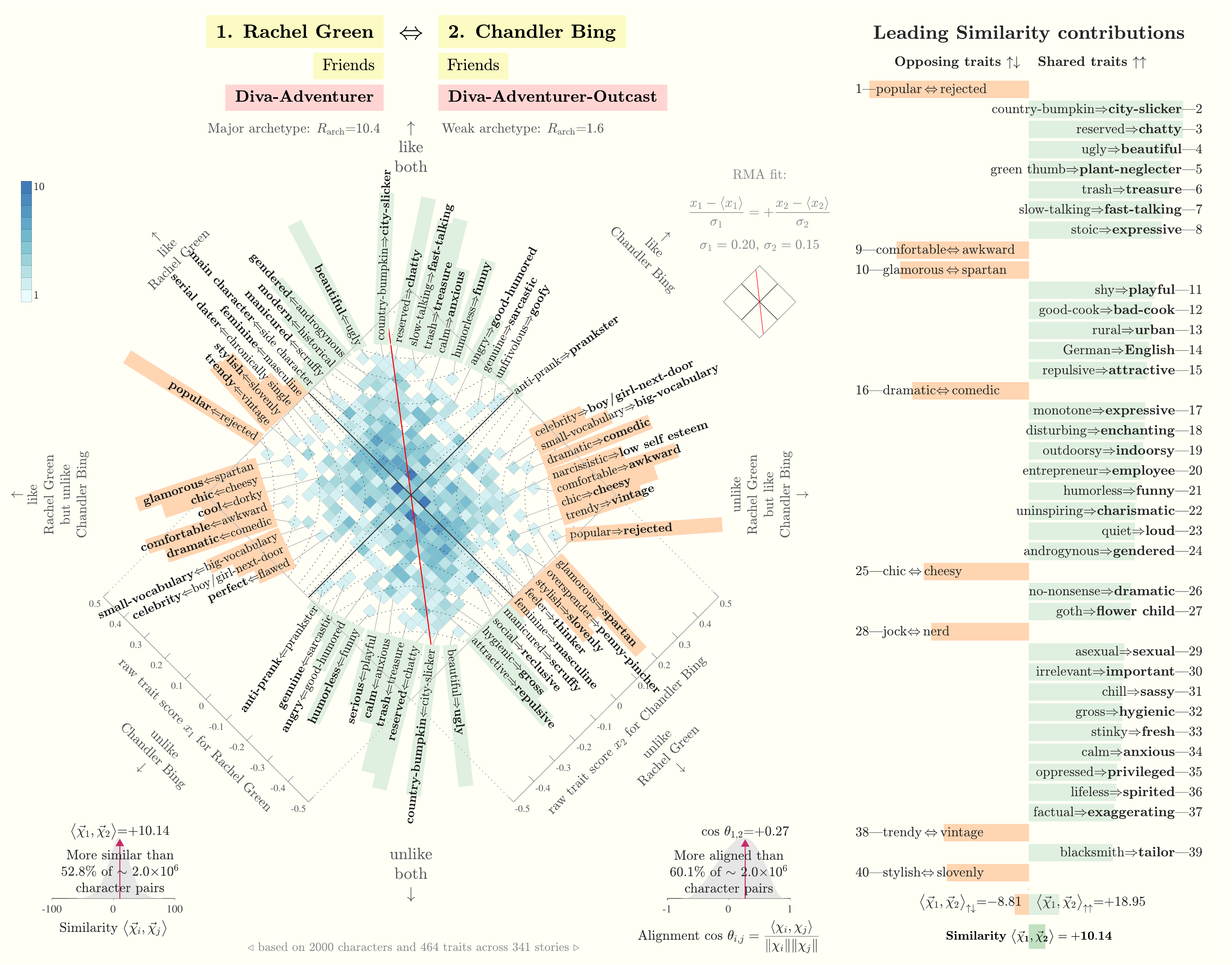}
      \caption{Trait-based comparison between \characterlinksimple{Friends-Rachel-Green}{Rachel Green} and \characterlinksimple{Friends-Chandler-Bing}{Chandler Bing}.}
      \label{fig:rachel_chandler_traits}
    \end{figure*}
\begin{figure*}[p]
  \centering
  \includegraphics[width=1.0\linewidth]{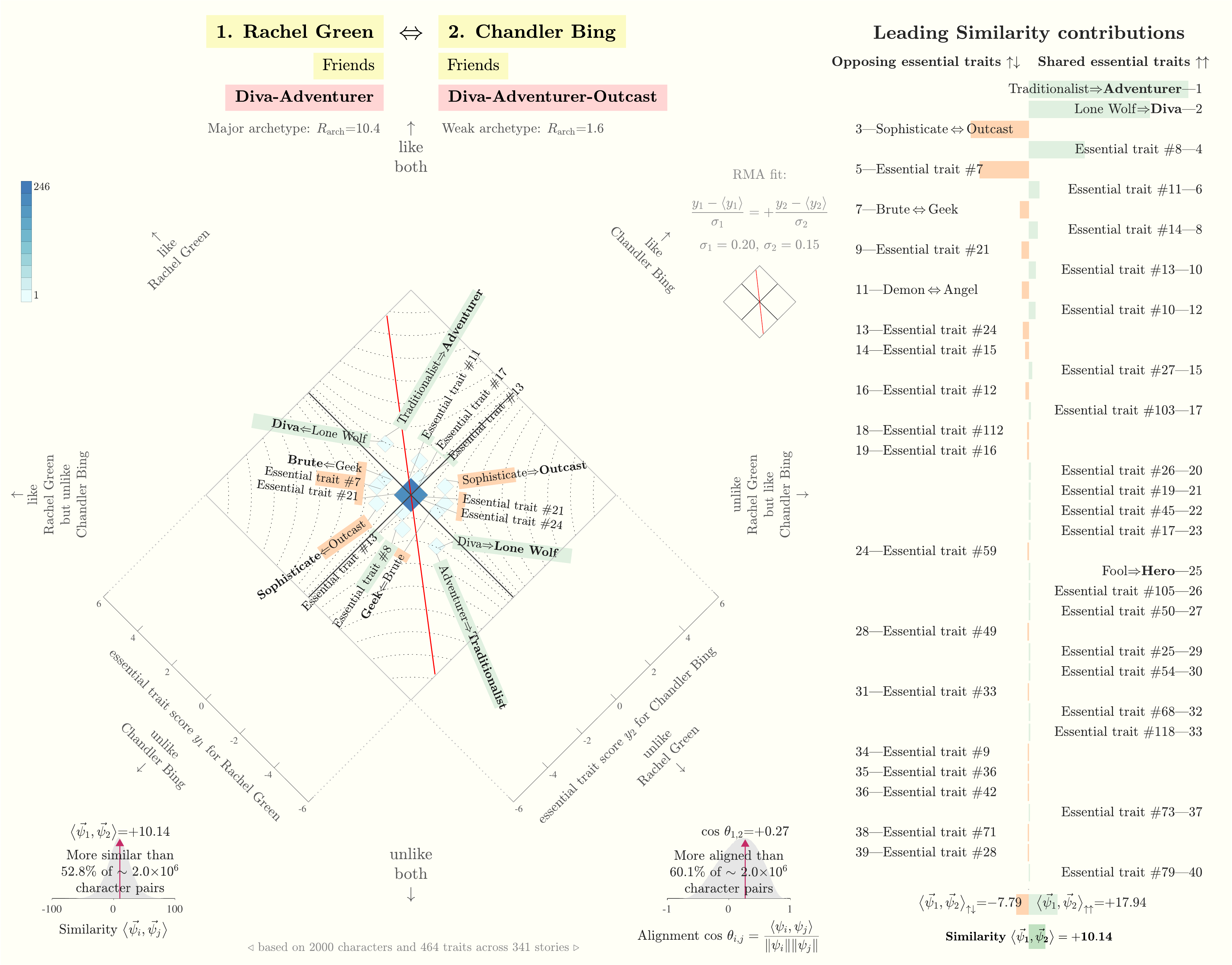}
  \caption{Archetype-based comparison between \characterlinksimple{Friends-Rachel-Green}{Rachel Green} and \characterlinksimple{Friends-Chandler-Bing}{Chandler Bing}.}
  \label{fig:rachel_chandler_archetype}
\end{figure*}
\clearpage

\begin{figure*}[p]
      \centering
      \includegraphics[width=1.0\linewidth]{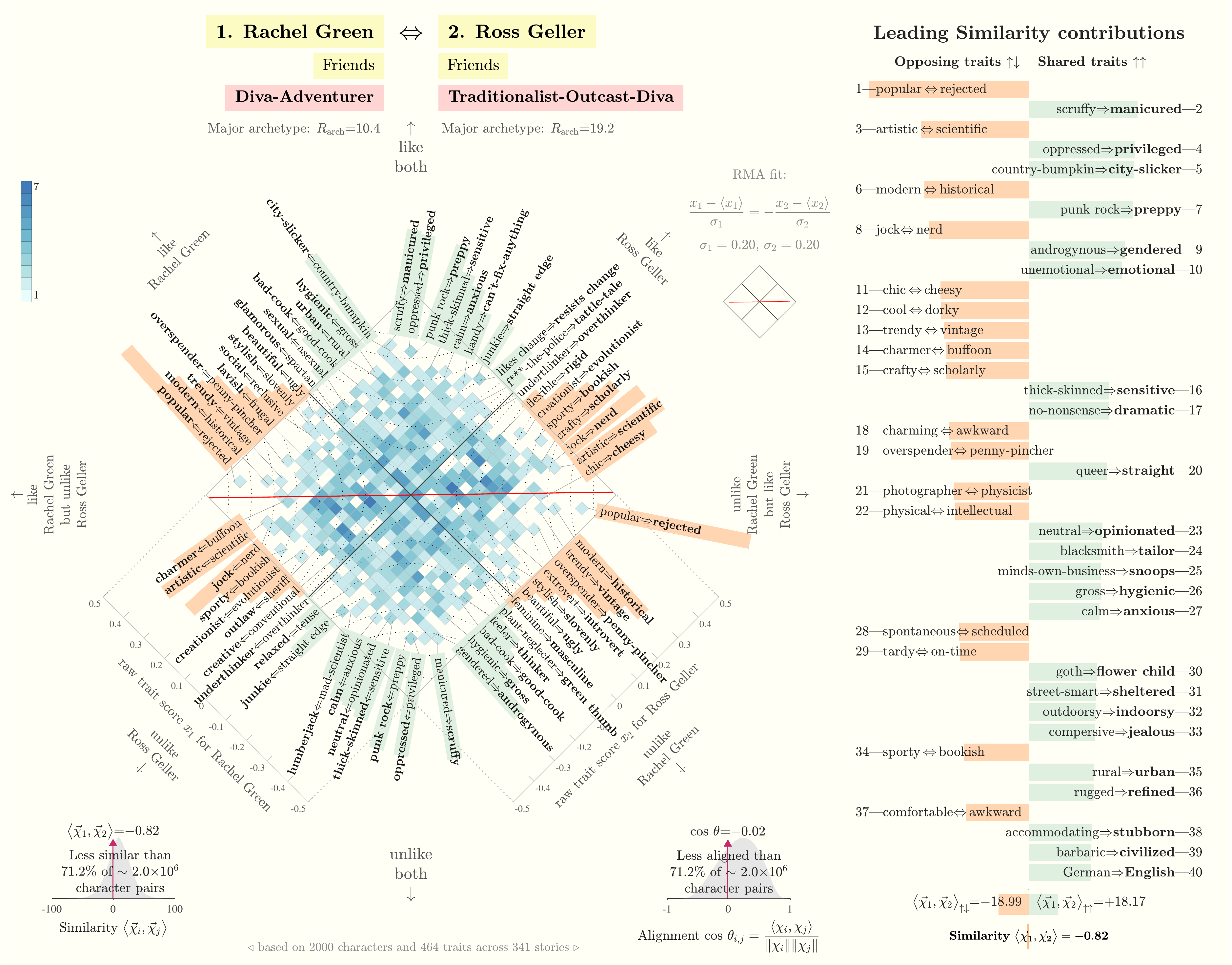}
      \caption{Trait-based comparison between \characterlinksimple{Friends-Rachel-Green}{Rachel Green} and \characterlinksimple{Friends-Ross-Geller}{Ross Geller}.}
      \label{fig:rachel_ross_traits}
    \end{figure*}
\begin{figure*}[p]
  \centering
  \includegraphics[width=1.0\linewidth]{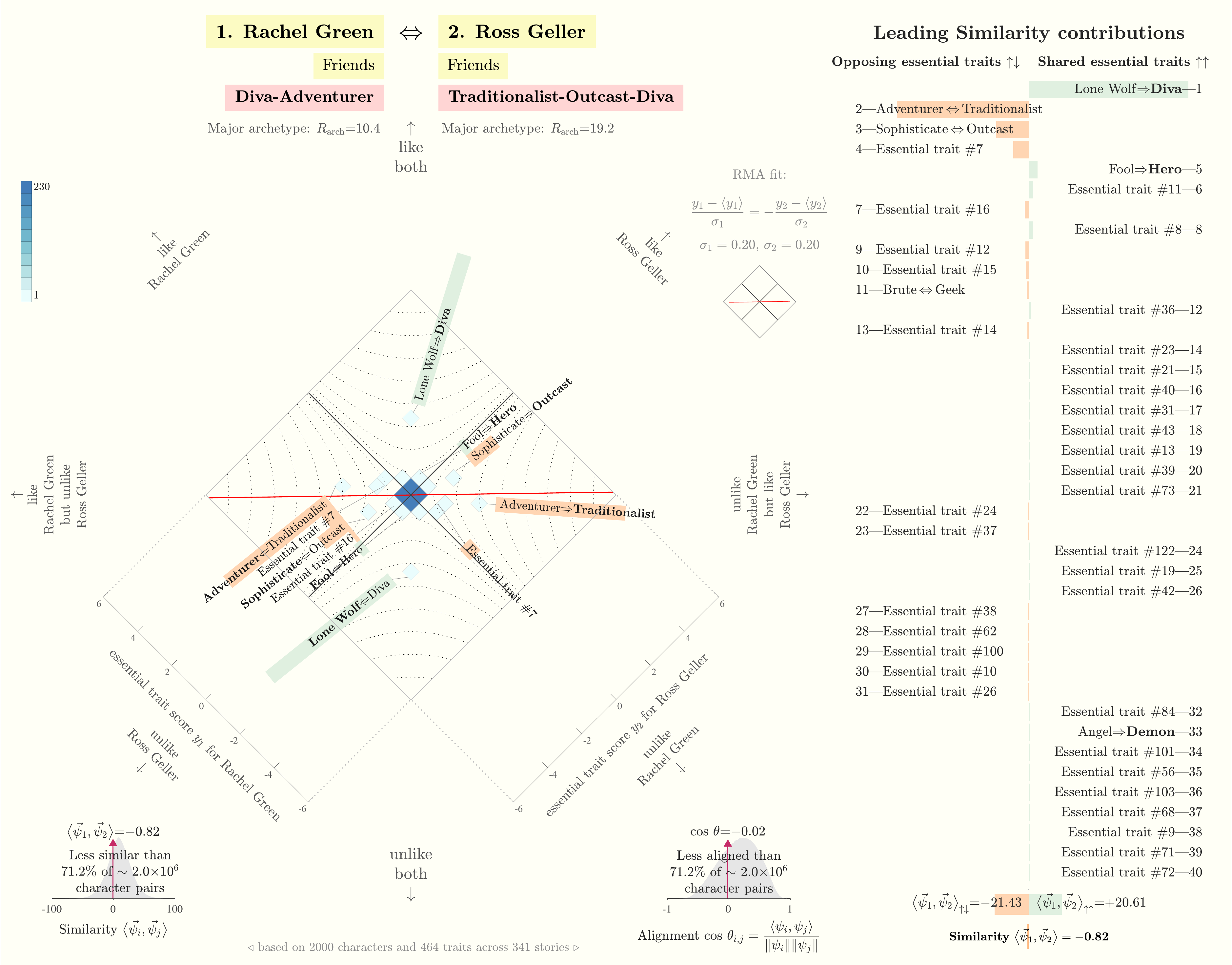}
  \caption{Archetype-based comparison between \characterlinksimple{Friends-Rachel-Green}{Rachel Green} and \characterlinksimple{Friends-Ross-Geller}{Ross Geller}.}
  \label{fig:rachel_ross_archetype_app}
\end{figure*}
\clearpage

\begin{figure*}[p]
      \centering
      \includegraphics[width=1.0\linewidth]{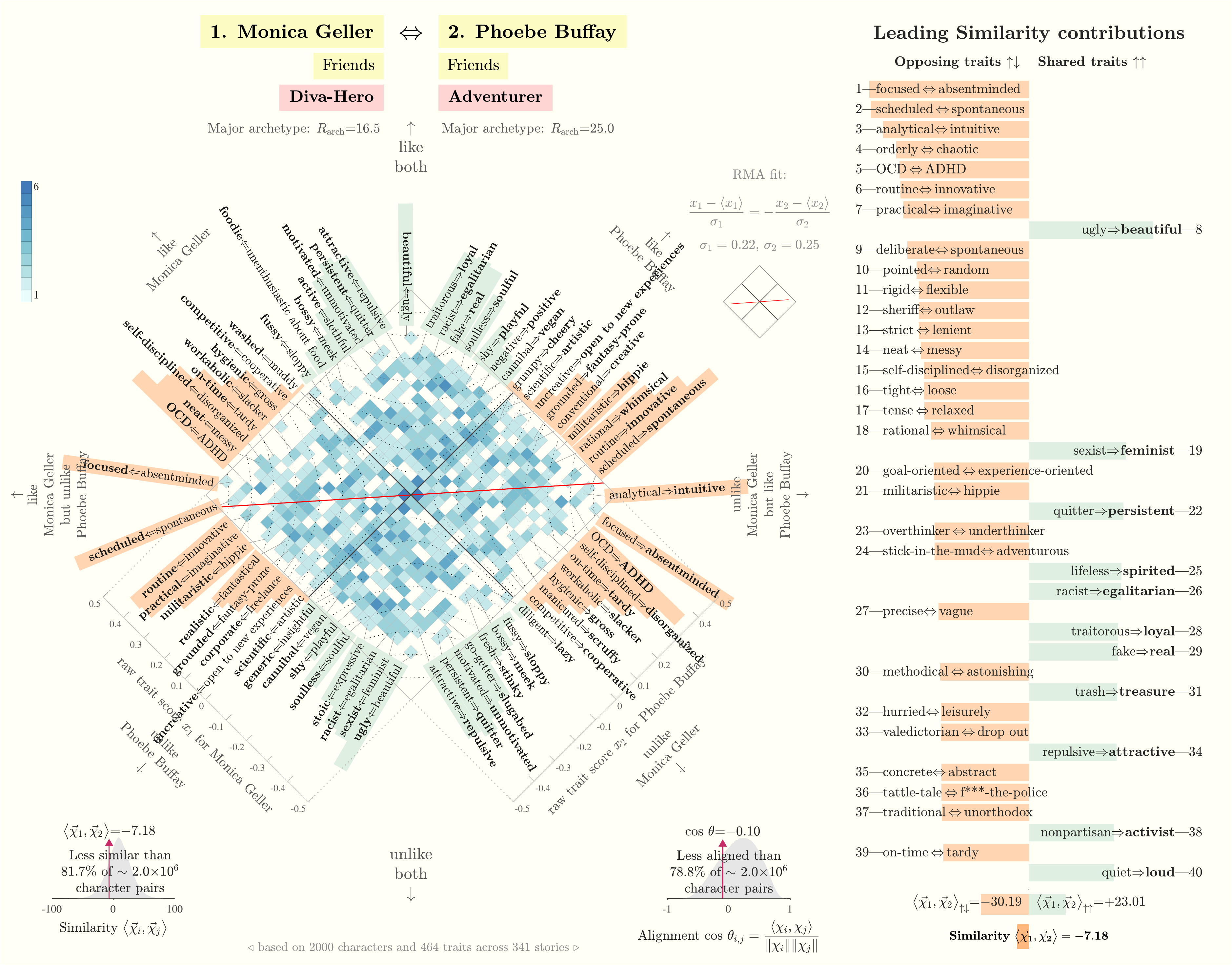}
      \caption{Trait-based comparison between \characterlinksimple{Friends-Monica-Geller}{Monica Geller} and \characterlinksimple{Friends-Phoebe-Buffay}{Phoebe Buffay}.}
      \label{fig:monica_phoebe_traits}
    \end{figure*}
\begin{figure*}[p]
  \centering
  \includegraphics[width=1.0\linewidth]{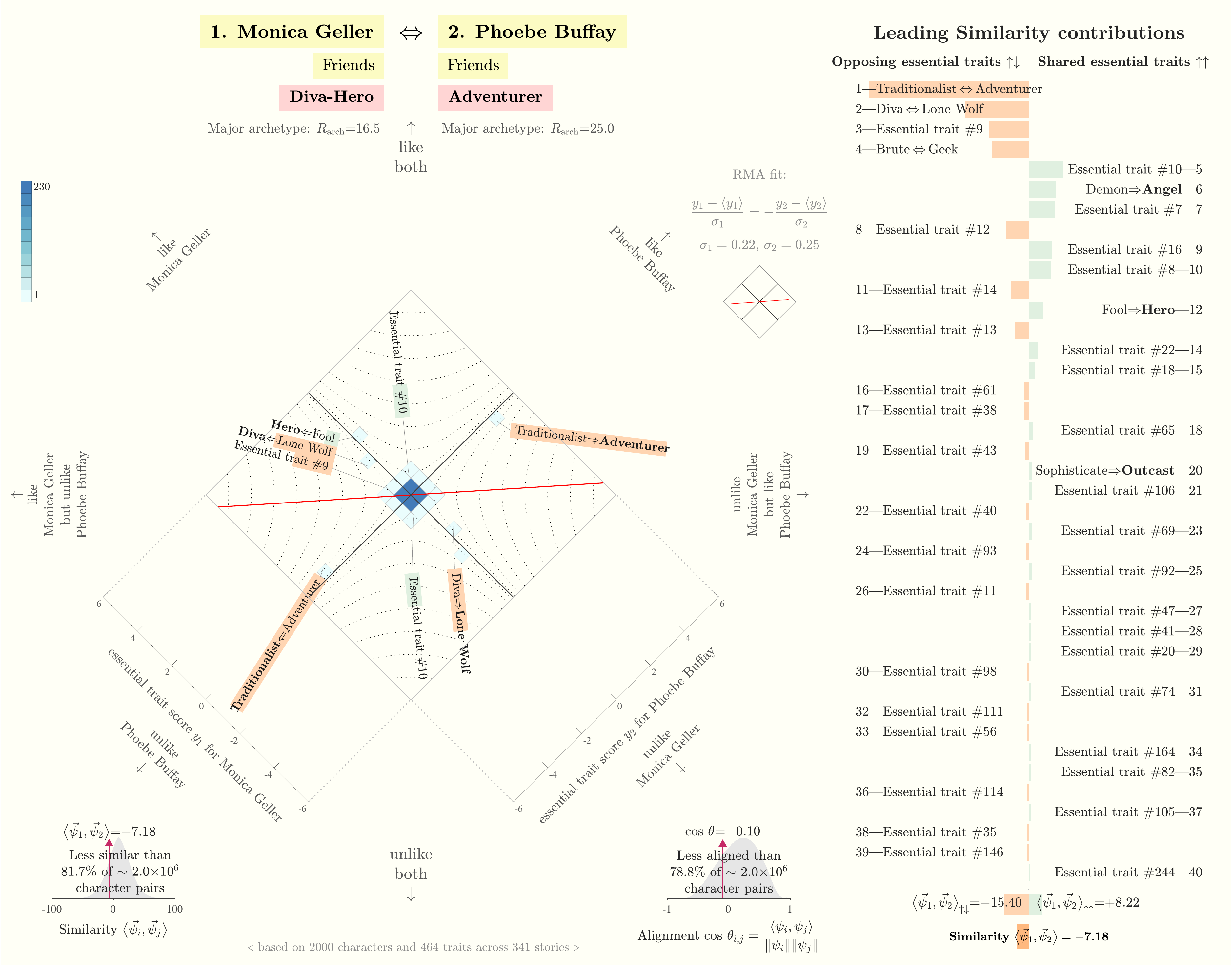}
  \caption{Archetype-based comparison between \characterlinksimple{Friends-Monica-Geller}{Monica Geller} and \characterlinksimple{Friends-Phoebe-Buffay}{Phoebe Buffay}.}
  \label{fig:monica_phoebe_archetype}
\end{figure*}
\clearpage

\begin{figure*}[p]
      \centering
      \includegraphics[width=1.0\linewidth]{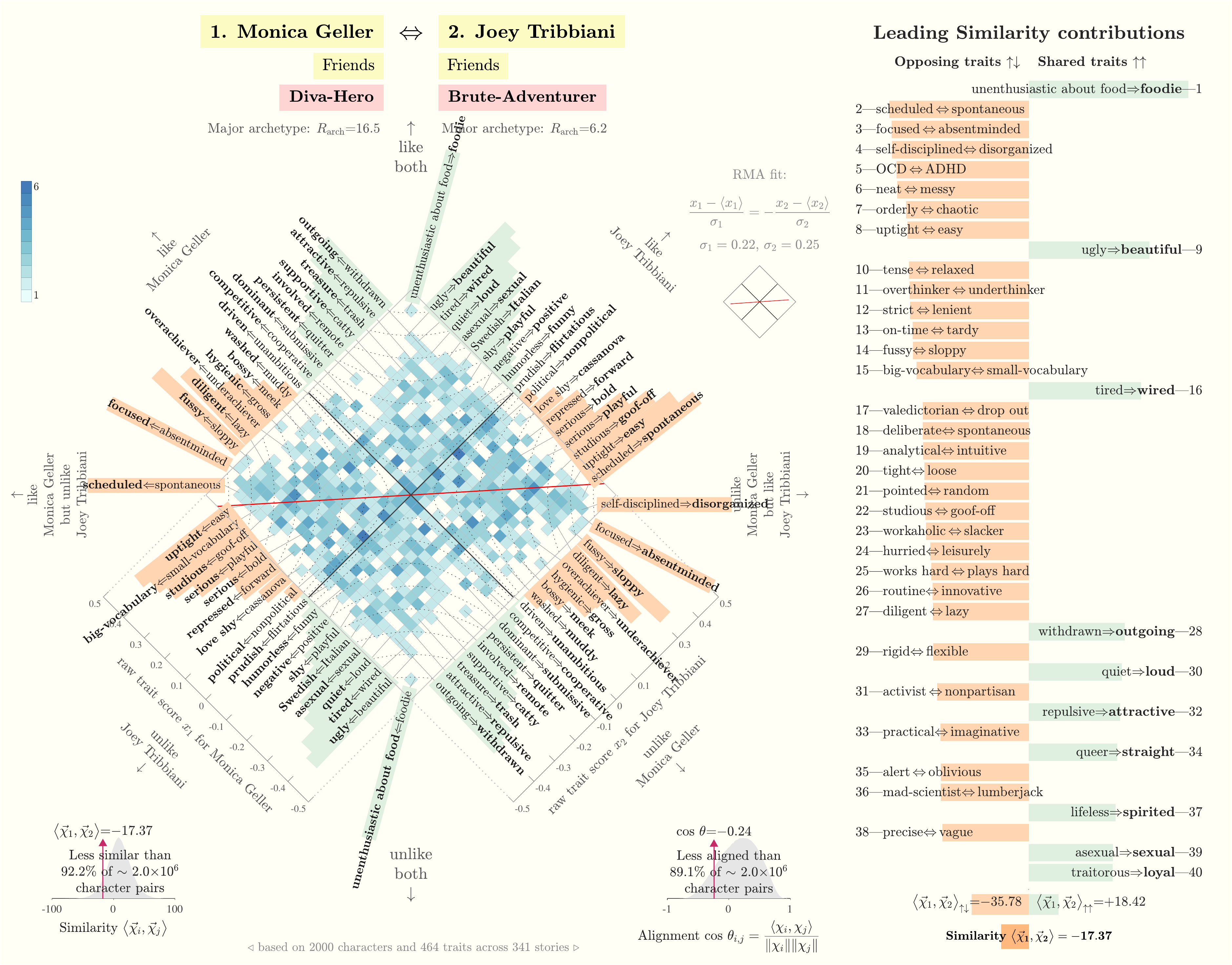}
      \caption{Trait-based comparison between \characterlinksimple{Friends-Monica-Geller}{Monica Geller} and \characterlinksimple{Friends-Joey-Tribbiani}{Joey Tribbiani}.}
      \label{fig:monica_joey_traits}
    \end{figure*}
\begin{figure*}[p]
  \centering
  \includegraphics[width=1.0\linewidth]{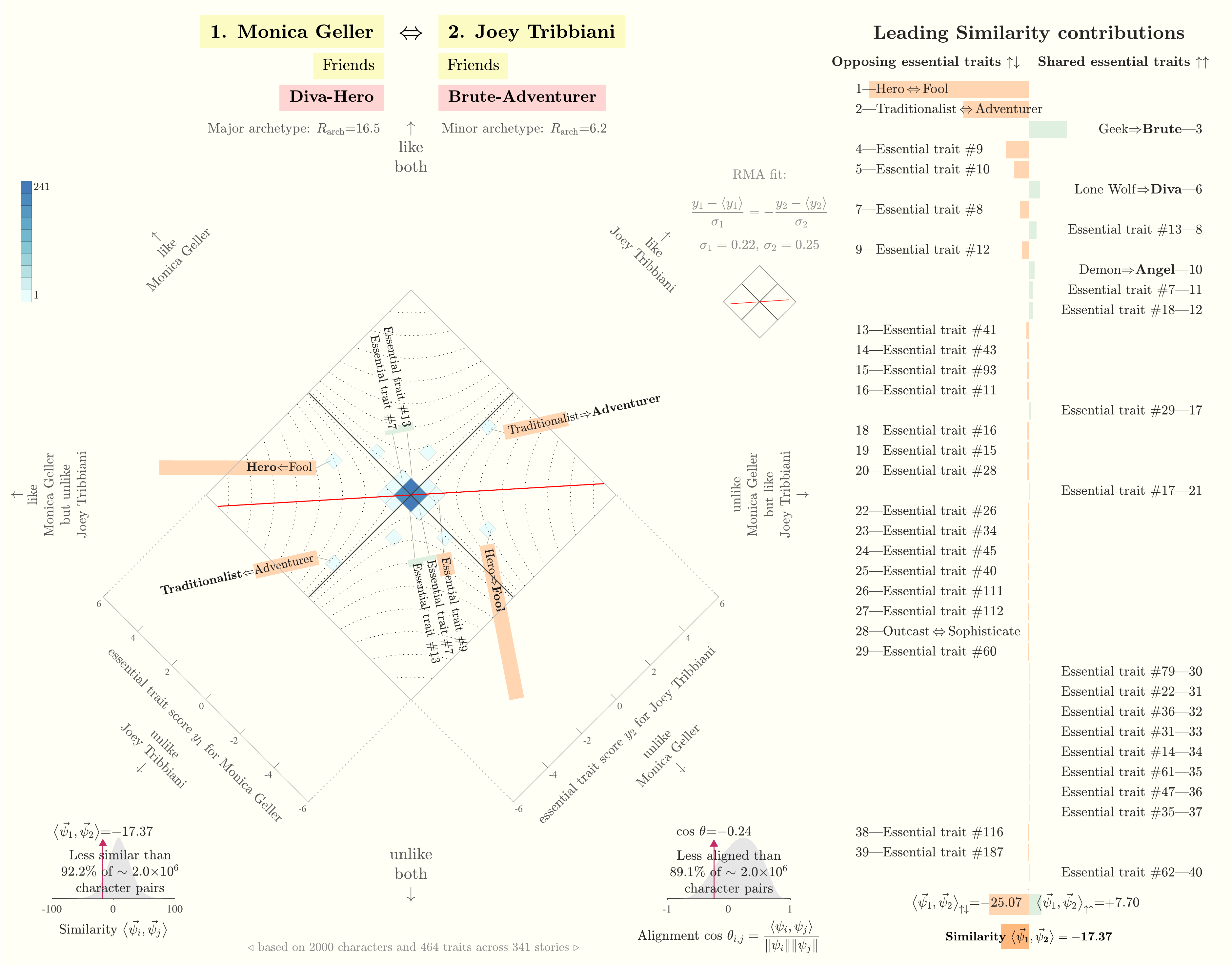}
  \caption{Archetype-based comparison between \characterlinksimple{Friends-Monica-Geller}{Monica Geller} and \characterlinksimple{Friends-Joey-Tribbiani}{Joey Tribbiani}.}
  \label{fig:monica_joey_archetype}
\end{figure*}
\clearpage

\begin{figure*}[p]
      \centering
      \includegraphics[width=1.0\linewidth]{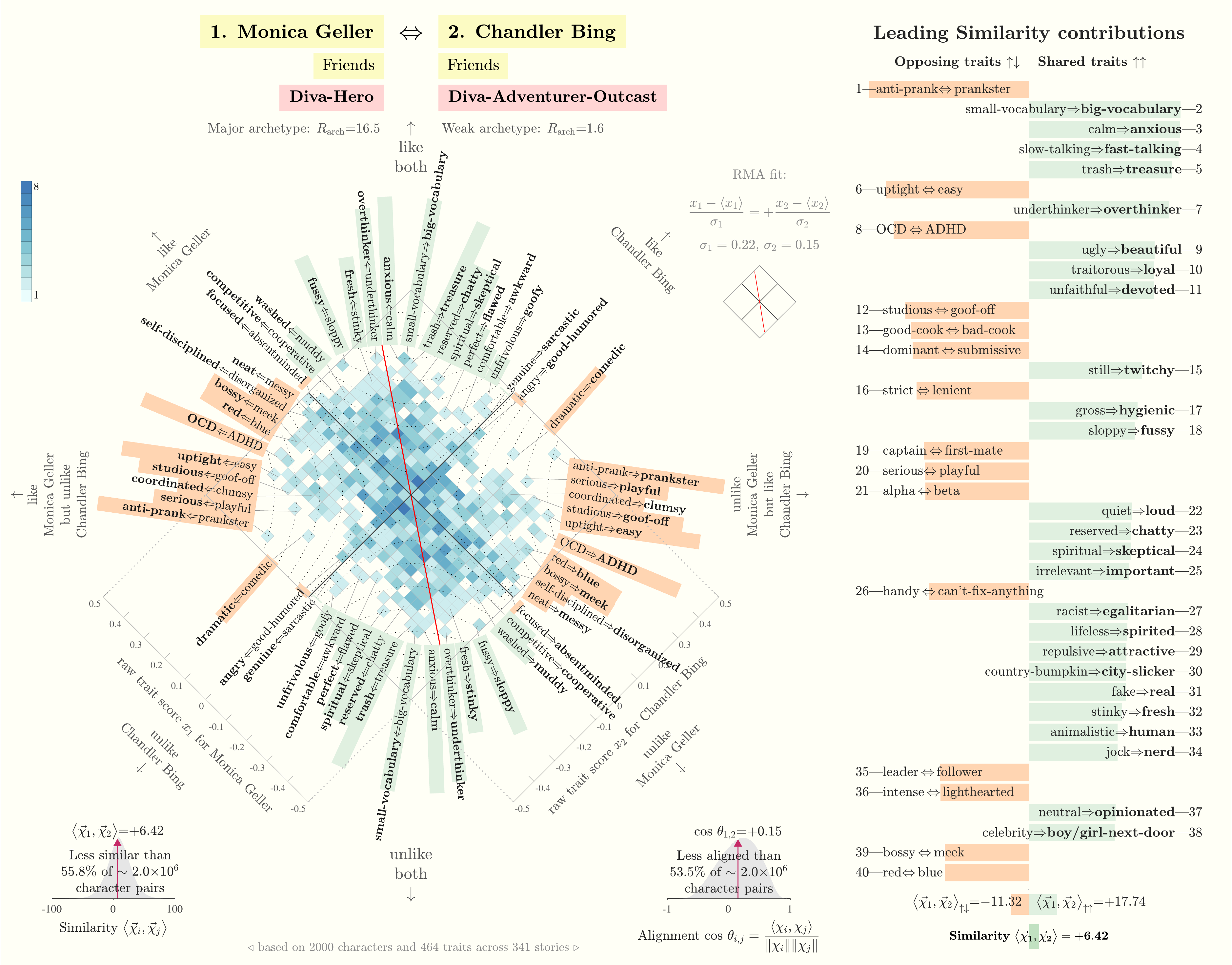}
      \caption{Trait-based comparison between \characterlinksimple{Friends-Monica-Geller}{Monica Geller} and \characterlinksimple{Friends-Chandler-Bing}{Chandler Bing}.}
      \label{fig:monica_chandler_traits}
    \end{figure*}
\begin{figure*}[p]
  \centering
  \includegraphics[width=1.0\linewidth]{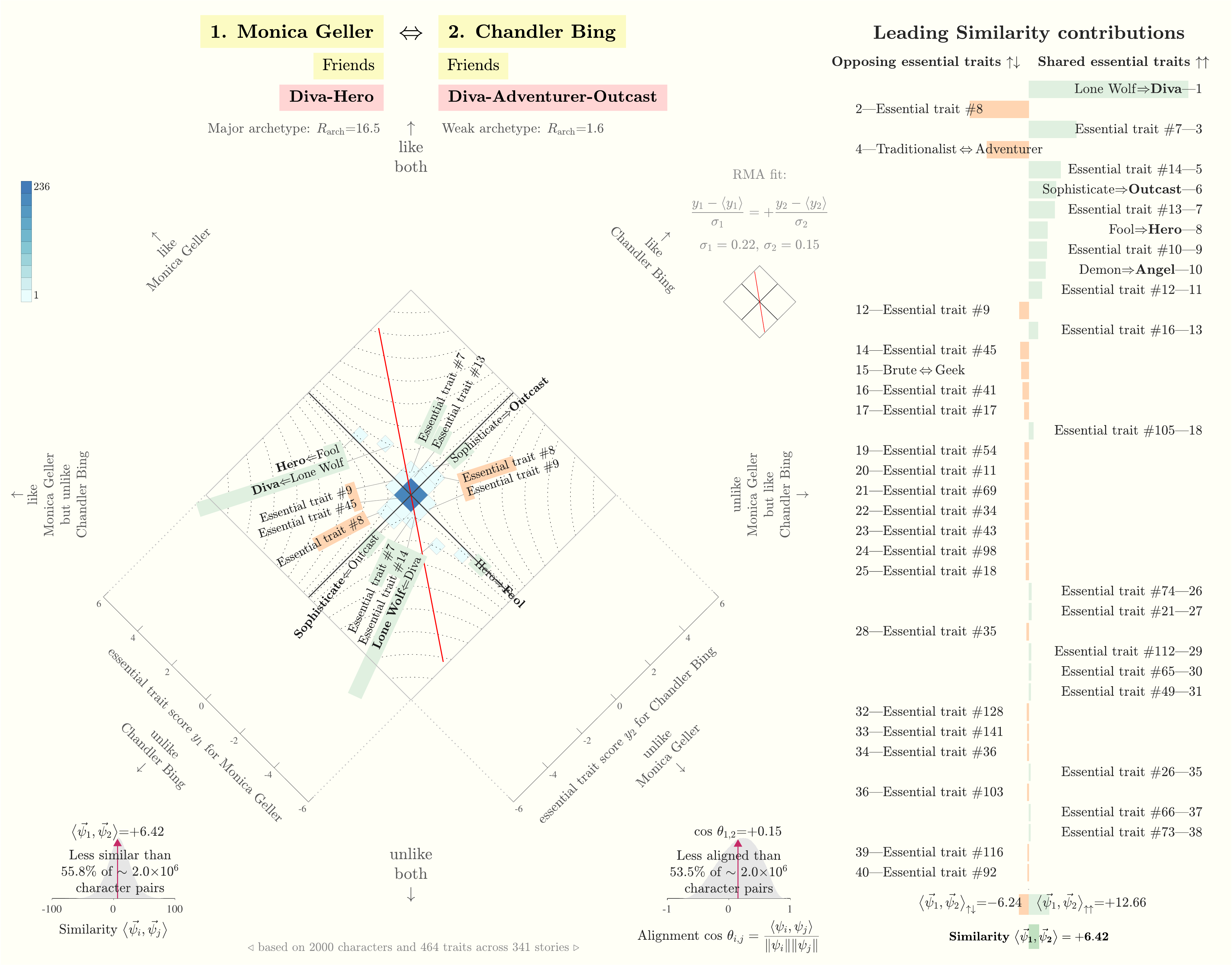}
  \caption{Archetype-based comparison between \characterlinksimple{Friends-Monica-Geller}{Monica Geller} and \characterlinksimple{Friends-Chandler-Bing}{Chandler Bing}.}
  \label{fig:monica_chandler_archetype_app}
\end{figure*}
\clearpage

\begin{figure*}[p]
      \centering
      \includegraphics[width=1.0\linewidth]{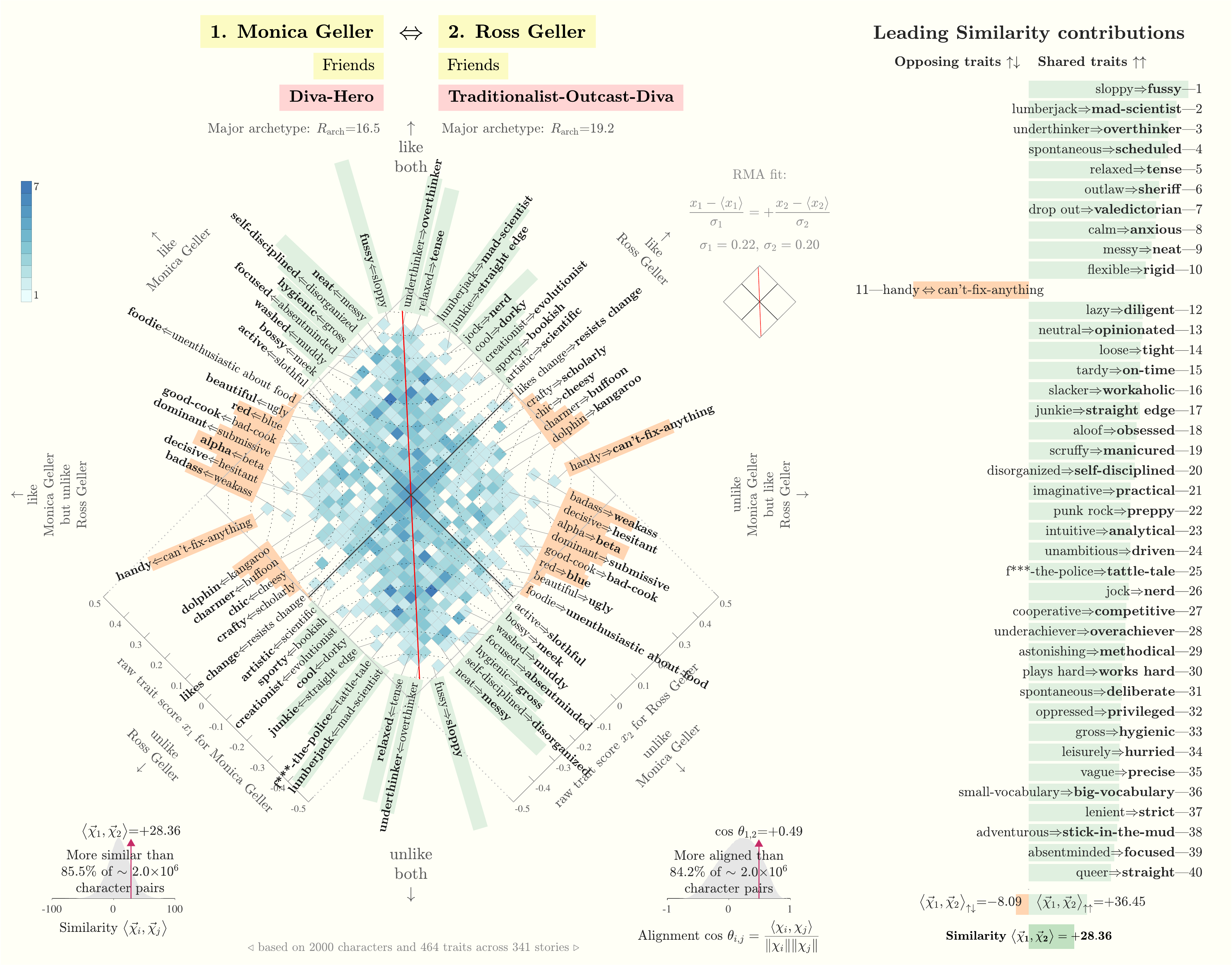}
      \caption{Trait-based comparison between \characterlinksimple{Friends-Monica-Geller}{Monica Geller} and \characterlinksimple{Friends-Ross-Geller}{Ross Geller}.}
      \label{fig:monica_ross_traits}
    \end{figure*}
\begin{figure*}[p]
  \centering
  \includegraphics[width=1.0\linewidth]{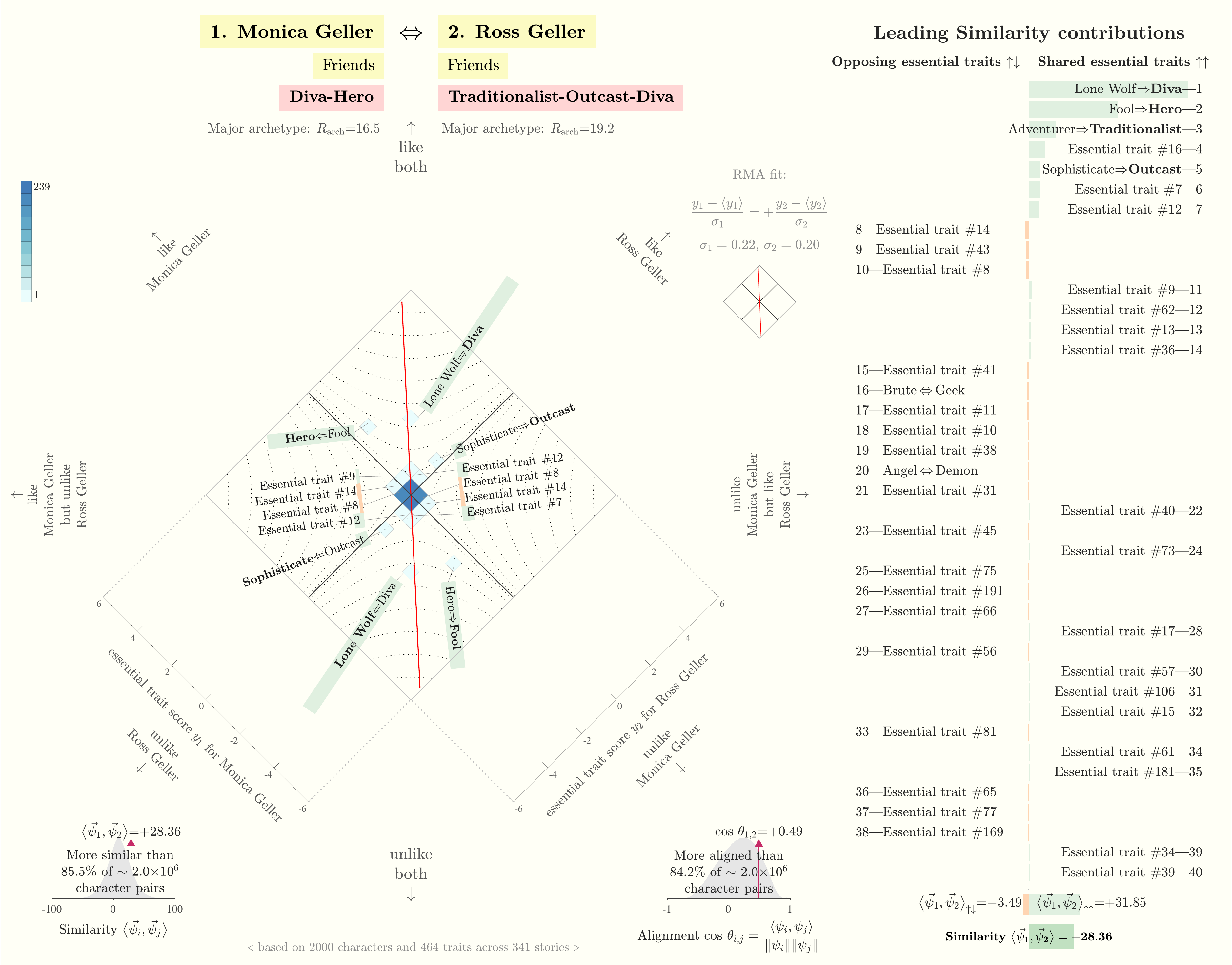}
  \caption{Archetype-based comparison between \characterlinksimple{Friends-Monica-Geller}{Monica Geller} and \characterlinksimple{Friends-Ross-Geller}{Ross Geller}.}
  \label{fig:monica_ross_archetype}
\end{figure*}
\clearpage

\begin{figure*}[p]
      \centering
      \includegraphics[width=1.0\linewidth]{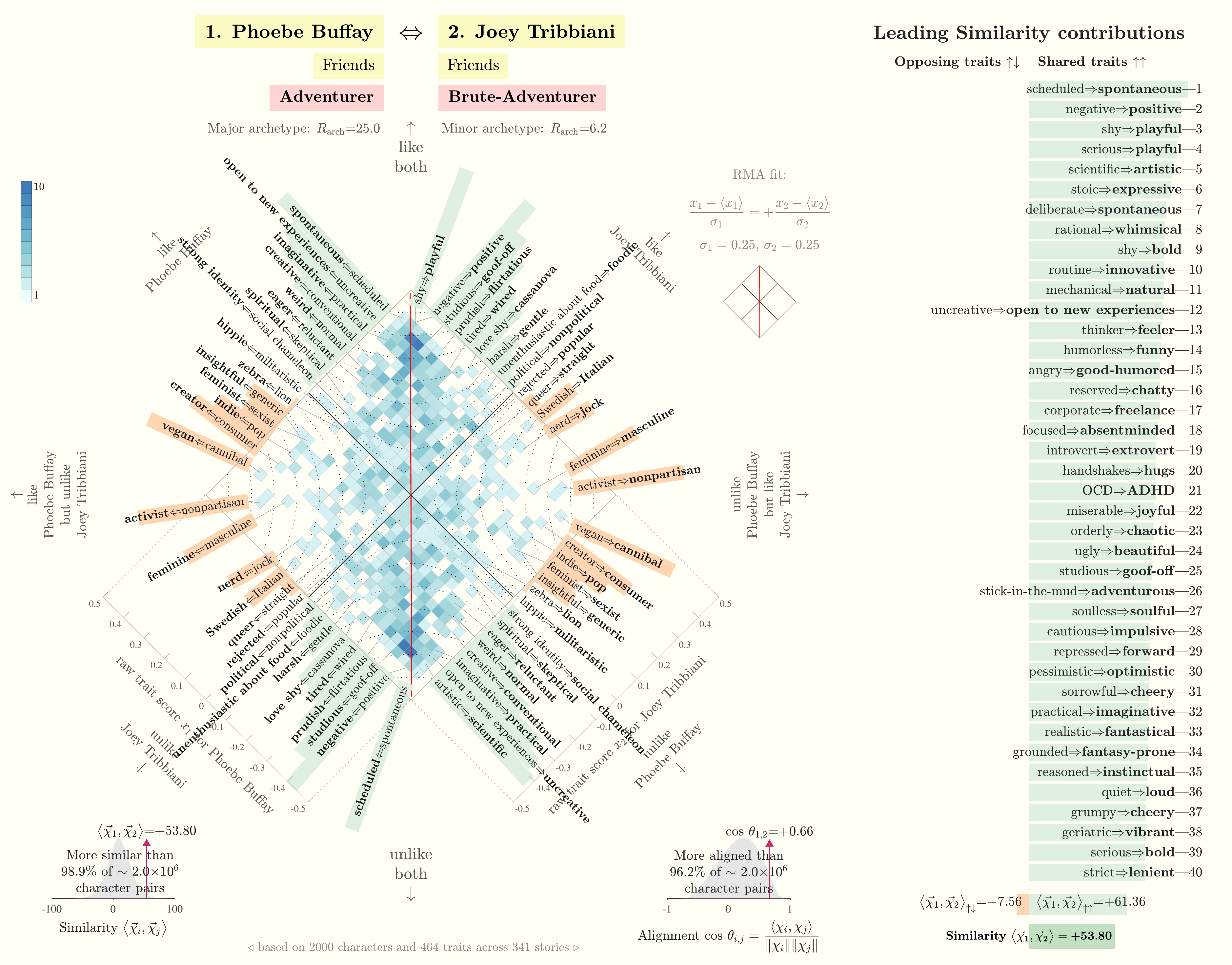}
      \caption{Trait-based comparison between \characterlinksimple{Friends-Phoebe-Buffay}{Phoebe Buffay} and \characterlinksimple{Friends-Joey-Tribbiani}{Joey Tribbiani} in the dominant essential subspace. The radial visualization shows their loadings across the leading essential dimensions. The right-hand panel lists the principal trait-level contributions to similarity. With an inner product of $+53.80$ and cosine alignment of $0.66$, the pair exhibits strong structural alignment within the ensemble.}
      \label{fig:phoebe_joey_traits}
    \end{figure*}
\begin{figure*}[p]
  \centering
  \includegraphics[width=1.0\linewidth]{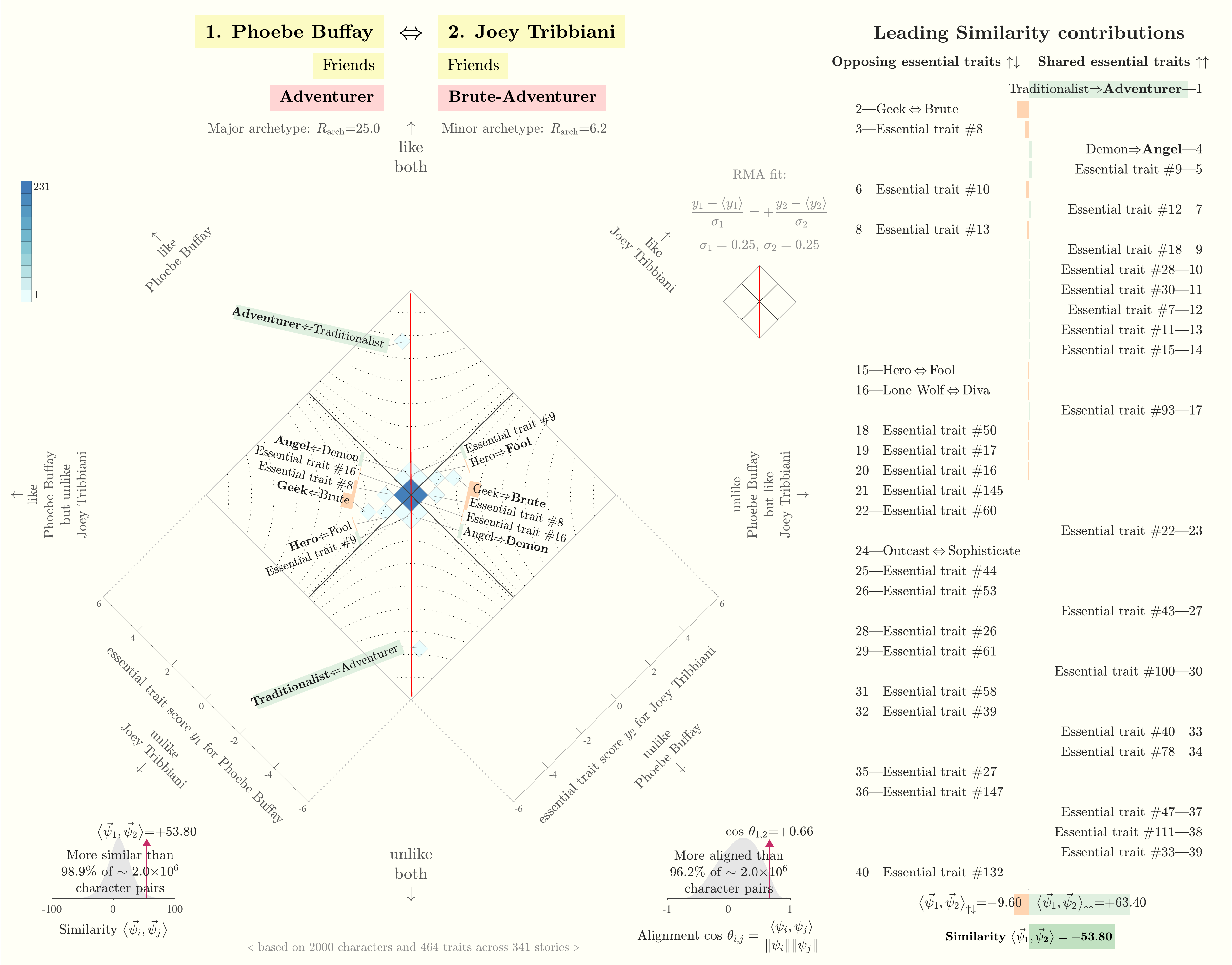}
  \caption{Archetype-based comparison between \characterlinksimple{Friends-Phoebe-Buffay}{Phoebe Buffay} and \characterlinksimple{Friends-Joey-Tribbiani}{Joey Tribbiani}.}
  \label{fig:phoebe_joey_archetype}
\end{figure*}
\clearpage

\begin{figure*}[p]
      \centering
      \includegraphics[width=1.0\linewidth]{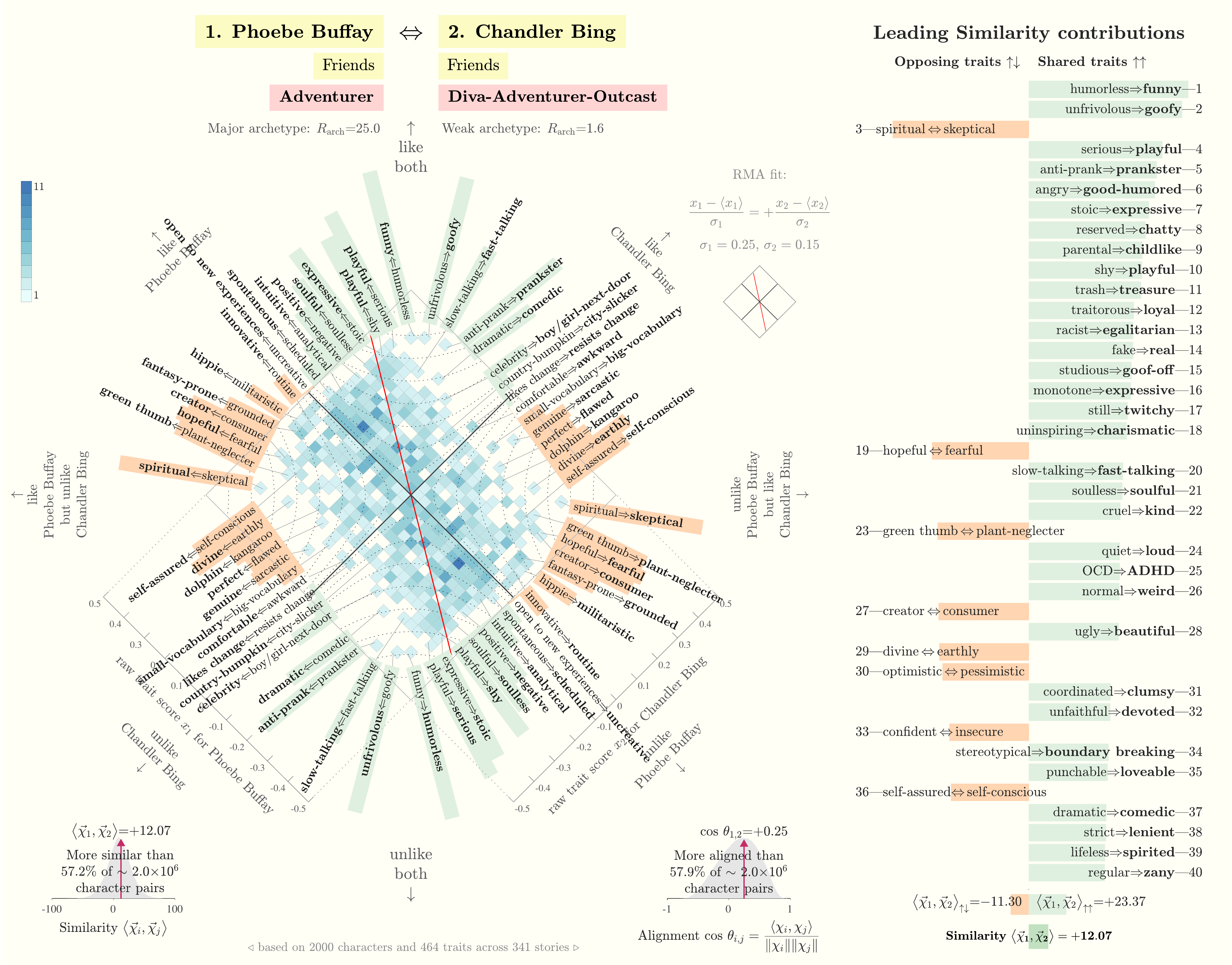}
      \caption{Trait-based comparison between \characterlinksimple{Friends-Phoebe-Buffay}{Phoebe Buffay} and \characterlinksimple{Friends-Chandler-Bing}{Chandler Bing}.}
      \label{fig:phoebe_chandler_traits}
    \end{figure*}
\begin{figure*}[p]
  \centering
  \includegraphics[width=1.0\linewidth]{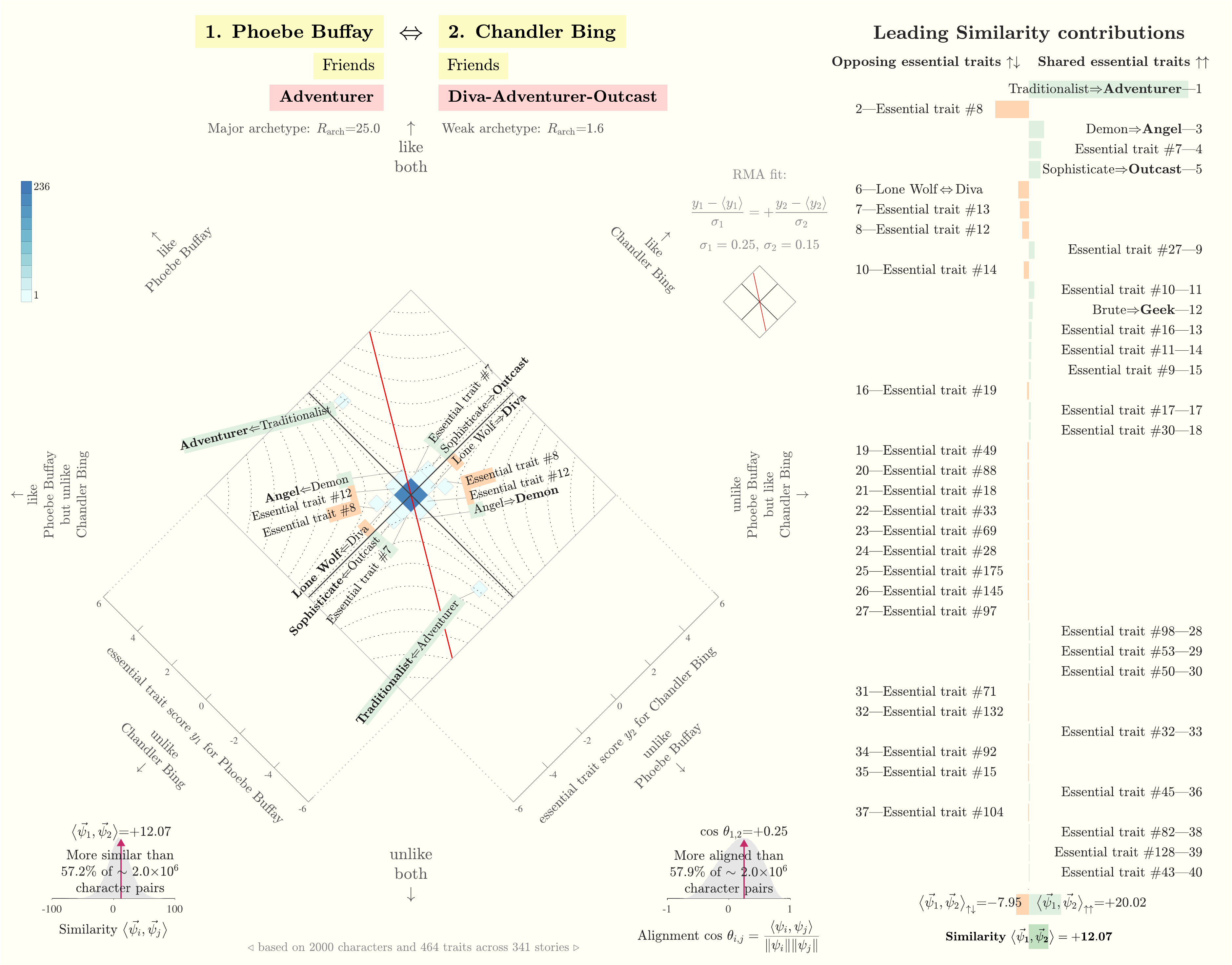}
  \caption{Archetype-based comparison between \characterlinksimple{Friends-Phoebe-Buffay}{Phoebe Buffay} and \characterlinksimple{Friends-Chandler-Bing}{Chandler Bing}.}
  \label{fig:phoebe_chandler_archetype}
\end{figure*}
\clearpage

\begin{figure*}[p]
      \centering
      \includegraphics[width=1.0\linewidth]{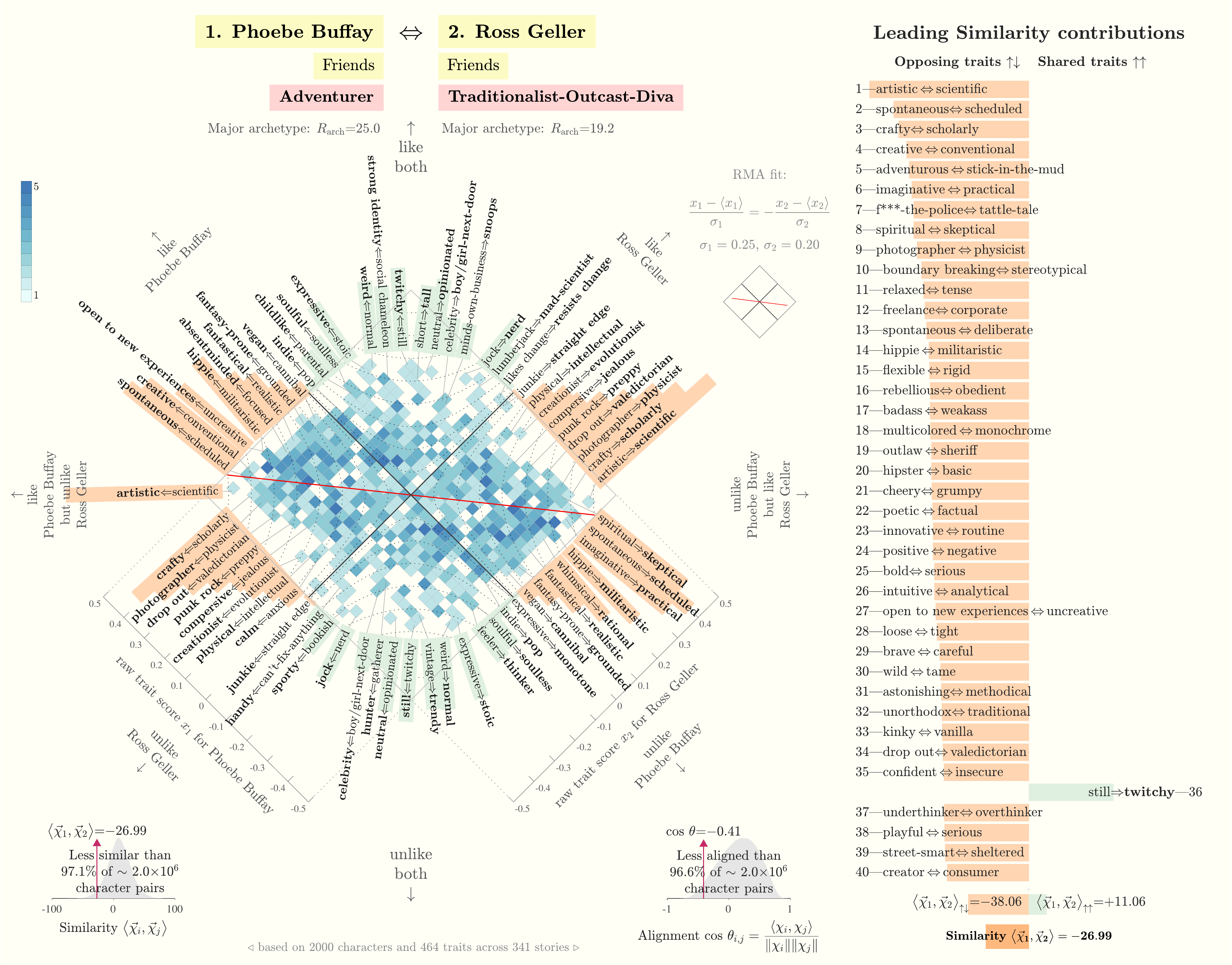}
      \caption{Trait-based comparison between \characterlinksimple{Friends-Phoebe-Buffay}{Phoebe Buffay} and \characterlinksimple{Friends-Ross-Geller}{Ross Geller}. The radial visualization displays their loadings across the dominant essential dimensions, while the right panel lists the leading trait-level contributions to similarity. The pair exhibits a negative inner product of $-26.99$ and cosine alignment of $-0.41$ (which is the cosine of the angle between these two character vectors), indicating structural divergence within the ensemble.}
      \label{fig:phoebe_ross_traits}
    \end{figure*}
\begin{figure*}[p]
  \centering
  \includegraphics[width=1.0\linewidth]{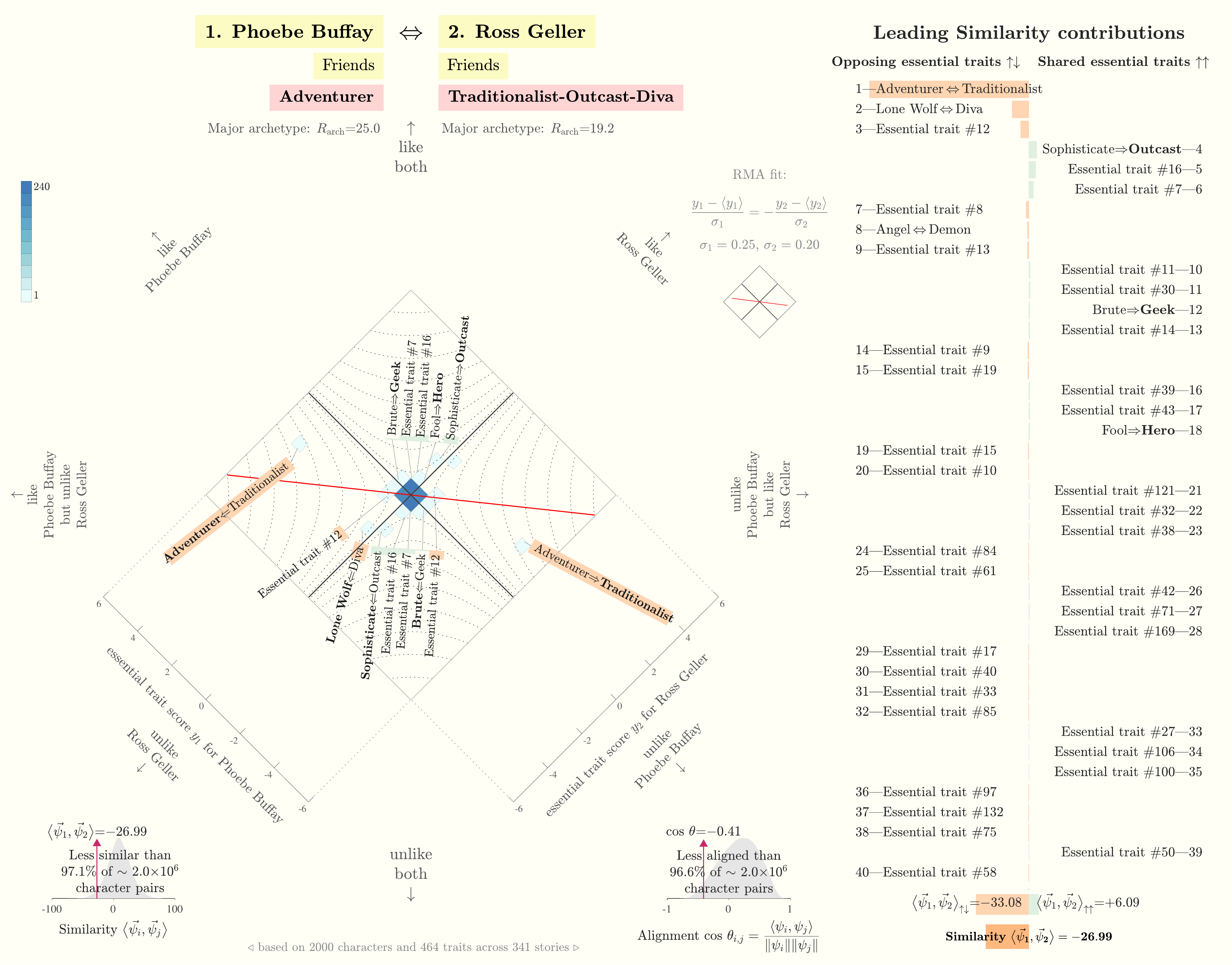}
  \caption{Archetype-based comparison between \characterlinksimple{Friends-Phoebe-Buffay}{Phoebe Buffay} and \characterlinksimple{Friends-Ross-Geller}{Ross Geller}.}
  \label{fig:phoebe_ross_archetype_app}
\end{figure*}
\clearpage

\begin{figure*}[p]
      \centering
      \includegraphics[width=1.0\linewidth]{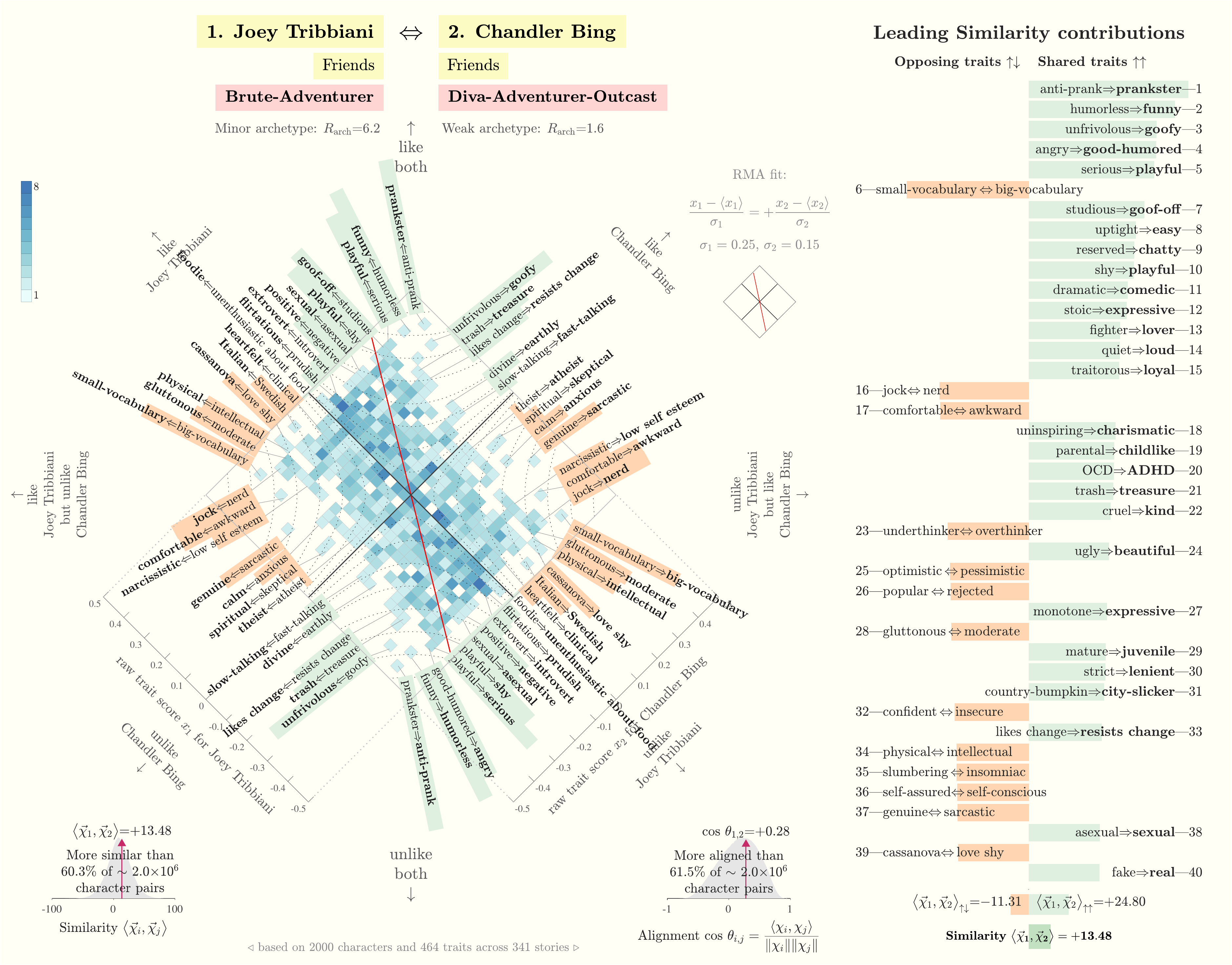}
      \caption{Trait-based comparison between \characterlinksimple{Friends-Joey-Tribbiani}{Joey Tribbiani} and \characterlinksimple{Friends-Chandler-Bing}{Chandler Bing}.}
      \label{fig:joey_chandler_traits}
    \end{figure*}
\begin{figure*}[p]
  \centering
  \includegraphics[width=1.0\linewidth]{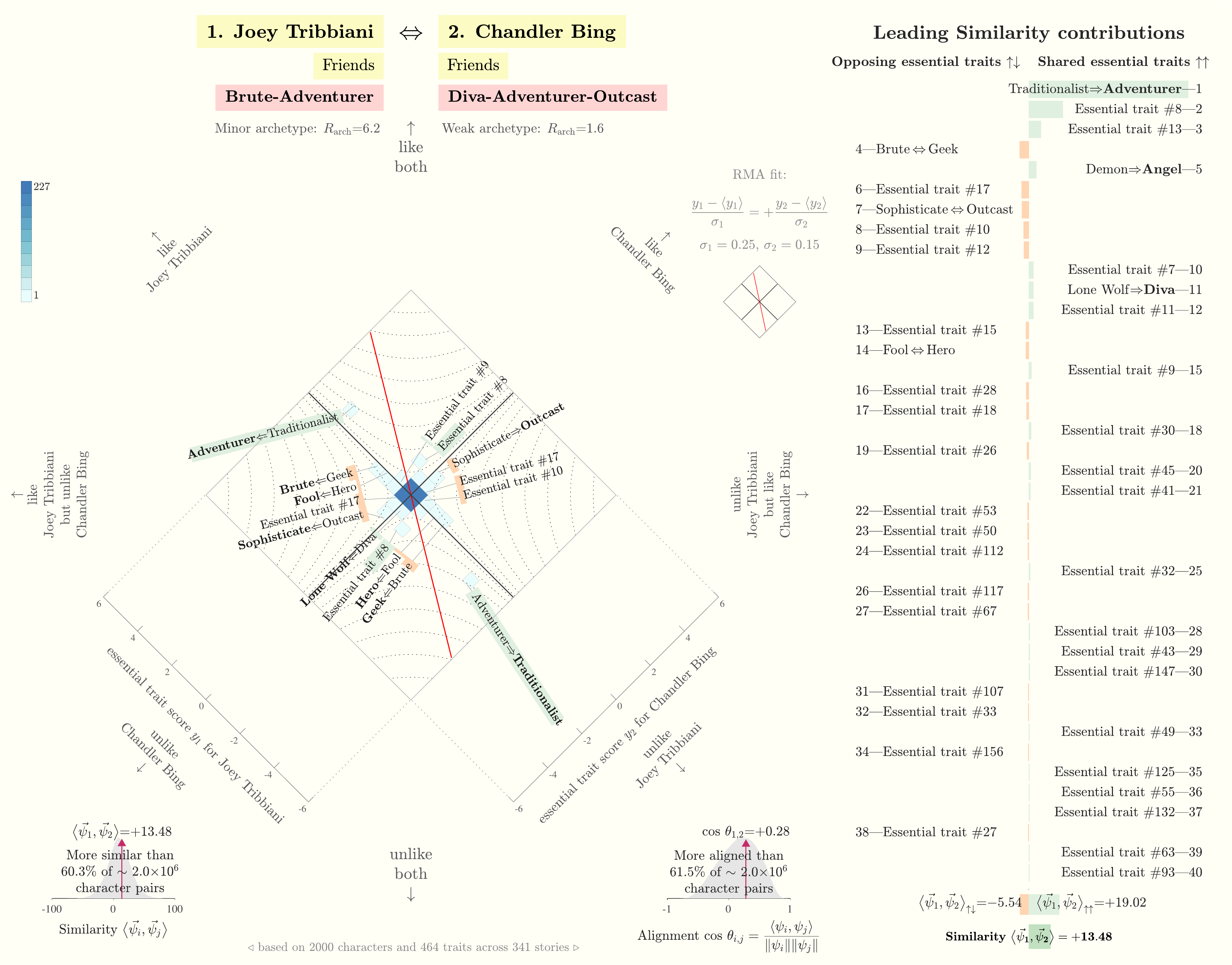}
  \caption{Archetype-based comparison between \characterlinksimple{Friends-Joey-Tribbiani}{Joey Tribbiani} and \characterlinksimple{Friends-Chandler-Bing}{Chandler Bing}.}
  \label{fig:joey_chandler_archetype}
\end{figure*}
\clearpage

\begin{figure*}[p]
      \centering
      \includegraphics[width=1.0\linewidth]{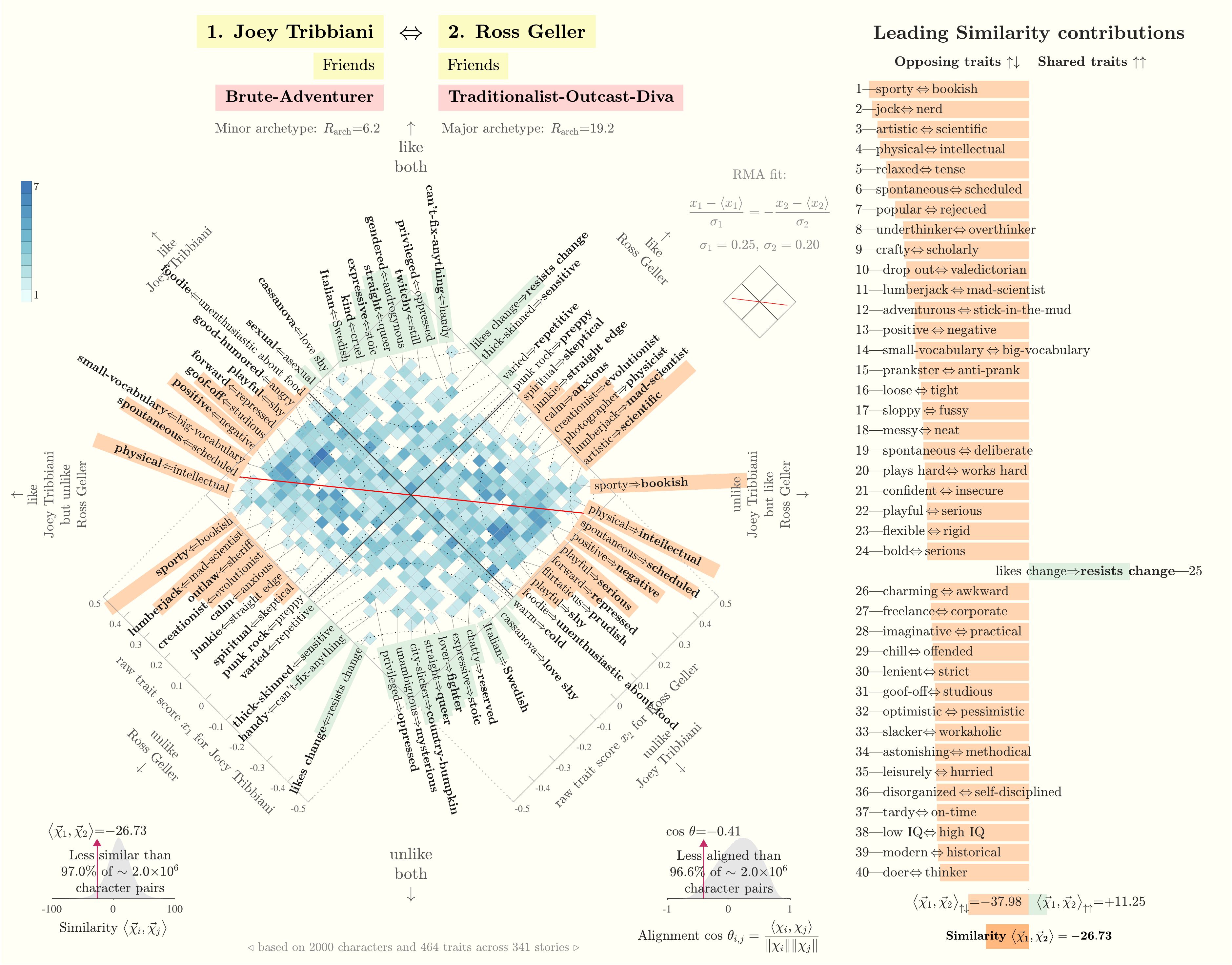}
      \caption{Trait-based comparison between \characterlinksimple{Friends-Joey-Tribbiani}{Joey Tribbiani} and \characterlinksimple{Friends-Ross-Geller}{Ross Geller}.}
      \label{fig:joey_ross_traits}
    \end{figure*}
\begin{figure*}[p]
  \centering
  \includegraphics[width=1.0\linewidth]{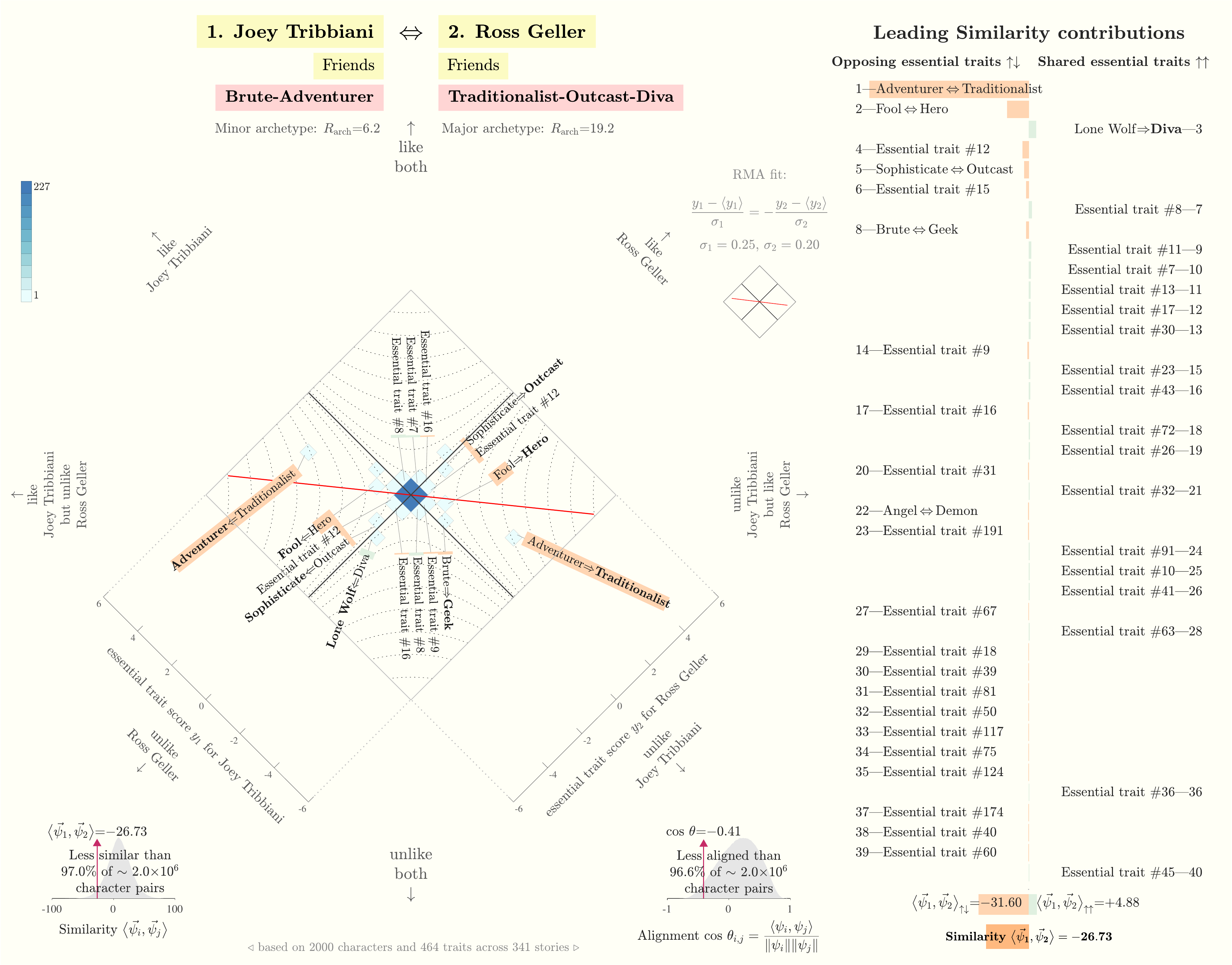}
  \caption{Archetype-based comparison between \characterlinksimple{Friends-Joey-Tribbiani}{Joey Tribbiani} and \characterlinksimple{Friends-Ross-Geller}{Ross Geller}.}
  \label{fig:joey_ross_archetype}
\end{figure*}
\clearpage

\begin{figure*}[p]
      \centering
      \includegraphics[width=1.0\linewidth]{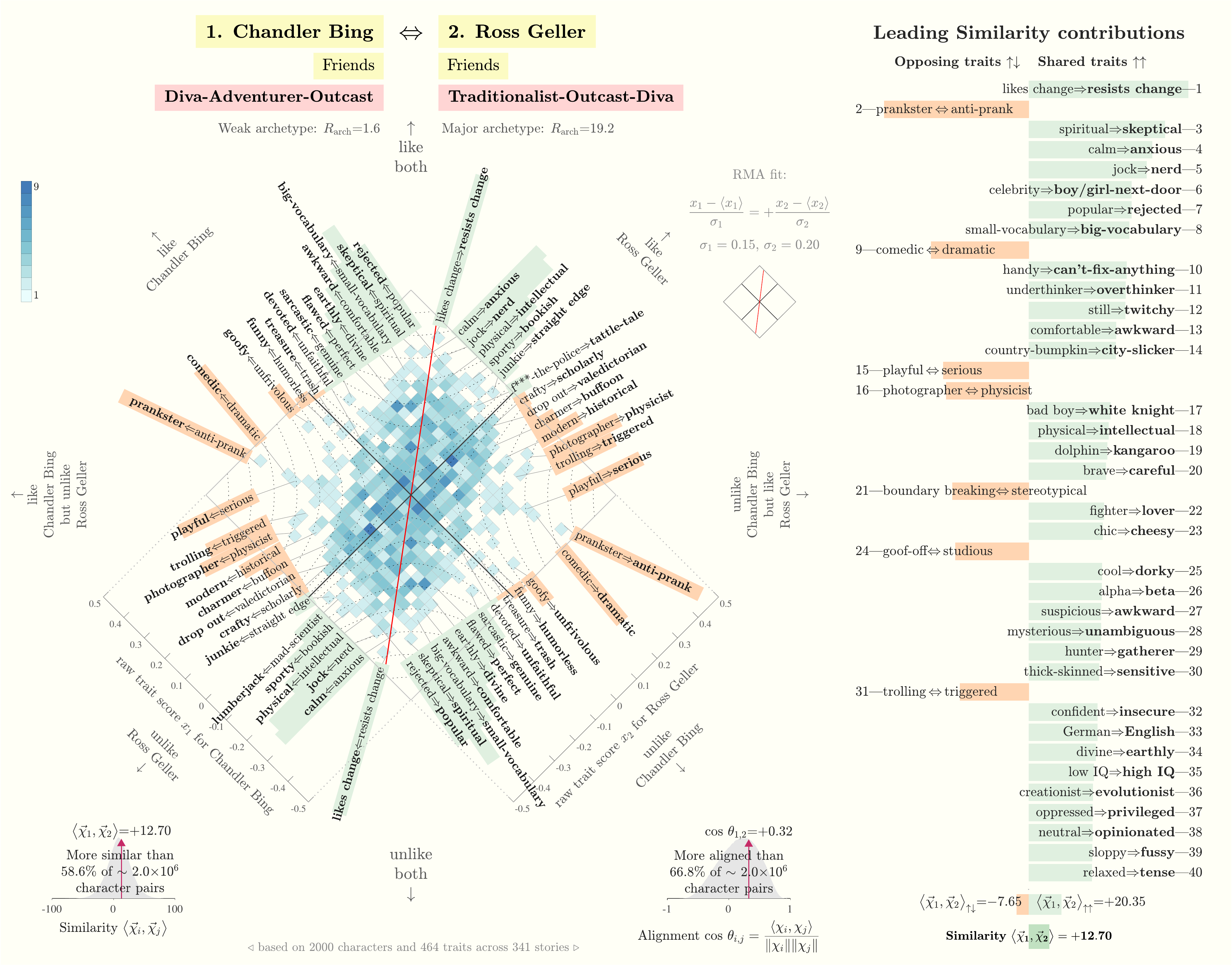}
      \caption{Trait-based comparison between \characterlinksimple{Friends-Chandler-Bing}{Chandler Bing} and \characterlinksimple{Friends-Ross-Geller}{Ross Geller}.}
      \label{fig:chandler_ross_traits}
    \end{figure*}
\begin{figure*}[p]
  \centering
  \includegraphics[width=1.0\linewidth]{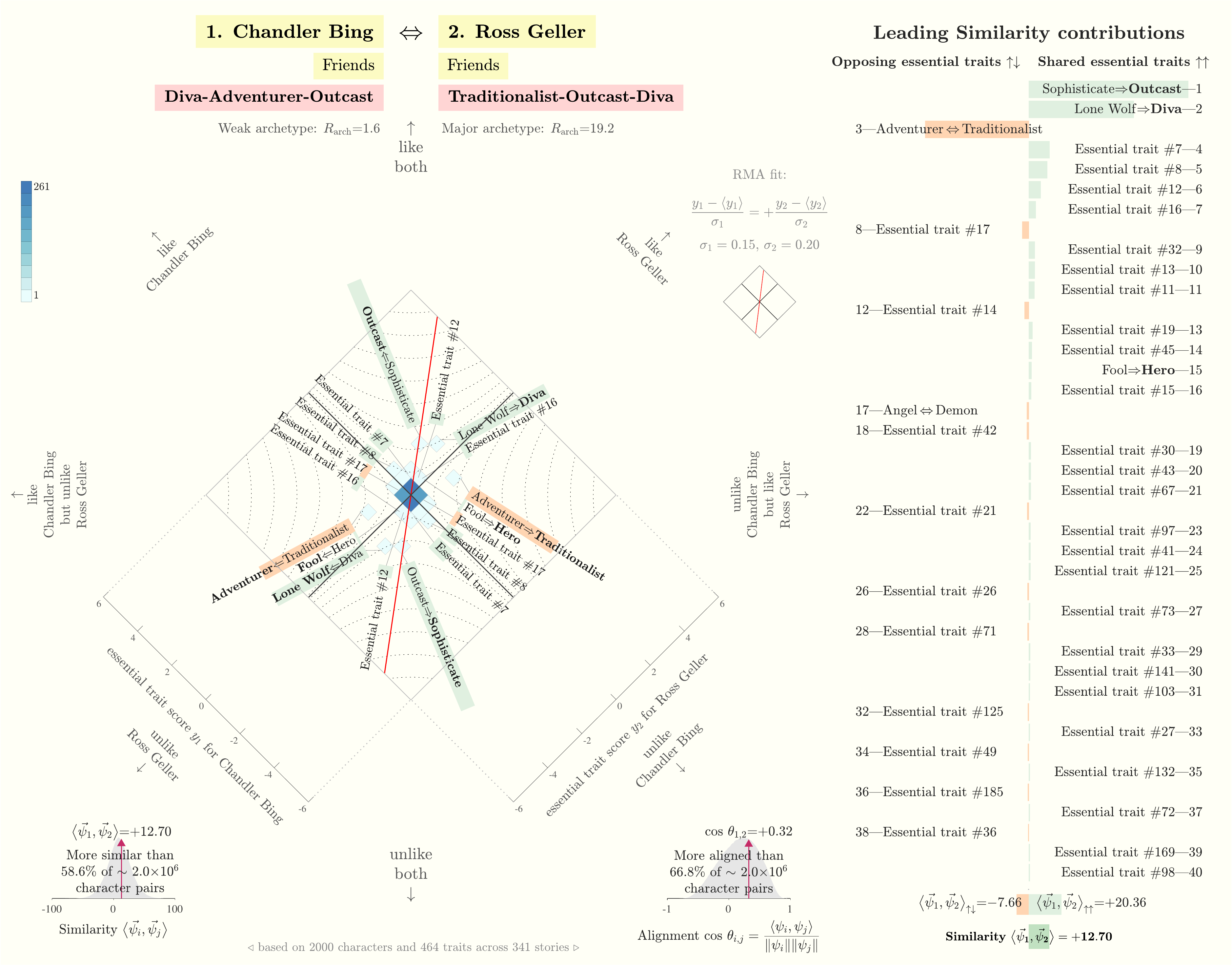}
  \caption{Archetype-based comparison between \characterlinksimple{Friends-Chandler-Bing}{Chandler Bing} and \characterlinksimple{Friends-Ross-Geller}{Ross Geller}.}
  \label{fig:chandler_ross_archetype}
\end{figure*}

\end{document}